\documentclass[a4paper,11pt]{article}
\usepackage{jcappub,bm,color} 
\usepackage{amssymb,amsfonts,slashed,amsthm,amsmath,graphicx, soul}
\usepackage{comment}
\usepackage{booktabs}
\usepackage{multirow}
\usepackage{makecell}
\usepackage{graphicx}
\usepackage{subcaption}
\usepackage{xcolor}
\usepackage{hyperref}
\usepackage{cleveref}

\makeatletter

\newcommand{\Rmnum}[1]{\expandafter\@slowromancap\romannumeral #1@}
\makeatother

\title{Decay and lifetime of oscillons coupled to an external scalar field: Insights from instability band analysis}

\author[a,b]{Siyao Li,}
\author[b,a,c]{Masahide Yamaguchi}
\author[d,e,f,g]{and Ying-li Zhang}

\affiliation[a]{\it Department of Physics, Institute of Science Tokyo,\\2-12-1 Ookayama, Meguro-ku, Tokyo 152-8551, Japan}
\affiliation[b]{\it Cosmology, Gravity and Astroparticle Physics Group, \\Center for Theoretical Physics of the Universe,\\ Institute for Basic Science, Daejeon 34126, Korea}
\affiliation[c]{\it Department of Physics and IPAP, Yonsei University, \\
50 Yonsei-ro, Seodaemun-gu, Seoul 03722, Korea}
\affiliation[d]{\it School of Physics Science and Engineering, Tongji University, Shanghai 200092, China}
\affiliation[e]{\it 
    Institute for Advanced Study of Tongji University, \\
    Shanghai 200092, China
}
\affiliation[f]{\it 
    Kavli Institute for the Physics and Mathematics of the Universe (WPI),\\
    The University of Tokyo Institutes for Advanced Study,
    The University of Tokyo, \\
    Chiba 277-8583, Japan
}
\affiliation[g]{\it 
    Center for Gravitation and Cosmology,
    Yangzhou University, \\
    Yangzhou 225009, China
}

\emailAdd{li.s.ap@m.titech.ac.jp}
\emailAdd{gucci@ibs.re.kr}
\emailAdd{yingli@tongji.edu.cn}
\date{}

\abstract{
Oscillons are long-lived, spherically symmetric solitons that can arise in real scalar field theories with potentials shallower than quadratic ones. 
They are considered to form via parametric resonance during the preheating stage after inflation and have extended lifetimes.
However, the estimation of their lifespan becomes complicated when taking into account the interactions between the inflaton field and other fields, as naturally expected in realistic reheating scenarios.
In this study, we investigate how the lifetime of a single oscillon is affected by the coupling to the external real scalar field.
By numerically computing the instability bands of the external field with the inhomogeneous oscillon profile as background, we show that the resonance behavior depends intricately on the coupling strength and shape of the oscillon. 
We analyze distinct instability mechanisms that dominate across different regimes of the coupling strength and oscillon shapes.
Especially, we show that the parametric resonance fails to occur when the oscillon size is too limited to drive enhancement of the external field.
Furthermore, our simulations show that as the oscillon loses energy, the exponential growth of the external field can terminate before the oscillon reaches its critical energy for collapse, which indicates that the external field does not necessarily lead to rapid destruction of oscillons even in the presence of strong coupling or with large amplitudes. 
These results suggest that oscillons can remain long-lived across a wide range of coupling strengths, with potential implications for their role in cosmological evolution.}

\begin{document}
\maketitle

\section{Introduction}\label{sec:introduction}
Cosmic inflation, driven by one or more scalar fields, offers the leading explanation for key observations such as the homogeneity and isotropy of the universe, as well as the origin of curvature perturbations in the observable universe \cite{guth_inflationary_1981, linde_new_1982, albrecht_cosmology_1982}.
While the slow-roll dynamics of inflation governs the exponential expansion, its post-inflationary fate determines the initial conditions for Big Bang Nucleosynthesis (BBN) \cite{bassett_inflation_2006,allahverdi_reheating_2010,amin_nonperturbative_2015}.
The inflation must eventually come to an end, and the universe must transition into a hot, radiation-dominated era, which is accomplished by a process known as \textit{reheating}. 
Before the final stage of reheating—characterized by thermalization and the production of Standard Model particles, the process is typically preceded by a non-perturbative phase known as \textit{preheating} \cite{kofman_reheating_1994,kofman_towards_1997, shtanov_universe_1995,micha_turbulent_2004}. 
During preheating, the inflaton field undergoes coherent oscillations near the minimum of its potential, leading to explosive particle production via \textit{parametric resonance}. 
This nonlinear phase can result in condensation of the inflaton field, and give rise to highly inhomogeneous, localized, oscillating non-topological solitons known as oscillons in a wide range of inflation models \cite{gleiser_pseudostable_1994, amin_inflaton_2010, amin_oscillons_2012,gleiser_generation_2011,lozanov_end_2014,kawasaki_i-ball_2014,lozanov_self-resonance_2018,antusch_properties_2019, kawasaki_oscillons_2021, piani_preheating_2023, aurrekoetxea_oscillon_2023}.
Besides, oscillons have also been observed in the context of cosmological phase transition \cite{copeland_oscillons_1995, adib_long-lived_2002, farhi_emergence_2008}.

It is well established that oscillons have very extended lifetimes due to an approximate particle number conservation \cite{kasuya_i-balls_2003,mukaida_correspondence_2014,mukaida_longevity_2017, olle_recipes_2021}.
Their slow decay through tiny classical radiation is investigated both in numerical and analytical methods \cite{gleiser_analytical_2008,fodor_computation_2009,fodor_radiation_2009,ibe_decay_2019,zhang_classical_2020,zhang_gravitational_2021,ibe_fragileness_2019}, as well as the quantum radiation \cite{hertzberg_quantum_2010, evslin_quantum_2023}.
This remarkable longevity enables oscillons to have a wide range of cosmological consequences, such as acting as seeds of primordial black holes (PBHs) \cite{cotner_primordial_2018, cotner_analytic_2019, widdicombe_black_2020,  kou_oscillon_2020, nazari_oscillon_2021} and contributing to the formation of non-linear structures in scalar field dark matter \cite{kolb_nonlinear_1994, olle_oscillons_2020, arvanitaki_large-misalignment_2020,kawasaki_oscillon_2020}.
They also play a significant role in enriching the reheating process, with their decay rates directly influencing the reheating temperature, which is crucial for many cosmological scenarios.
Moreover, recent studies have revealed that substantial amounts of gravitational waves can be produced both during the formation of oscillons at preheating and through their subsequent decay \cite{zhou_gravitational_2013, antusch_gravitational_2017,amin_gravitational_2018, lozanov_gravitational_2019, hiramatsu_gravitational_2021, lozanov_enhanced_2023,lozanov_universal_2025}.

Since reheating requires the inflaton field to couple to other degrees of freedom, such as scalars and fermions, it is natural to explore oscillon formation and evolution in the presence of such couplings.
Oscillons with external coupling have been found and discussed in the literature in various contexts \cite{farhi_oscillon_2005, graham_electroweak_2007, hertzberg_quantum_2010, gleiser_generation_2011, kawasaki_decay_2014, van_dissel_symmetric_2022, cyncynates_nonperturbative_2022, piani_ephemeral_2025}.
In particular, four-point interactions of the form $g \phi^2 \chi^2$ between the inflaton and other scalar fields are frequently considered, motivated by multi-field inflationary models such as chaotic inflation \cite{linde_chaotic_1983, dimopoulos_n-flation_2008}, hybrid inflation \cite{linde_hybrid_1994, garcia-bellido_preheating_1998}, string theory inflation \cite{baumann_inflation_2015}, etc.
For instance, lattice simulation in Ref.~\cite{gleiser_generation_2011} has demonstrated that two-field oscillons can form in hybrid inflation models, which are found to have lifetimes up to five times longer than their single-field counterparts.
Ref.~\cite{van_dissel_symmetric_2022} has also investigated multi-field oscillons, in which the external scalar fields have identical self-couplings to $\phi$, finding that the lifetimes of the composite oscillons remain comparable with that of the single-field ones, albeit slightly reduced under attractive external couplings and  slightly extended under repulsive ones.

On the other hand, external couplings open additional decay channels for oscillon via particle production into the external field, through both perturbative decay and parametric resonance.
It is pointed out in Ref.~\cite{hertzberg_quantum_2010} that the product particles escape away instead of forming composite oscillons when the spectator field lacks appropriate self-interaction.
A modified Floquet analysis has been developed there to compute the Floquet exponents of the external scalar field with inhomogeneous background of individual oscillon profiles, which dominates the decay rate of oscillon through this channel.
Later, Ref.~\cite{kawasaki_decay_2014} has applied this approach to compute the Floquet exponents  for a scalar field tri-linearly coupled to oscillons with Gaussian-type profiles in a monomial potential model.
More recently, Ref.~\cite{shafi_formation_2024} has performed a (3+1)-dimensional lattice simulation of oscillon formation in an expanding universe, focusing on E- and T-model $\alpha$-attractor potentials.
Their results demonstrate the oscillons can form in suitable regions of parameter space and subsequently lose a fraction—or even all—of their energy through the external resonance at late times after the formation.
In addition, they report an inverse power-law dependence of the lifetime of the oscillon population on the strength of the external coupling.

Motivated by these findings, we aim to gain a more detailed understanding of how external couplings affect oscillon decay and to interpret the above lattice results by analyzing the external decay rate of individual oscillons in more detail.
In this work, we consider a quartic interaction, $g \phi^2 \chi^2$, between an external scalar field $\chi$ without nonlinear self-interactions and oscillon field $\phi$ in sextic polynomial potential. 
We perform simulations with spherically symmetric configurations to compute the growth rate of the $\chi$ field within its instability bands, where it undergoes exponential growth, on the spatially inhomogeneous oscillon profile.
We analyze how the growth rate of the $\chi$ field depends on the coupling constant $g$ and the shape of oscillon profile.
In addition to extracting a particle escape rate from the Floquet exponents obtained in the homogeneous background case, which is consistent with the conclusions in Ref.~\cite{hertzberg_quantum_2010, kawasaki_decay_2014}, we discover an indirect resonance arising with a very small growth rate due to mode mixing caused by the inhomogeneous oscillon profiles under attractive interaction, $g<0$, which has not been found to our best knowledge.
Finally, we perform full simulations of the coupled evolution of both the oscillon and the external field.
Our results confirm that oscillons are not necessarily destroyed entirely by the external resonance, which is well explained by our analysis of instability bands and is consistent with the results presented  in Ref.~\cite{shafi_formation_2024}.

The rest of the paper is organized as follows.
In section \ref{sec:single-field oscillon}, we revisit the single-field oscillons in a sextic polynomial potential.
In section \ref{sec:instability bands}, we perform a simulation under spherical symmetry to compute the instability bands of the external scalar field $\chi$, coupled to oscillon through $g\phi^2 \chi^2$, and compare them with the Floquet chart in a homogeneous background.
In section \ref{sec:two-field simulation}, we perform a full simulation of two fields and present the results of oscillon decay and its consequence.
In section \ref{sec:conclusion}, we summarize and conclude our work.

\section{Oscillons in single scalar field model}\label{sec:single-field oscillon}
The model we consider in this paper is the following,
\begin{align}
\begin{split}
    \mathcal{L} = &\frac{1}{2}\partial_\mu \phi \partial^\mu \phi -V(\phi) + \frac{1}{2}\partial_\mu \chi \partial^\mu \chi - \mathcal{V}(\chi) - g \phi^2 \chi^2,\label{eq:full lagrangian}\\
    V(\phi) &=  \frac{1}{2} m^2 \phi^2 + V_{\textrm{nl}}(\phi),~~\mathcal{V}(\chi) = \frac{1}{2} m_\chi^2 \chi^2.
\end{split}
\end{align}
Here both $\phi$ and $\chi$ are real scalar fields.
The equations of motion derived from the Lagrangian in the Minkowski background are
\begin{align}
    \Ddot{\phi} - \nabla^2 \phi + m^2\phi +\frac{d V_{\textrm{nl}}}{d \phi} + 2g \chi^2\phi= 0,\label{eq:full phi eom}\\
    \Ddot{\chi} - \nabla^2 \chi + m_\chi^2 \chi + 2g\phi^2\chi= 0.\label{eq:full chi eom}
\end{align}
In this work, we consider that oscillons can form in the $\phi$ sector since $V(\phi)$ contains self-interaction terms, while $\chi$ sector does not support soliton solutions due to the absence of non-linear terms when the coupling $g$ is turned off.
Motivated by the fact that preheating after inflation can lead to oscillon formation, $V(\phi)$ can take other shapes preferred by inflation models with a global minimum at $\phi = 0$.
We are taking the following polynomial potential as an example for this study,
\begin{align}
     V_{\textrm{nl}}(\phi) = - \lambda \phi^4 + g_6\phi^6,\label{eq:phi6 potential}
\end{align}
where $\lambda, g_6 >0$, and there should be $V(\phi) \geq 0$ to ensure the existence of a unique vacuum at $\phi = 0$.

\subsection{Oscillon solutions from Q-ball analogy and adiabatic condition}\label{subsec:approximate conservation and adiabatic condition}

Before turning on the external coupling to $\chi$, we would like to review some important properties that have been well studied in the single-field model, i.e., when $g=0$.

Oscillons are considered as localized scalar field condensation under spherical symmetry.
To understand the solutions and stability of oscillons, it is useful to begin with an analogy to the Q-ball, which is a spherically symmetric and localized soliton solution in a complex scalar field model with $U(1)$ symmetry.
The stability of a Q-ball is associated with the conserved Noether charge, corresponding to the $U(1)$ symmetry of a complex scalar field, $\Phi$, 
\begin{align}
    Q= i \int d^3 x (\dot{\Phi}\Phi^* - \Phi\dot{\Phi^*} ).
\end{align}
Given a charge $Q_0$, a Q-ball solution should be the configuration minimizing the total energy. 
By introducing a Lagrange multiplier $\omega$, the problem becomes finding $\Phi$ and the value of $\omega$ that minimize the following energy,
\begin{equation}
    \begin{aligned}
        E_\omega = E + \omega(Q_0 - Q)= \int d^3x [|\dot{\Phi} + i\omega\Phi|^2 + |\nabla \Phi|^2 + V(|\Phi|) - \omega^2 |\Phi|^2] +\omega Q_0. \label{eq:Lagrange multiplier}
    \end{aligned}
\end{equation}
The solution minimizing the first term in Eq.~\eqref{eq:Lagrange multiplier} should be in the form of 
\begin{align}
    \Phi(t,r) = \frac{1}{\sqrt{2}}e^{-i\omega t}\psi(r),
\end{align}
where $\psi(r)$ is a real field described by the following equation in spherical coordinates derived by the variational of $E_\omega$,
\begin{align}
    \frac{d^2\psi}{dr^2} + \frac{2}{r} \frac{d\psi}{dr} - \left[ ( m^2-\omega^2) \psi +\frac{\partial V_{\textrm{nl}}(\psi)}{\partial \psi}\right] = 0. \label{eq:qball spherical eom}
\end{align}
The boundary condition for a Q-ball solution should require the regularity at the origin and vacuum outside the Q-ball, 
\begin{equation}
\label{eq:qball bc}
    \frac{d\psi(r)}{dr} \bigg|_{r=0} =  0,~~
    \psi(r\to \infty) \to 0.
\end{equation}
This is clearly a boundary value problem, which can be simply solved by shooting method.
However, since we are looking for a spatially localized solution which should approach zero faster than $1/r$ to achieve a finite energy value by the integral in Eq.~\eqref{eq:Lagrange multiplier}, there are conditions for $\omega$ value to allow such kind of solutions.
If we replace the spatial coordinate $r$ by time $t$, field $\psi$ by position $x$, Eq.~\eqref{eq:qball spherical eom} and \eqref{eq:qball bc} now describe a particle motion starting from stationary and stopping at the origin with a time-dependent friction and potential $V_\omega(x) = - \left[ \frac{1}{2}( m^2-\omega^2) x^2 +V_{\textrm{nl}}(x)\right]$.
The necessary conditions on $\omega$ value for the existence of such a bounce solution, i.e., a localized solution for profile $\psi(r)$ are
\begin{align}
    \min \left[\frac{2V(\psi)}{\psi^2} \right] < \omega^2 < m^2.\label{eq:localized potential condition}
\end{align}
The condition $\omega^2 < m^2$ can be understood in the following way: at $r\to \infty$, $\psi \to 0$ and $V(\psi) \to \frac{1}{2} m^2 \psi^2$ where the potential is dominated by the quadratic term, thus $V_\omega(\psi) \to \frac{1}{2} (\omega^2 - m^2)\psi^2$.
Then the Eq.~\eqref{eq:qball spherical eom} is asymptotically $\partial_r^2 \psi + 2\partial_r \psi /r +(\omega^2 - m^2) \psi = 0$ and has solution $\psi(r \to \infty) \sim e^{\pm i \sqrt{\omega^2 -m^2} r} /r$.
This solution decays too slowly to be localized because it will give an infinite energy, which is obviously not desired for solitons.

For a $\omega$ within the above range, we can obtain $\psi(r)$ by solving the equation of Eq.~\eqref{eq:qball spherical eom} using the shooting method.
Then the corresponding charge $Q$ and energy $E$ can be computed as 
\begin{align}
    Q &= 4\pi \omega \int dr~ r^2\psi^2(r),
    ~~E = 4\pi \int dr~ r^2 \left[ \frac{1}{2} \left( \frac{d \psi}{dr}\right)^2 + V(\psi) + \frac{1}{2} \omega^2 \psi^2 \right].\label{eq:spherical Q-ball Q and E}
\end{align}

In contrast to Q-balls, oscillons in real scalar field theories lack a $U(1)$ symmetry and hence do not possess a conserved Noether charge.
But it has been shown that an approximate $U(1)$ symmetry can be identified under non-relativistic condition, when the real scalar field $\phi$ corresponds to the real part of a complex scalar field $\Phi$, i.e. $\phi = \Re[\Phi]$ \cite{mukaida_correspondence_2014, mukaida_longevity_2017, blaschke_oscillons_2025, blaschke_oscillons_2025-1}.
The non-relativistic condition is satisfied when the potential $V(\phi)$ is dominated by the mass term, 
\begin{align}
     \frac{d^2 V_{\textrm{nl}}(\phi)}{d \phi^2} \ll m^2,\label{eq:adiabatic poptential condition}
\end{align}
so that the complex field can be decomposed as 
\begin{align}
    \Phi(t,\mathbf{x}) = e^{-i\omega t} \Psi(t,\mathbf{x}) + \delta\Phi(t,\mathbf{x}), ~~\omega\simeq m
\end{align}
Here $\omega$ is a frequency smaller than the mass $m$, whose deviation from $m$ results from the gradient energy of inhomogeneous $\Psi$. 
Meanwhile, the non-relativistic limit indicates that the gradient energy of $\Psi$ is subdominant compared to the kinetic energy of the oscillation, so that the $\omega$ is a frequency very close to $m$ with $0<m-\omega \ll m$.
Additionally, the kinetic energy of $\Psi$ should also be subdominant, which means it can only vary with respect to a timescale much longer than the period of the leading oscillation.
$\delta \Phi$ contains other rapid oscillation modes with oscillation timescale being much faster than $m^{-1}$, which should also be small.
We note that, despite being assumed as small, $\delta \Phi$ is an important difference of this complex field solution from the Q-ball solution.
It appears as a result of the approximation of the $U(1)$ symmetry here in contrast to the exact $U(1)$ symmetry of Q-ball.
We will see in the next section that these rapid oscillation modes actually include propagating modes emitting energy from the localized oscillons and result in the slow time dependence of the profile 2$\psi(t,r) = \Re{[\Psi(t, r)]}$ (also the frequency $\omega$), which makes oscillons decay spontaneously.
Therefore, an oscillon solution can be approximated as the real part of the Q-ball solution in a form of 
\begin{align}
    \phi(t,r) \simeq 2\psi(t,r) \cos{(\omega t)} + \xi(t,r),\label{eq:full phi with oscillon and perturbation}
\end{align}
the factor of 2 is taken for future convenience.
The assumption that the amplitude of higher modes is subdominant can be justified by numerical Fourier transformation of oscillons \cite{salmi_radiation_2012}.
In order to look for an oscillon solution in a short timescale comparable with the period  , we can neglect the slow time dependence of the profile $\psi(t, r)$ and assume 
\begin{align}
    \phi(t,r) \simeq \phi_{osc} + \xi (t,r),~~~\phi_{osc} \approx 2\psi(r) \cos{(\omega t)}.\label{eq:single frequency ansatz}
\end{align}
This is a good approximation for oscillons when the potential $V(\phi)$ respects the conditions in Eq.~\eqref{eq:localized potential condition} and Eq.~\eqref{eq:adiabatic poptential condition}.
These inequalities depict a potential 
dominated by the quadratic term but becomes shallower than a purely quadratic potential within a certain range.
For the sextic potential in Eq.~\eqref{eq:phi6 potential}, this requires $\lambda>0, g_6>0$.
Though it becomes steeper than the quadratic at large $\phi$ where $\phi^6$ term is dominant, there should always be a range near the origin satisfying the non-relativistic condition, where we can find oscillon solutions with amplitudes smaller than $\sqrt{\frac{\lambda}{g_6}}$.

Under the non-relativistic condition (in other words, adiabatic condition), it is possible to define a conserved quantity corresponding to the approximate $U(1)$ symmetry as \cite{kasuya_i-balls_2003}
\begin{align}
    I \equiv \frac{1}{\omega}\int d^3x \overline{\dot{\phi}^2},\label{eq:adiabatic charge}
\end{align}
where the overline denotes taking the time average for a period of oscillation, defined as $\overline{y(t)} \equiv \frac{1}{T}\int_0^Tdt~y(t)$.
Substituting the spherical symmetric oscillon ansatz in Eq.~\eqref{eq:single frequency ansatz}, we obtain
\begin{align}
    I = 8\pi\omega \int dr ~r^2 \psi^2(r),\label{eq:spherical adiabatic charge}
\end{align}
which is exactly half of the Noether charge of Q-ball in Eq.~\eqref{eq:spherical Q-ball Q and E}(the factor is 4 times enlarged due to the factor of 2 in the ansatz), while the other half should be restored in the imaginary part of a Q-ball.
So we can consider this quantity as the charge of oscillons.

The time-averaged energy of an oscillon can be computed from the profile by
\begin{align}
    \overline{E} &= \int d^3x \left[ \frac{1}{2} \overline{\dot{\phi^2}} + \frac{1}{2} \overline{(\nabla\phi)^2} + \overline{V(\phi)} \right] \nonumber\\
    &=4\pi \int dr~r^2 \left[\omega^2 \psi^2 + \left( \frac{\partial\psi}{\partial r} \right)^2 + m^2 \psi^2 + V_{\textrm{eff}}(\psi) \right],
    ~~V_{\textrm{eff}}(\psi) \equiv \overline{V_{\textrm{nl}}(\phi)}.\label{eq:time-average energy from oscillon profile}
\end{align}

\subsection{Oscillon profiles} \label{subsec:single field oscillon profile}
In the single-field model, the equation of motion under spherical symmetry is
\begin{align}
    \Ddot{\phi} - \frac{\partial^2 \phi}{\partial r^2} - \frac{2}{r}\frac{\partial \phi}{\partial r} +m^2 \phi +\frac{d V_{\textrm{nl}}}{d \phi} = 0.\label{eq:spherical eom of phi}
\end{align}
From this section, we take the sextic potential in Eq.~\eqref{eq:phi6 potential} to illustrate the specific oscillon profiles and computations.

Substituting the ansatz in Eq.~\eqref{eq:single frequency ansatz} into the equation of motion in Eq.~\eqref{eq:spherical eom of phi} after multiplying by a $\cos(\omega t)$ and taking the time average over one period, we can derive the equation for the oscillon profile $\psi(r)$,
\begin{align}
    \frac{d^2\psi}{dr^2} + \frac{2}{r} \frac{d\psi}{dr} - \left[ ( m^2-\omega^2) \psi +\frac{1}{2}\frac{\partial V_{\textrm{eff}}(\psi)}{\partial \psi}\right] = 0,\label{eq:eom oscillon profile}\\
    V_{\textrm{eff}}(\psi) = \overline{V_{\textrm{nl}}(\phi)} = - 6\lambda \psi^4 + 20g_6\psi^6, \label{eq:Veff phi6}
\end{align}
where we use $\overline{\cos^4{(\omega t )}} = 3/8$, $\overline{\cos^6{(\omega t )}} = 5/16$.
This equation can be derived by Lagrange multiplier method utilizing conserved charge defined in Eq.~\eqref{eq:adiabatic charge} to the time-averaged energy given by Eq.~\eqref{eq:time-average energy from oscillon profile} as well.
With the boundary conditions applied for regularity at the origin and vacuum outside oscillon
\begin{equation}
    \frac{d\psi(r)}{dr} \bigg|_{r=0} =  0,~~
    \psi(r\to \infty) \to 0,\label{eq:oscillon bc}
\end{equation}
we arrive at a shooting problem similar to that of the Q-ball profile in Eq.~\eqref{eq:qball spherical eom} with just a different potential.
So the existence of a localized solution should require the same condition of $\omega$ as Eq.~\eqref{eq:localized potential condition} for a certain $\lambda$ and $g_6$.
Though the oscillon profiles in $1+1$ dimensions can be solved analytically order by order with small amplitude expansion \cite{amin_flat-top_2010}, an analytical solution is difficult to obtain in $3+1$ dimensions.
Given a value of $\omega$ within the allowed range, we can solve the boundary value problem numerically.
We note that the boundary value problem may have more than one solution, as in the case of the the ``excited" oscillons found in Ref.~\cite{van_dissel_oscillon_2023}, but in this work we are only interested in the 'fundamental' oscillons with nodeless profiles, which can be uniquely found for a given $\omega$ value.

\begin{figure}[t]
    \centering
    \begin{subfigure}[b]{0.48\textwidth}
        \centering        \includegraphics[width=\textwidth]{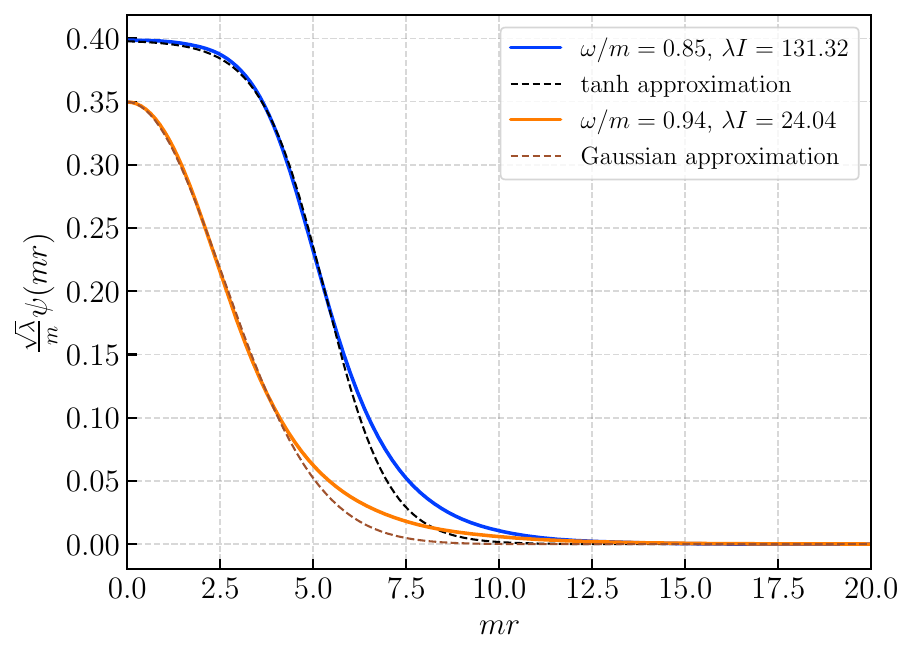}
    \end{subfigure}
    \hfill
    \begin{subfigure}[b]{0.48\textwidth}
        \centering        \includegraphics[width=\textwidth]{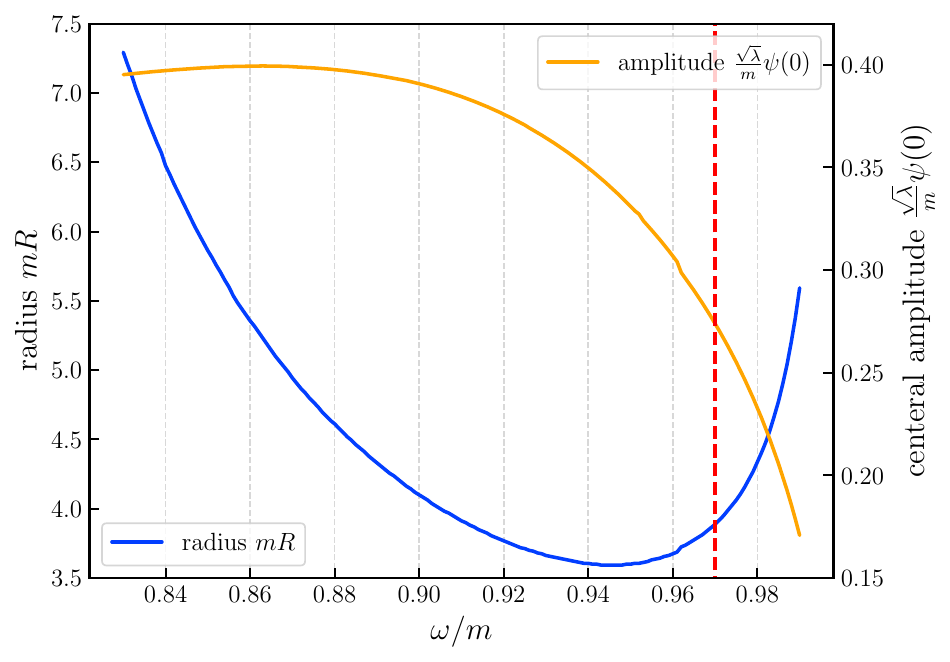}
    \end{subfigure}
    \caption{Left panel: Oscillon profiles solved numerically for $\omega/m= 0.85$ and $\omega/m= 0.94$, with $m^2 g_6/\lambda^2 = 16/15$. 
    The corresponding charges $I$ are computed by the integral of the profile as given by Eq.~\eqref{eq:spherical adiabatic charge}. 
    Right panel: The dependence of the oscillon profile shape on the fundamental frequency $\omega$, characterized by radius and amplitude at the center. 
    It is shown that neither the radius nor the amplitude is a monotonic function of $\omega$, which implies the shape change of oscillon profile is not a monotonic pattern during the whole life of a large oscillon with an initial large charge. 
    The red dashed lines are the critical values of $\omega_{\text{death}}/m = 0.97$ and $\lambda I_{\text{death}} = 20.07$ for ``energetic death'', beyond which the oscillon solution is no more stable against perturbations. 
    We define the end of the oscillon lifetime as the moment when oscillon reaches this critical value, as used in Section \ref{subsec:decay and lifetime}.}
    \label{fig:oscillon profile}
\end{figure}
\begin{figure}[t]
    \centering
    \includegraphics[width=8cm]{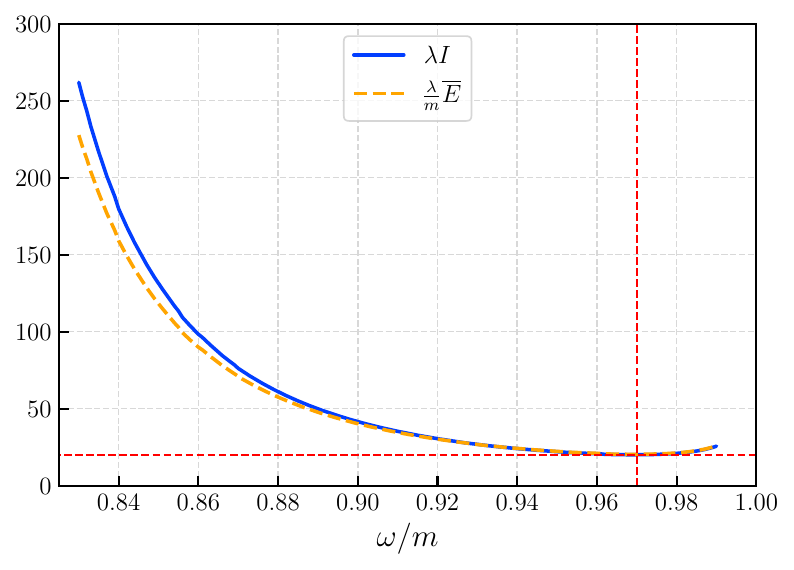}
    \caption{The relation between the charge and energy of an oscillon and its fundamental frequency $\omega$, obtained by solving Eq.~\eqref{eq:eom oscillon profile} for different values of $\omega$ with $m^2 g_6/\lambda^2 = 16/15$, and integrating Eq.~\eqref{eq:spherical adiabatic charge} and Eq.~\eqref{eq:time-average energy from oscillon profile}. 
    The red dashed lines indicate the critical values of $\omega_{\text{death}}/m = 0.97$ and $\lambda I_{\text{death}} = 20.07$ for ``energetic death''.}\label{fig:energy-omega relation}
\end{figure}

With the normalization explicitly given in App.~\ref{app:normalization}, we solve the profile $\psi(r)$ numerically by taking the only free parameter $m^2 g_6/\lambda^2=16/15$. 
From the numerical solution examples shown in Figure~\ref{fig:oscillon profile}, we see that for an $\omega$ close to $m$, the profile is in a thick-wall regime and can be well approximated by a Gaussian function 
\begin{align}
    \psi(r) \simeq \psi(0)e^{-r^2/R^2},
\end{align}
while a hyperbolic tangent function $\psi(r) \simeq \psi(0)[\tanh{(-ar + b)}+1]$ is a good approximation for a ``flat-top'' oscillon in thin-wall regime with small $\omega$.
These two cases can be seen as the thin-wall and thick-wall approximation respectively for the effective potential of the bounce solution in Eq.~\eqref{eq:eom oscillon profile} due to the change of the linear term coefficient value.

We can characterize the shape of the oscillon profiles with two parameters, the amplitude $\psi(0)$ and the radius $R$ defined as the $r$ value where $\psi(r)$ falls to $\psi(0)/e$. 
Figure \ref{fig:oscillon profile} shows the variation of the profile shapes depending on $\omega$ in the right panel.
We can see that the radius and amplitudes are both nonmonotonic functions of $\omega$.
For ``flat-top" oscillons with small $\omega$(around $\omega/m <0.9$ in the figure) and large charge and energy, as $\omega$ increases from a small value (corresponding to the decrease of $I$ and $\overline{E}$), the depletion of charge and energy mainly results in the reduction of the oscillon radius, while the amplitude is hardly affected (even slightly growing in height).
When the energy and charge become further smaller($\omega$ becomes larger), the amplitude becomes more sensitive to the charge and the profile transitions to a Gaussian function.
Noting that when $\omega$ exceeds $m$, the sign of the linear term in Eq.~\eqref{eq:eom oscillon profile} reverses, leading to a qualitative change in the solutions from localized configurations to plane waves. 
Consequently, as $\omega$ approaches the lower bound of $m$ from below, the oscillon profile tends to rapidly broaden and decrease in amplitude, signaling a transition toward a non-localized regime.

With the profile numerically solved, we can compute the corresponding charge and energy of the oscillon by integrating Eq.~\eqref{eq:spherical adiabatic charge} and Eq.~\eqref{eq:time-average energy from oscillon profile}.
The relation of charge and energy of oscillons with the fundamental frequency $\omega$ is plotted in Figure~\ref{fig:energy-omega relation}.
It has been shown that the oscillon solution in three dimensions with $\omega_{\textrm{death}} < \omega < m$ is not stable against infinitesimal perturbations, where $\omega_{\textrm{death}}$ is defined at \cite{lee_nontopological_1992, multamaki_analytical_2000}
\begin{align}
    \frac{d\overline{E}(\omega)}{d\omega} \bigg|_{\omega_{\textrm{death}}} = 0,
\end{align}
which can be understood as the Vakhitov and Kolokolov stability condition \cite{vakhitov_stationary_1973} as well.
Therefore, based on the interests of this work, we are only concerned about the oscillons with $\omega < \omega_{\textrm{death}}$ and there is a monotonic relation between the fundamental frequency and energy (also charge). This implies that the oscillon frequency and profile are uniquely determined by the energy (or charge) occupied in the localized region at formation in the early universe.
Then in the decay process, when its energy (charge) reaches the critical value $\overline{E}_{\textrm{death}}$($I_{\textrm{death}}$) corresponding to $\omega_{\textrm{death}}$, the oscillon is forced to fall apart rapidly into dissipative waves, which is called ``energetic death'' \cite{cyncynates_structure_2021}.
In the following part of this paper, we consider the ``energetic death" as the end of an oscillon lifetime.

\subsection{Oscillon decay and lifetime}\label{subsec:decay and lifetime}
We now analyze the decay process and lifetime of an oscillon.
Because the conservation of the adiabatic charge in the real scalar field we discussed in Sec.~\ref{subsec:approximate conservation and adiabatic condition} is not a strict law, as given in Eq.~\eqref{eq:full phi with oscillon and perturbation}, there should always be a perturbation $\xi(t,r)$ around the oscillon solutions, through which the oscillon decays spontaneously\footnote{One exception is found in Ref.~\cite{ibe_fragileness_2019} to be a particular logarithmic potential, where an oscillon with a Gaussian profile is the exact solution. Yet those oscillons still decay as a result of perturbation resonance.}.
A semi-analytical method for estimating oscillon decay rate through classical radiation by solving the perturbation $\xi$ in terms of profile $\psi(r)$ is developed in Ref.~\cite{ibe_decay_2019}, and later improved by Ref.~\cite{zhang_classical_2020} through taking into account the effective mass term of the perturbation.

The equation of motion for $\xi$ can be obtained by substituting Eq.~\eqref{eq:full phi with oscillon and perturbation} into the equation of motion for $\phi$,
\begin{align}
\begin{split}
    -V_{\textrm{nl}}'(\phi)  &= (\Box + m^2)\phi\\
    &= (\Box + m^2) 2\psi(r)\cos{(\omega t)} + (\Box + m^2) \xi,
\end{split}
\end{align}
then substituting Eq.~\eqref{eq:eom oscillon profile} and expanding to $\xi$, the equation of motion for perturbation $\xi$ is obtained
\begin{align}
    (\Box + m^2 + V_{\textrm{nl}}''(\phi_{osc})) \xi = V'_{\rm eff}(\psi)\cos{(\omega t)} - V_{\textrm{nl}}'(\phi_{osc}), \label{xi eom}
\end{align}
where $V_{\textrm{nl}}''(\phi_{osc})$ plays a role of effective mass with time and spatial dependence.
For the sextic potential we take, $V_{\textrm{eff}}(\psi)$ is given in Eq.~\eqref{eq:Veff phi6} and $V_{\textrm{nl}}'(\phi_{osc})$ is computed as 
\begin{align}
    V_{\textrm{nl}}'(\phi_{osc}) = -32 \lambda \psi^3 \cos^3{(\omega t )} + 192 g_6 \psi^5 \cos^5{(\omega t )}.
\end{align}
If the effective mass $V''(\phi_{osc})$ is negligible by assuming the oscillon amplitude is small compared to the mass $m^2$, it is possible to solve the $\xi(t,\textbf{x})$ with the Green's function as done in Ref.~\cite{mukaida_longevity_2017, ibe_decay_2019}.
Ref.~\cite{zhang_classical_2020} improved upon this method by including $V''(\phi_{osc})$ term, finding that for the $\phi^6$ potential the results remain largely unchanged, though their approach yields significant improvements for other types of potentials.
So we neglect it here for simplicity, thus, the solution at $r \to \infty$,is given by
\begin{align}
    \xi(t,r) &= -\frac{1}{2\pi} \left([-4\lambda\Xi_3(\kappa_3) + 30g_6 \Xi_5(\kappa_3)] \frac{\cos (3\omega t - \kappa_3 r)}{r} + 6g_6 \Xi_5(\kappa_5) \frac{\cos (5\omega t - \kappa_5 r)}{r} \right),\label{eq:perturbation solution}\\\nonumber
    \kappa_j &= \sqrt{(j\omega)^2-m^2}(j>1),~~~~\Xi_n(\kappa_j) = 4\pi \int dr \psi^n(r) \frac{r \sin (\kappa_j r)}{\kappa_j}, 
\end{align}
where $\Xi_n(\kappa_j)$ represents the contribution to the mode $j\omega$ from the $\cos^n(\omega t)$ term in $V'_{nl}$ (See Ref.~\cite{ibe_decay_2019, zhang_classical_2020} for more general, detailed derivation and discussions).
Thus, $\xi(t,r)$ can be obtained numerically as well as $\psi(r)$.
Then the energy decay rate of the oscillon through $\xi$ is computed as 
\begin{align}
    \Gamma_\xi \equiv \frac{1}{\overline{E}} \Bigg|\overline{\frac{dE}{dt}}\Bigg| = \frac{4\pi r^2 |\overline{T_{0r}}|}{\overline{E}}= 4\pi r^2 \frac{|\overline{\partial_0 \xi \partial_r \xi}|}{\overline{E}},\label{eq:gamma of single field}
\end{align}
where $\overline{E}$ is computed by Eq.~\eqref{eq:time-average energy from oscillon profile}.
And the lifetime of an oscillon can be estimated by 
\begin{align}
    \tau(\overline{E}_\text{ini}) = -\int_{\overline{E}_\text{ini}}^{\overline{E}_{\textrm{death}}}\frac{d\overline{E}}{\overline{E} \Gamma_\xi}.\label{eq:single field tau}
\end{align}

We also perform a numerical simulation of a single oscillon evolution.
The normalization we used for the numerical simulation is listed in Appendix \ref{app:normalization}.
Since we are considering spherical symmetry, the three-dimensional space can be effectively simulated by the radial equation in a one-dimensional box.
We numerically solve the nonlinear radial equation of motion given in Eq.~\eqref{eq:spherical eom of phi} on a box size of $r_{box} = 192m^{-1}$ with 3072 grids. 
But in the following part of this paper, we compute all the physical quantities only within a radius of $r_{max} = 40m^{-1}$ (this is large enough compared with oscillon radius) and the rest part is used to eliminate the unphysical waves reflected by the boundary of the simulation box with a finite size.
We use the adiabatic damping boundary condition proposed and used in Ref.~\cite{gleiser_long-lived_2000,gleiser_generation_2011}.
We write the detailed steps in the Appendix \ref{app:ADB}.

The initial conditions are taken as
\begin{align}
    \widetilde{\phi}(0, \widetilde{r}) &= 2\widetilde{\psi}(\widetilde{r}),\\
    \dot{\widetilde{\phi}}(0,\widetilde{r}) &= 0,
\end{align}
where $\psi(r)$ is the solution of Eq.~\eqref{eq:normalized profile eom} obtained by shooting method. 
The time evolution in Eq.~\eqref{eq:spherical eom of phi} is performed by a fourth-order symplectic integrator method \cite{yoshida_construction_1990} with time steps $\Delta t = 0.01m^{-1}$ and the spatial derivatives are calculated by the fourth-order central difference method.

\begin{figure}[t]
\centering
    \begin{subfigure}[b]{0.48\textwidth}
        \centering        \includegraphics[width=\textwidth]{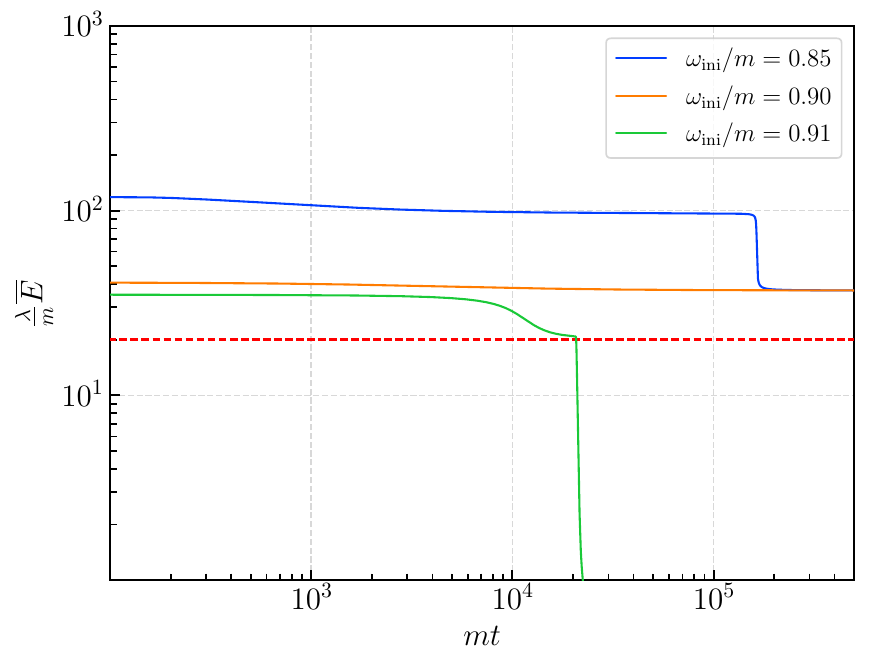}
    \end{subfigure}
    \hfill
    \begin{subfigure}[b]{0.48\textwidth}
        \centering        \includegraphics[width=\textwidth]{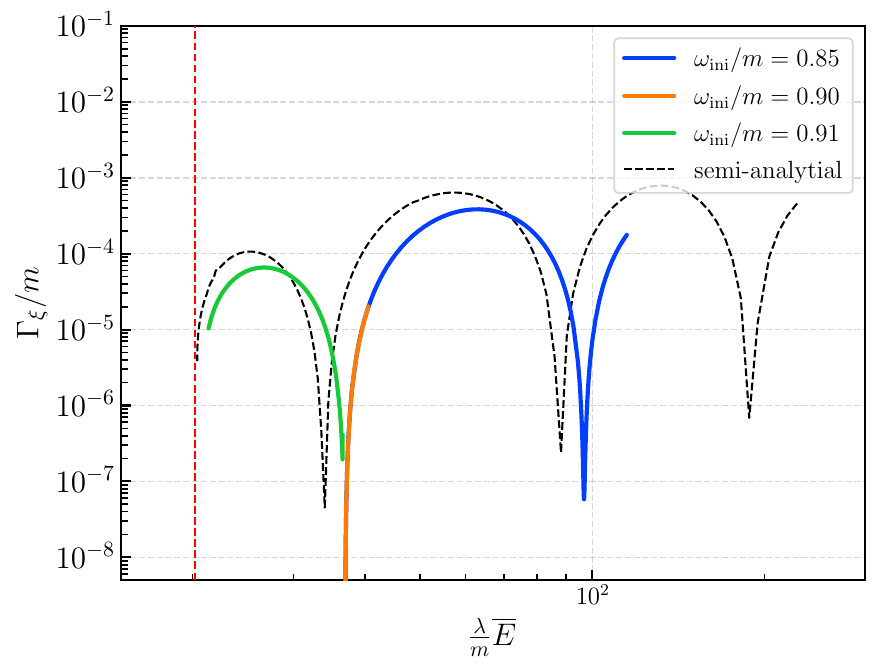}
    \end{subfigure}
    \caption{Left panel: numerical decay of the single-field oscillon charge with time, starting from different initial $\omega$ with $g_6/\lambda = 16/15$. The red dashed line denotes the critical value $\overline{E}_{\textrm{death}}$ of the ``energetic death". Right panel: the energy decay rate $\Gamma_\xi$ at different energy $E$, obtained from numerical simulation.}\label{fig:single field oscillon decay}
\end{figure}

During the numerical simulation, we monitor the time-dependent charge and energy by using
\begin{align}
    \widetilde{\overline{I}}( \widetilde{t}) &= \frac{1}{\widetilde{\omega} T_{\textrm{ave}}} \int_{\widetilde{t}}^{\widetilde{t} +T_{\textrm{ave}}} d\widetilde{t} \int_0^{\widetilde{r}_{max}} d\widetilde{r}~ 4\pi \widetilde{r}^2\dot{\widetilde{\phi}}^2,\label{eq:numerical I}\\
    \widetilde{\overline{E}}(\widetilde{t}) &= \frac{1}{T_{\textrm{ave}}} \int_{\widetilde{t}}^{\widetilde{t}+ T_{\textrm{ave}}} d\widetilde{t} \int_0^{\widetilde{r}_{max}} d\widetilde{r}~ 4\pi \widetilde{r}^2 \left(\frac{1}{2} \dot{\widetilde{\phi}}^2 + \frac{1}{2} (\partial_{\widetilde{r}} \widetilde{\phi})^2 + V(\widetilde{\phi}) \right),
\end{align}
where the instant frequency $\widetilde{\omega}$ is measured by using the time interval between the maximum of field amplitude at the center of oscillon.
We take $T_{\textrm{ave}} = 100m^{-1}$ unless otherwise stated, which is a timescale much larger than the oscillation period and small enough compared to the timescale of the adiabatic change of $I$.
Then we compute the energy decay rate by definition, $\Gamma_\xi = \dot{\overline{E}}/\overline{E}$ with $\dot{\overline{E}}$ obtained by the fourth-order difference.

Figure~\ref{fig:single field oscillon decay} demonstrates the numerical decay of single-field oscillons.
Instead of running the simulation for extremely long time, we show the results of simulations starting from different initial $\omega$, which corresponds to different initial charges and energies of oscillons.
The later segment of the blue curve transitions smoothly into the initial segment of the orange curve, which also approaches the green line after longer time evolution.
It can be seen that the oscillon undergoes a number of stepped platforms, like discrete ``energy levels" during the decay process, and the oscillon stays at each platform for a long time and falls quickly between the platforms.
As shown in the right panel, the platforms where oscillon stays long correspond to the dips of the energy decay rate $\Gamma_\xi$.
The essence of the small $\Gamma_\xi$ at certain energy values is the zero points of the leading modes of the perturbation $\xi(t,r)$ given in Eq.~\eqref{eq:perturbation solution}, in the particular potential we consider, the mode of $\kappa_3$. 
As soon as the energy of oscillon reaches the critical value $\overline{E}_{\textrm{death}}$, denoted by the dashed line, the rest charge and energy decay in an instant and the oscillon dies out.

\section{Coupling to an external scalar field: Instability bands of the external field on oscillon background}\label{sec:instability bands}
Now we introduce the external coupling to another real scalar field $\chi$.
The Lagrangian is given in the Eq.~\eqref{eq:full lagrangian}, we consider the case for $g \neq 0$.

Ref.~\cite{shafi_formation_2024} has studied the formation and decay of oscillons with the same external coupling in an expanding universe by performing lattice simulation.
Their simulations have shown that in the presence of external coupling, a substantial number of oscillons can still form within an appropriate parameter space, and subsequently decay into the external field fluctuations via parametric resonance at late times after their formation — though still much sooner than the typical lifetime of single-field oscillons.
Motivated by these findings, we aim to analyze the instability bands of the external resonance, especially the dependence on the external coupling strength and oscillon shapes.
As mentioned in Sec.\ref{sec:introduction}, different from the simulation of a population of oscillons in Ref.~\cite{shafi_formation_2024}, we will be focusing on the decay of an individual oscillon with the presence of the external scalar field in flat spacetime.

Still assuming spherical symmetry, the equations of motion of two fields are
\begin{align}
    \Ddot{\phi} - \frac{\partial^2 \phi}{\partial r^2} - \frac{2}{r}\frac{\partial \phi}{\partial r} +m^2 \phi +\frac{d V_{\textrm{nl}}}{d \phi} + 2g\chi^2\phi= 0\\
    \Ddot{\chi} - \frac{\partial^2 \chi}{\partial r^2} - \frac{2}{r}\frac{\partial \chi}{\partial r} + m_\chi^2 \chi + 2 g \phi^2 \chi=0.
\end{align}
Taking the single frequency approximation for oscillon configuration of $\phi$, $\phi(t,r) = 2\psi(r)\cos{\omega t}$, the equations of motion for $\chi(t,r)$ becomes
\begin{align}
    \Ddot{\chi} - \frac{\partial^2 \chi}{\partial r^2} - \frac{2}{r}\frac{\partial \chi}{\partial r} + m_\chi^2 \chi+ 8g\psi^2(r) \cos^2{(\omega t)} \chi = 0.\label{eq:chi eom}
\end{align}
This equation is the same as the equation for preheating, except for the spatial dependence of the oscillating field amplitude $\psi(r)$.
Fourier transform of this equation becomes
\begin{align}
    \Ddot{\chi_k} + (k^2 + m_\chi^2)\chi_k + 8 g \cos^2 (\omega t) \int \frac{d^3 k'}{(2 \pi)^3} \Psi(\mathbf{k}-\mathbf{k}') \chi_{k'} = 0, \label{eq:Eom chik}
\end{align}
where $\chi_k$ and $\Psi(k)$ are the Fourier transforms of $\chi(x)$ and $\psi^2 (x)$, respectively.
The last term comes from the convolution of $\psi(r)^2$ and $\chi(r)$, which contains a mixture of different modes of $\chi$ with various $k$ values.

It is known that the parametric resonance occurs at the early stage of preheating when the amplitude of oscillation is large and leads to certain modes of the product field exponentially growing \cite{kofman_towards_1997}.
The exponential growth can be interpreted by the collective decay with Bose stimulation effect, indicating the production rate of $\chi$ particles is enhanced by the number of $\chi$ particles.
However, unlike the homogeneous oscillation appearing in the preheating, the inhomogeneous amplitude $\psi(r)$ of oscillons results in the modes mixing in Eq.~\eqref{eq:Eom chik} and makes analytical solution difficult.

\subsection{Neglecting inhomogeneous oscillon profile}
If we neglect the inhomogeneity of $\psi(r)$ and take the amplitude at the oscillon center, $\psi(r) \simeq \psi_0 \equiv \psi(r=0)$, this equation can be simplified as identical to the preheating equation,
\begin{align}
     \Ddot{\chi_k} + (k^2 + m_\chi^2)\chi_k + 8 g \psi_0^2 \cos^2 (\omega t) \chi_k = 0, \label{eq:Mathieu's eq with homogeneous}
\end{align}
which can be organized to the form of a standard Mathieu's equation,
\begin{align}
    &\chi_k^{''} + (A_k + 2 q \cos{(2z)})\chi_k  = 0,\\
    A_k &= \frac{k^2 + m_\chi^2}{\omega^2} + 2q,~~ q = \frac{2g\psi_0^2}{\omega^2},~~ z = \omega t,\label{eq:standard Mathieu's eq}
\end{align}
where prime denotes the derivative with respect to $z$.
The behavior of this equation, which is exactly the same equation for reheating, is well-investigated in the literature.
According to the Floquet's theorem \cite{floquet_sur_1883,mclachlan_theory_1947}, a Mathieu's equation should admit a complex solution in the form of,
\begin{align}
    \chi_k(z) = \mathcal{P}_{+}(z) e^{\mu z} + \mathcal{P}_{-}(z) e^{- \mu z}, \label{eq:floquet theorem}
\end{align}
where $\mathcal{P}_{\pm}(z)$ are periodic functions with a period of $\pi$ in respect to $z$, and $\mu$ is generally a complex number called the Floquet exponent.
If $\mu$ has a nonzero real part, resonance occurs and part of the solution can be exponential growth with time $z$, while the solution will be stably oscillating if $\mu$ is purely imaginary.

From the study of the Mathieu's equation, when $|q| \ll 1$ and $A_k>0$, the resonance happens in narrow bands around $A_k \simeq n^2,~n=1,2,...$.
Since small $|q|$ corresponds to coupling $g$ small enough that the perturbative treatment of the interaction works.
Then the $n$-th narrow band has a width of $\Delta k \sim q^n$ in momentum space and corresponds to production of a $\chi$ particle pair with momentum $k^2 + m_\chi^2 \simeq (n m)^2$  from the decay of $n$ $\phi$ particles \cite{lozanov_expansion_2017}.
The most pronounced and widest narrow band is the first one around $A_k \simeq 1$ with $\mu_{\textrm{max}} \approx |q|/2$, which can be derived in the limit of $|q| \ll 1$ as shown in App~\ref{app:mathieu's eq}.
When $|q|$ becomes large, the broad resonance happens much more efficiently and leads to a burst of energy transfer from $\phi$ to $\chi$. 
The values of exponent $\mu$ within this range can only be computed by numerical Floquet analysis.

\begin{figure}[t]
    \begin{subfigure}[t]{0.48\textwidth}
        \centering        \includegraphics[width=\textwidth]{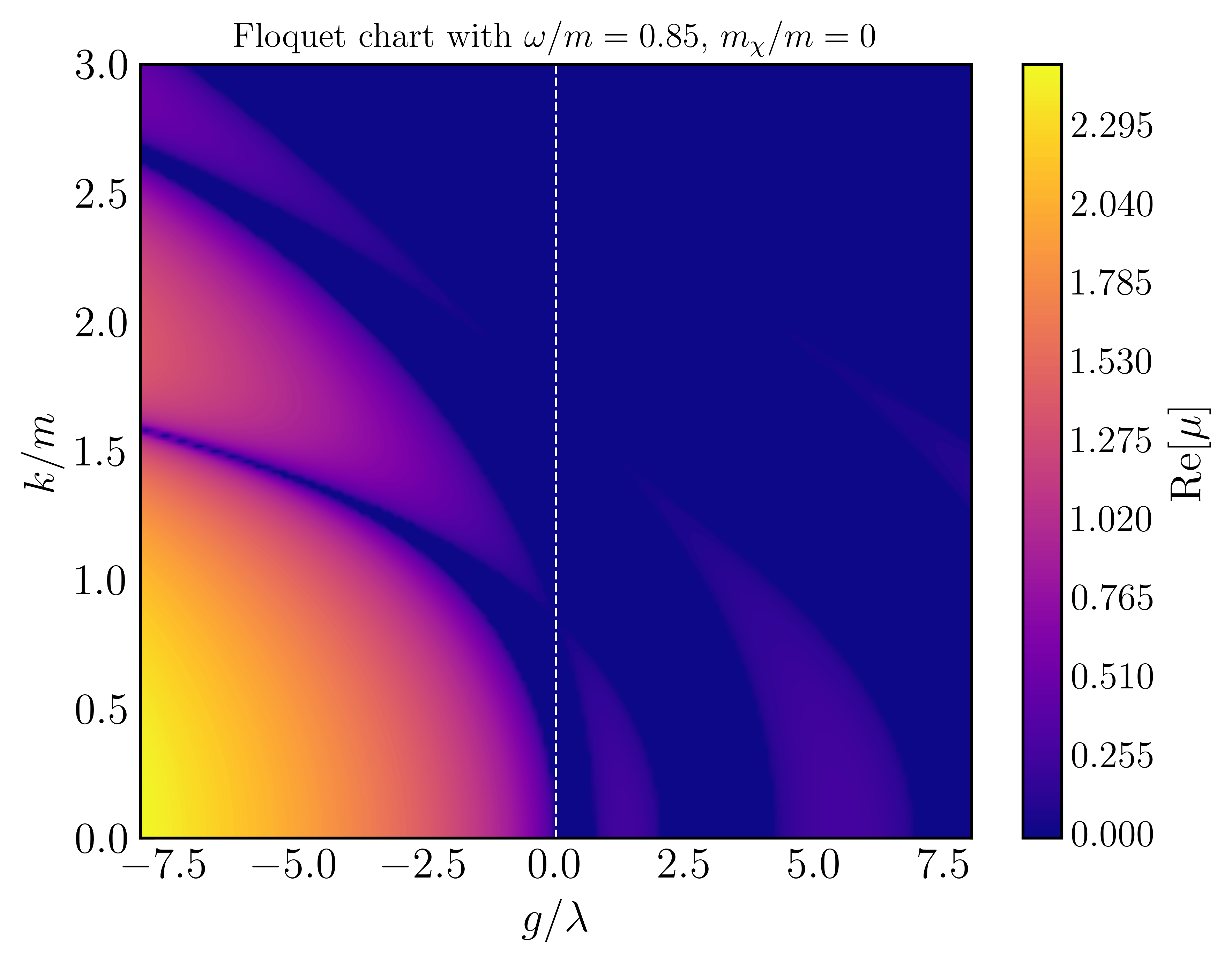}
    \end{subfigure}
    \hfill
    \begin{subfigure}[t]{0.48\textwidth}
        \centering        \includegraphics[width=\textwidth]{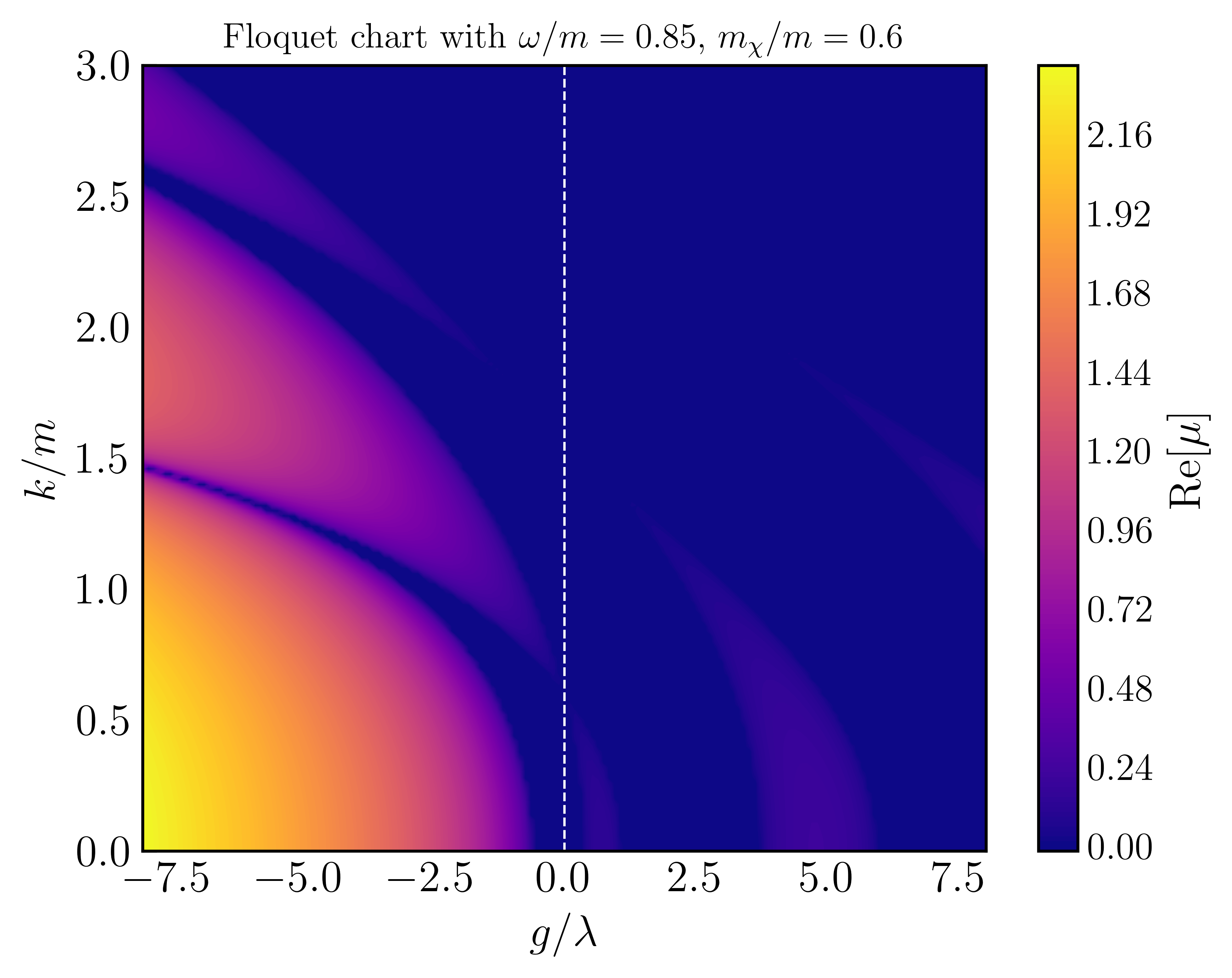}
    \end{subfigure}
\caption{Floquet charts for Mathieu's equation in Eq.~\eqref{eq:standard Mathieu's eq} neglecting oscillon inhomogeneity in terms of $k$ and $g$, with the center amplitude of an oscillon whose $\omega/m = 0.85$ is taken for $\psi_0$. The left and right panel are when $m_\chi/m = 0$ and  $m_\chi/m = 0.6$, respectively. The dashed line is where $g=0$. }\label{fig:Floquet_chart_homogeneous}
\end{figure}
\begin{figure}[b]
    \centering        \includegraphics[width=0.5\textwidth]{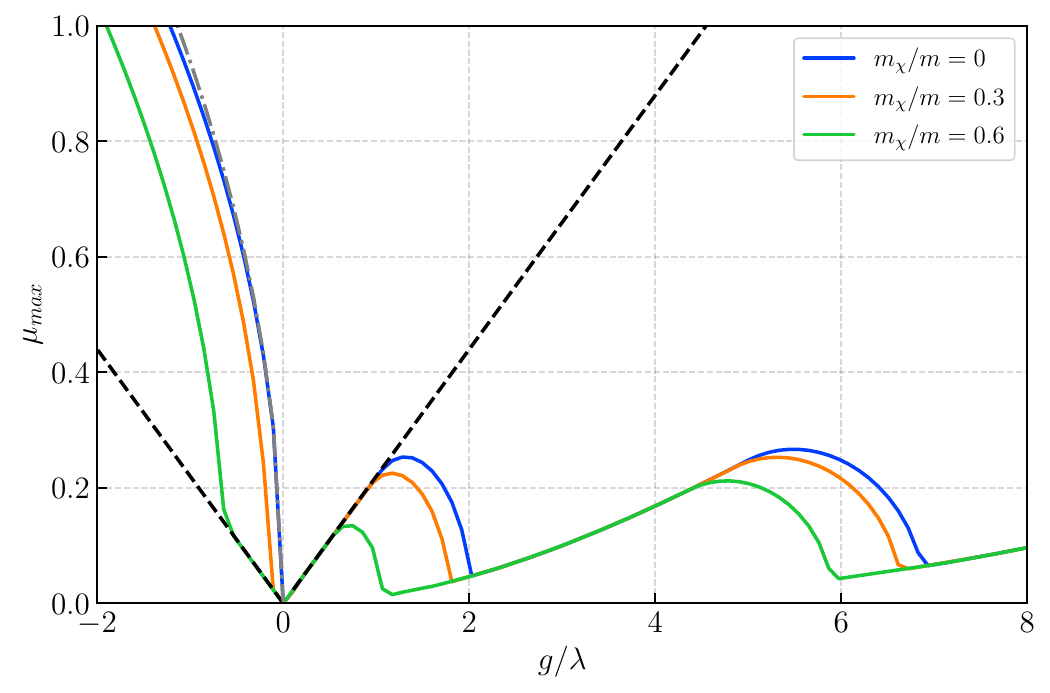}
\caption{The maximum value of the real part of Floquet exponents $\mu_{\max} \equiv \max{(\Re{(\mu)})}$ among all the modes with various $k$, which is the growth rate of the mode that grows fastest, for various values of coupling strength $g/\lambda$ and $m_\chi/m$. The center amplitude $\psi_0 (\simeq 0.4 m/\sqrt{\lambda})$ and frequency of an oscillon with $\omega/m = 0.85$ are used in the computation. The dashed black line is the linear relation in the first narrow band, $\mu_{\max} = |q|/2$, and the grey dot-dashed line is the square root relation in the tachyonic band for $m_\chi/m = 0$, $\mu_{\max} = 2 \psi_0\sqrt{|g|}/\omega$.}\label{fig: homogeneous mu max}
\end{figure}

We note that the equation of $\chi_k$ given in Eq.~\eqref{eq:Mathieu's eq with homogeneous} differs from a standard Mathieu's equation whose $A_k$ and $q$ are free parameters independent of each other.
The dependence here is a result of the constant contribution of $\cos^2$.
On the other hand, the two parameters are independent for a three-point interaction between $\phi$ and $\chi$, $\mathcal{L}_{\textrm{int}} \supset \phi \chi^2$, and the equation behaves closer to a standard Mathieu's equation, whose Floquet chart is symmetric about $g=0$ and has no tachyonic bands due to positive definite $A_k$.

Figure \ref{fig:Floquet_chart_homogeneous} shows the results of Floquet analysis for Eq.~\eqref{eq:standard Mathieu's eq} with various values of parameters $A_k$ and $q$ in terms of $k$ and $g$.
It is apparently far different from the Floquet chart for the normal standard Mathieu's equation shown in App.~\ref{app:mathieu's eq} due to the dependent relation of the parameters, $A_k$ and $q$.
The horizontal axis in the left panel, where $k=0, m_\chi = 0$, corresponds to the dividing line $A =2q$ in the standard Floquet chart.
By comparing the left and right panels, we can see that for a larger $m_\chi$, the Floquet chart in the right panel of Fig.~\ref{fig:Floquet_chart_homogeneous} is shifted downwards along the $k$ axis, because the contribution of $m_\chi$ in the Mathieu's equation, Eq.~\eqref{eq:standard Mathieu's eq}, can be equivalently seen as an increased value of $k$.
The attractive interaction ($g<0$) and the repulsive interaction ($g>0$) have very different instability band shapes.
For a positive $g$, there are only narrow bands corresponding to the standard narrow bands passed through by the straight line $A_k = 2q$, and the shapes of the narrow bands are elongated and thinned due to the tilting of the horizontal axis.
Thus, in the first narrow band, we have 
\begin{align}
    \mu_{\textrm{max}} \approx |q|/2 = \frac{\psi_0^2}{\omega^2} g,~g>0 \label{eq:narrow band mu relation}
\end{align}
in the range of small $|g|$, where $\mu_{\textrm{max}} \equiv \max(\Re{(\mu)})$ is the exponent of the fastest growing mode.
Further, the narrow bands in the range of positive $g$ become smaller as $\chi$ field becomes heavier, as shown in Fig.~\ref{fig: homogeneous mu max} and will disappear for a heavy enough $\chi$, .

For a negative $g$, in addition to the narrow bands at large $k$, the broad bands appear due to tachyonic instability when $A_k < 0$, i.e., when 
\begin{align}
    g < -\frac{k^2 + m_\chi^2}{4 \psi_0^2}.
\end{align}
In this case, as discussed in App.~\ref{app:mathieu's eq}, when $|q| \ll \sqrt{|A_k|}$, the effect of the negative $A_k$ term leading to a non-oscillating exponential solution dominates over the oscillating term in Eq.~\eqref{eq:standard Mathieu's eq}, so the solution has $\Re{(\mu)} \gg \Im{(\mu)}$.
And there is
\begin{align}
    \Re{(\mu)} \approx \sqrt{|A_k|}  = \frac{1}{\omega}\sqrt{|k^2 + m_\chi^2 + 4\psi_0^2 g|},
\end{align}
in the region of small $|g|$.
For massless $\chi$, $k=0$ mode is always in the tachyonic band, thus, all the range of negative $g$ falls into tachyonic instability band with $\mu_{\textrm{max}} \propto \frac{2\psi_0}{\omega}\sqrt{|g|}$.
For massive $\chi$, the first narrow band at large $k$ gives the dominant growing modes until $k=0$ mode enters the tachyonic band at $g<-\frac{m_\chi^2}{4\psi_0^2}$. 
Thus, as Fig.~\ref{fig: homogeneous mu max} shows, the dependence of $\mu_{max}$ on $g$ when $g<0$ is
\begin{align}
    \mu_{max} \approx
    \begin{cases}
         |q|/2 = -\frac{\psi_0^2}{\omega^2} g,~~~~~-\frac{m_\chi^2}{4\psi_0^2}\lesssim g <0\\
        \frac{1}{\omega}\sqrt{|m_\chi^2 + 4\psi_0^2 g|}, ~~~~~g\lesssim-\frac{m_\chi^2}{4\psi_0^2}.\label{eq:tachyonic band mu relation}
    \end{cases}
\end{align}

We note that the Floquet exponent $\mu$ is defined in Eq.~\eqref{eq:floquet theorem} in terms of $z$, the actual growth of $\chi$ modes we observe in terms of $t$ is $\chi_k(t) \propto e^{ \Re{(\widetilde{\mu})} t}$, $\widetilde{\mu} = \mu\omega $.

\subsection{The effect of inhomogeneous oscillon profiles}\label{subsec: bgfix}

In the previous subsection, the resonance band of $\chi$ was computed under the approximation that the radial spatial dependence of oscillon profile is neglected.
However, once the inhomogeneity of $\psi(r)$ is taken into account, because of the mixture of $\chi$ modes, a normal Floquet analysis can not solve the equation of $\chi$ in the Fourier space, Eq.~\eqref{eq:Eom chik}.
A modified Floquet analysis was developed to include the inhomogeneous $\psi(r)$ in Ref.~\cite{hertzberg_quantum_2010}, and used in Refs.~\cite{kawasaki_decay_2014, antusch_parametric_2016} to compute a three-point coupling and six-point coupling, respectively.

In this work, we conduct a numerical simulation under the normalization written in App.~\ref{app:normalization}. 
We first evolve $\widetilde{\chi}(\widetilde{t}, \widetilde{r})$ on the background of an oscillon solution by fixing $\widetilde{\phi}(\widetilde{t},\widetilde{r}) = 2\widetilde{\psi} (\widetilde{r}) \cos{(\widetilde{\omega} \widetilde{t})}$ by hand for every time step, where $\widetilde{\psi}(\widetilde{r})$ is the numerically obtained profile for various given $\widetilde{\omega}$ of single-field oscillons in Sec.~\ref{subsec:single field oscillon profile}.

The initial condition for $\widetilde{\chi}$ is taken as a Gaussian function,
\begin{align}
    \begin{split}
        \widetilde{\chi}(0,\widetilde{r}) &= \chi_0 e^{-\widetilde{r}^2/\widetilde{R}_\chi^2},\\
        \dot{\widetilde{\chi}}(0,\widetilde{r}) &= 0,
    \end{split}
\end{align}
where we take $\chi_0 = 0.1$, $\widetilde{R}_\chi=3$. 
At the beginning of our simulation, most components of the initial Gaussian $\chi$ field dissipate, except for the modes acquiring exponential enhancement by any mechanism.
After a short time evolution, the $\chi$ field gets relaxed to a real configuration on the given $\phi$ background, usually determined by the mode growing quickest.
So the parameter choices of initial input for $\widetilde{\chi}$ have no effect on the growth rate, but only on the time length of a first relaxation stage, which is cut off for further growth rate fitting.
Here we assume the resonance can cause $\widetilde{\chi}$ to grow into a localized oscillon-like structure thanks to the nontrivial spatial dependence background of $\widetilde{\phi}$.
Actually, a more complete simulation should start from a random fluctuation of $\widetilde{\chi}$ in three-dimensional space.
But as the three-dimensional simulation in Ref.~\cite{van_dissel_symmetric_2022} shows, $\widetilde{\chi}$ can eventually grow from the random fluctuation into the bump configuration if it falls into an instable band.
So it causes no effect that we skip the growth phase from the noise to avoid the singular behavior of random fluctuation at the origin in our radial simulation.

Then we evolve the equation of $\widetilde{\chi}(\widetilde{t}, \widetilde{r})$ in Eq.~\eqref{eq:chi eom} with the same numerical setup and method in Sec.~\ref{subsec:decay and lifetime}.
We calculate the energy of $\widetilde{\chi}$ sector by
\begin{align}
    \widetilde{\overline{E}}_\chi(\widetilde{t}) = \frac{1}{T_{\textrm{ave}}} \int_{\widetilde{t}}^{\widetilde{t}+T_{\textrm{ave}}} d\widetilde{t} \int_0^{\widetilde{r}_{max}} d\widetilde{r}~ 4\pi \widetilde{r}^2 \left(\frac{1}{2} \dot{\widetilde{\chi}}^2 + \frac{1}{2} (\partial_{\widetilde{r}} \widetilde{\chi})^2 + \frac{1}{2} \widetilde{m}_\chi^2 \widetilde{\chi}^2 \right), \label{eq:chi energy}
\end{align}
where we still take $T_{\textrm{ave}}=100m^{-1}$.
We assume the fastest-growing mode dominates the energy of $\chi$, thus we fit the Floquet exponent $\Re{(\mu)}$ by $\widetilde{\overline{E}}_\chi \propto e^{2\Re{(\mu)} \widetilde{\omega}\widetilde{t}}$, where $\widetilde{\omega}$ is the frequency of the background oscillon.
We note that $\widetilde{\overline{E}}_\chi$ is evaluated within a boxsize of $r_{max}$, therefore, if $\chi$ field falls in a stable band where $\Re{(\mu)}=0$, $\widetilde{\overline{E}}_\chi$ will decrease with time because the energy of the initial input will propagate to the outside of the box by dissipative modes of $\chi$.
For illustrative purpose, we take $\Re{(\mu)}=0$ when $\widetilde{\overline{E}}_\chi$ keeps decreasing after an initial relaxation time in the simulation.

\begin{figure}[t]
    \centering
    \includegraphics[width=\textwidth]{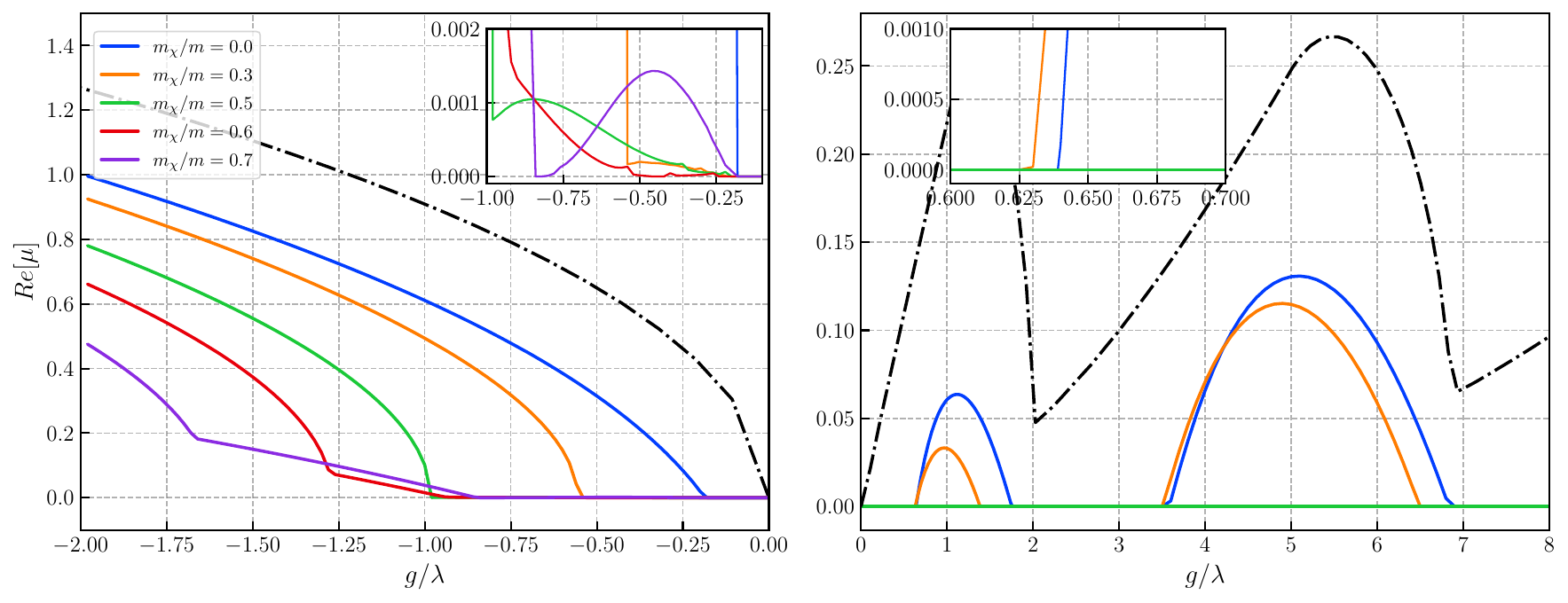}
    \caption{The growth rate of $\chi$ field obtained in the numerical simulation with fixed oscillon background for various $m_\chi$ with the background of an osillon with $\omega/m = 0.85$. The black dashed line is the $\mu_{\max}$ obtained by the homogeneous Floquet analysis for $m_\chi=0$ in the last subsection for comparison. }\label{fig:massive mu compare}
\end{figure}
Figure \ref{fig:massive mu compare} shows the results of $\Re{(\mu)}$ obtained by the numerical simulation with inhomogeneous oscillon solution as the background for various values of $m_\chi$.
The results of homogeneous Floquet analysis obtained in the last subsection are plotted in dashed lines for illustrative purpose.
It obviously shows that compared to the relation of $\mu_{max}$ and $g$ in Eq.~\eqref{eq:narrow band mu relation}, \eqref{eq:tachyonic band mu relation}, the growth rate of $\chi$ field is reduced due to the inhomogeneous oscillon profile with finite radius taken into account.
As discussed in Ref.~\cite{hertzberg_quantum_2010, kawasaki_decay_2014}, this reduction can be understood as the escape rate of the $\chi$ particles from the oscillon with finite size.
Though we describe the oscillons with classical field $\phi$, they can be seen as collections of numerous $\phi$ particles in coherent states, and the exponential growth of $\chi$ can be considered as explosive production of $\chi$ particles inside the oscillon.
Then the $\chi$ particles produced with non-zero momentum escape away from the oscillons.
If the escape rate is faster than the production rate, no growth of $\chi$ field can be seen in the simulation, which does not capture the energy propagated out of the box.
The escape rate can be estimated by $\Gamma_{\text{escape}} \sim v_\chi/R \sim \frac{p_\chi}{E_\chi R}$, where $p_\chi = \sqrt{E_\chi^2 - m_{\chi,\textrm{eff}}^2}$, $E_\chi = m$, and $R$ is the radius of the background oscillon.

Then we can define a critical value of $g \sim g_{\text{cri}}^\pm$, beyond which the dependence of $\Re(\mu)$ on $g$ begins to align with the predictions of the Floquet analysis in Eqs.~\eqref{eq:narrow band mu relation} and \eqref{eq:tachyonic band mu relation} reduced by the escape rate.
In the case of three-point coupling, $\mathcal{L}_{\textrm{int}} \supset \phi \chi^2$, the dependence of the critical coupling $g_{\text{cri}}$ on the mass of $\chi$ is found as $g_{\text{cri}} \propto v_\chi =\sqrt{1 - 4 m_\chi^2/m^2}$ until reaching its lower bound at $m_\chi/m \approx 0.5$ ~\cite{kawasaki_decay_2014}.
Nevertheless, whereas the unstable bands in the three-point coupling case, governed by independent $a$ and $q$, exhibit regular shapes that are largely insensitive to $m_\chi$ within a typical range, the situation is markedly different for the four-point coupling considered here. 
The presence of varying narrow bands and the emergence of tachyonic bands make determining $g_{\text{cri}}$ significantly more complicated.

\begin{figure}[bt]
    \begin{subfigure}[t]{0.48\textwidth}
        \centering        \includegraphics[width=\textwidth]{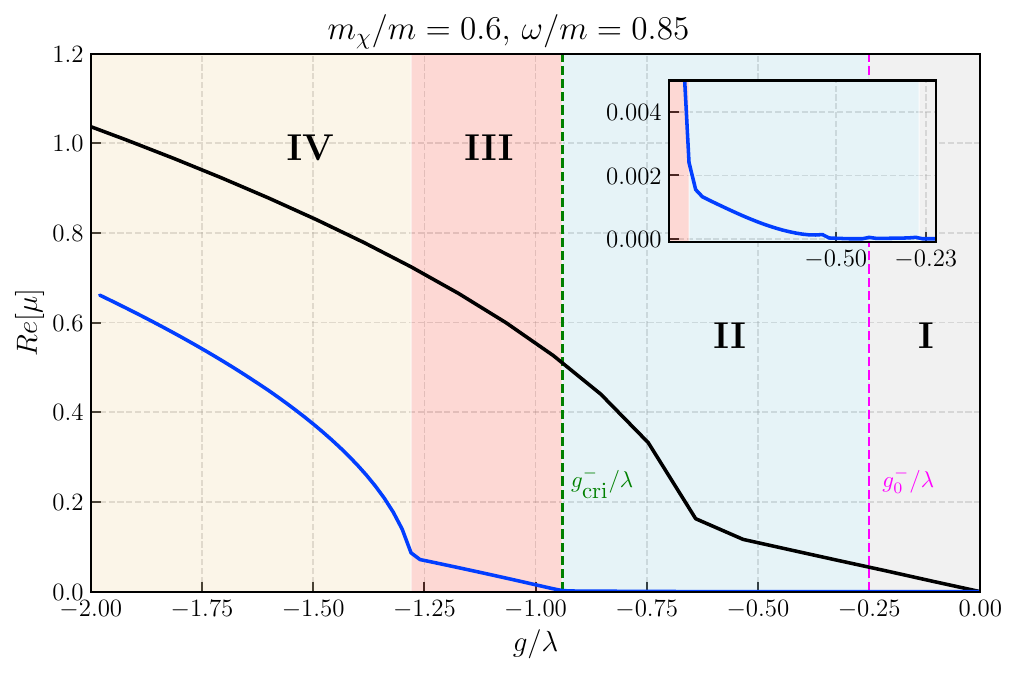}
    \end{subfigure}
    \hfill
    \begin{subfigure}[t]{0.48\textwidth}
        \centering        \includegraphics[width=\textwidth]{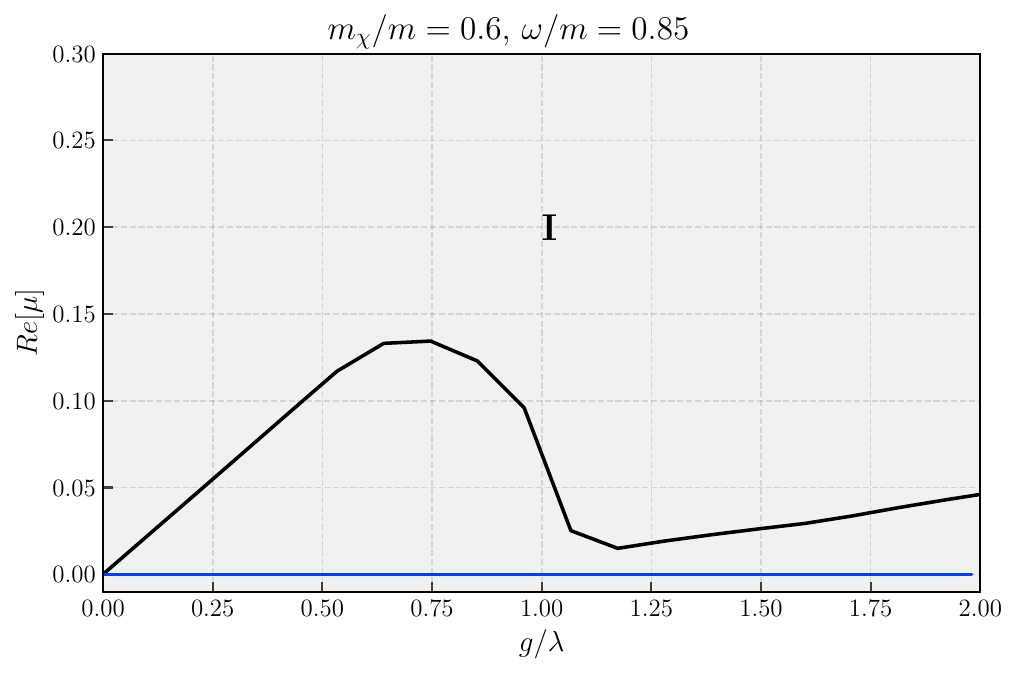}
    \end{subfigure}
    \begin{subfigure}[t]{0.48\textwidth}
        \centering        \includegraphics[width=\textwidth]{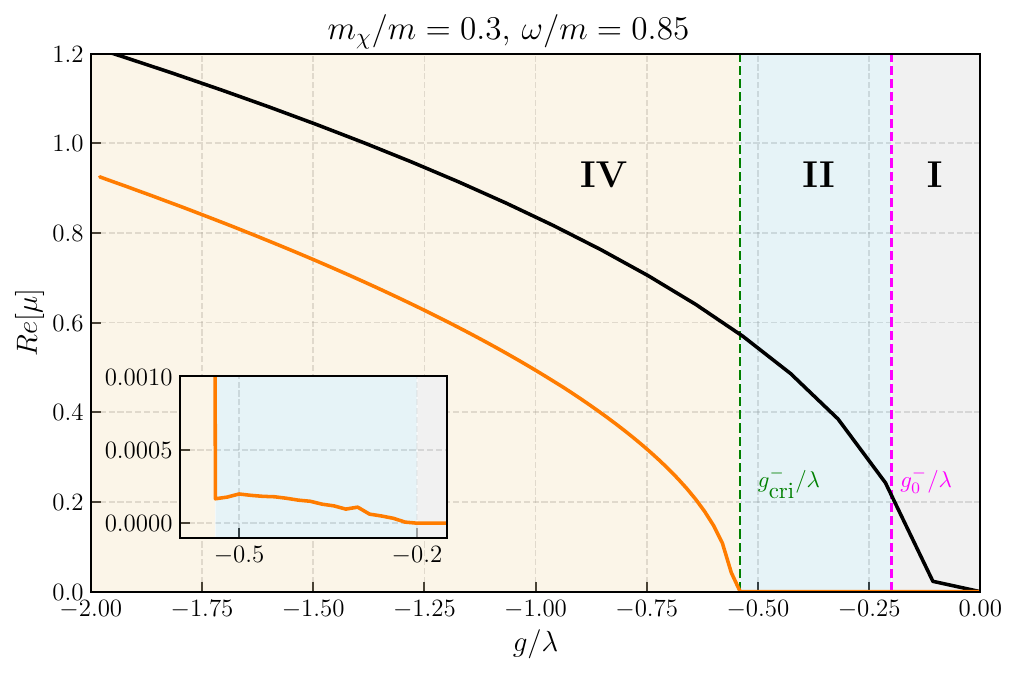}
    \end{subfigure}
    \hfill
    \begin{subfigure}[t]{0.48\textwidth}
        \centering        \includegraphics[width=\textwidth]{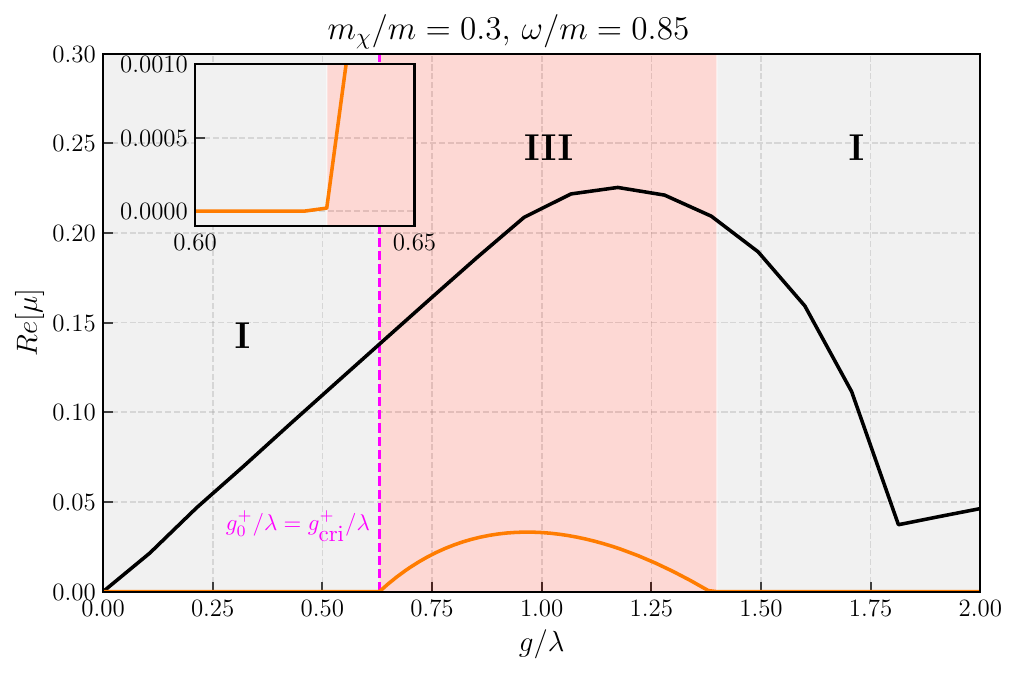}
    \end{subfigure}
    \vfill
    \caption{The blue and orange lines show the exponents of $\chi$ field growth obtained by numerical simulation with the background fixed as an oscillon for $m_\chi/m =0.6$ and $m_\chi/m = 0.3$, respectively. 
    The black curves in the top panel show the Floquet exponents $\mu_{\textrm{max}}$ obtained by the homogeneous Floquet analysis for $m_\chi/m =0.6$ and for $m_\chi/m = 0.3$ in the bottom panel.
    An oscillon with $\omega/m = 0.85$ is used for all the curves. 
    We categorize four distinct ranges of $g$ according to the primary mechanism responsible for the exponential growth of the $\chi$ field.  
    In Region $\text{\Rmnum{1}}$, the growth rate $\Re(\mu)$ of the $\chi$ field is zero, as no enhancement exceeds the particle escape rate. 
    Region $\text{\Rmnum{2}}$ corresponds to instability bands induced by a new mechanism due to the inhomogeneous oscillon profile (see text for details). 
    This region is characterized by a tiny order of growth rate, which may result from mode mixing between the bound-state mode of $\chi$-which are largely confined and do not readily escape-the escaping modes within the resonance bands. In Region $\text{\Rmnum{3}}$, narrow parametric resonance becomes the dominant source of exponential growth, while tachyonic instability excites the bound state rapidly in Region $\text{\Rmnum{4}}$. Two critical values of $g$ are defined as follows: the growth rate $\Re{[\mu]}$ is nonzero when $|g|>|g_0^\pm|$, and $\Re(\mu) \simeq \Gamma_{\textrm{escape}}$ at $|g|>g_{\text{cri}}^\pm$.} \label{fig:mu_regions}
\end{figure}

To discuss different mechanisms responsible for $\Re{(\mu)}$ behavior in different coupling ranges, we categorize four ranges for negative $g$ as shown in the left panel of Fig.~\ref{fig:mu_regions}.
The Region $\text{\Rmnum{1}}$ is where the coupling is too weak for the produced $\chi$ particles to accumulate.
The Region $\text{\Rmnum{2}}$ corresponds to a very small growth rate of order $\Re(\mu) \sim \mathcal{O}(10^{-4})$, the origin of which we will discuss in more detail later.
The Region $\text{\Rmnum{3}}$ is where parametric resonance in the first narrow band serves as the dominant enhancement to overcome the particle escape rate.
In this region, the value of $g_{cri}$ can be estimated as $\Re{(\mu)} \simeq \Gamma_{\text{escape}}$, 
\begin{align}
  \frac{\psi_0^2}{\omega}|g| \simeq  \frac{\sqrt{m^2 - m_\chi^2 - 4g\psi_0^2}}{m R},
\end{align}
where we take the time-averaged effective mass of $\chi$ for particle velocity estimation. And this gives 
\begin{align}
    g_{\text{cri}}^\pm \sim \frac{(\omega/m)}{\psi_0^2 R^2} \left(-2\frac{\omega}{m} \pm \sqrt{(m^2-m_\chi^2)R^2 + 4 \left(\frac{\omega}{m} \right)^2} \right). \label{eq:g_cri}
\end{align}
As the bare mass $m_\chi$ increases, the $|g_{\text{cri}}^\pm|$ becomes smaller because the escape velocity of $\chi$ particles decreases.
This can be seen consistently with the tendency of $g_{\text{cri}}^+$ values of $m_\chi/m <0.5$ in the right panel of Fig.~\ref{fig:massive mu compare} and $g_{\text{cri}}^-$ of $m_\chi/m >0.5$ in the left panel, for instance.
In the figure, when $m_\chi/m >0.2$, the shrinking shape of the narrow band on the positive semi-axis of $g$ due to increasing $m_\chi$ starts to become the major factor in determining the value of $g_{\text{cri}}^+$ and shifts it significantly.
When $m_\chi/m \leq 0.5$ and $g < 0$, the narrow band is not long enough to overcome the escape rate before the tachyonic band takes over, so the Region $\text{\Rmnum{3}}$ disappears and it enters the Region $\text{\Rmnum{4}}$ directly with $g_{\text{cri}}^-$.

\begin{figure}[t]
    \begin{subfigure}[t]{0.48\textwidth}
        \centering        \includegraphics[width=\textwidth]{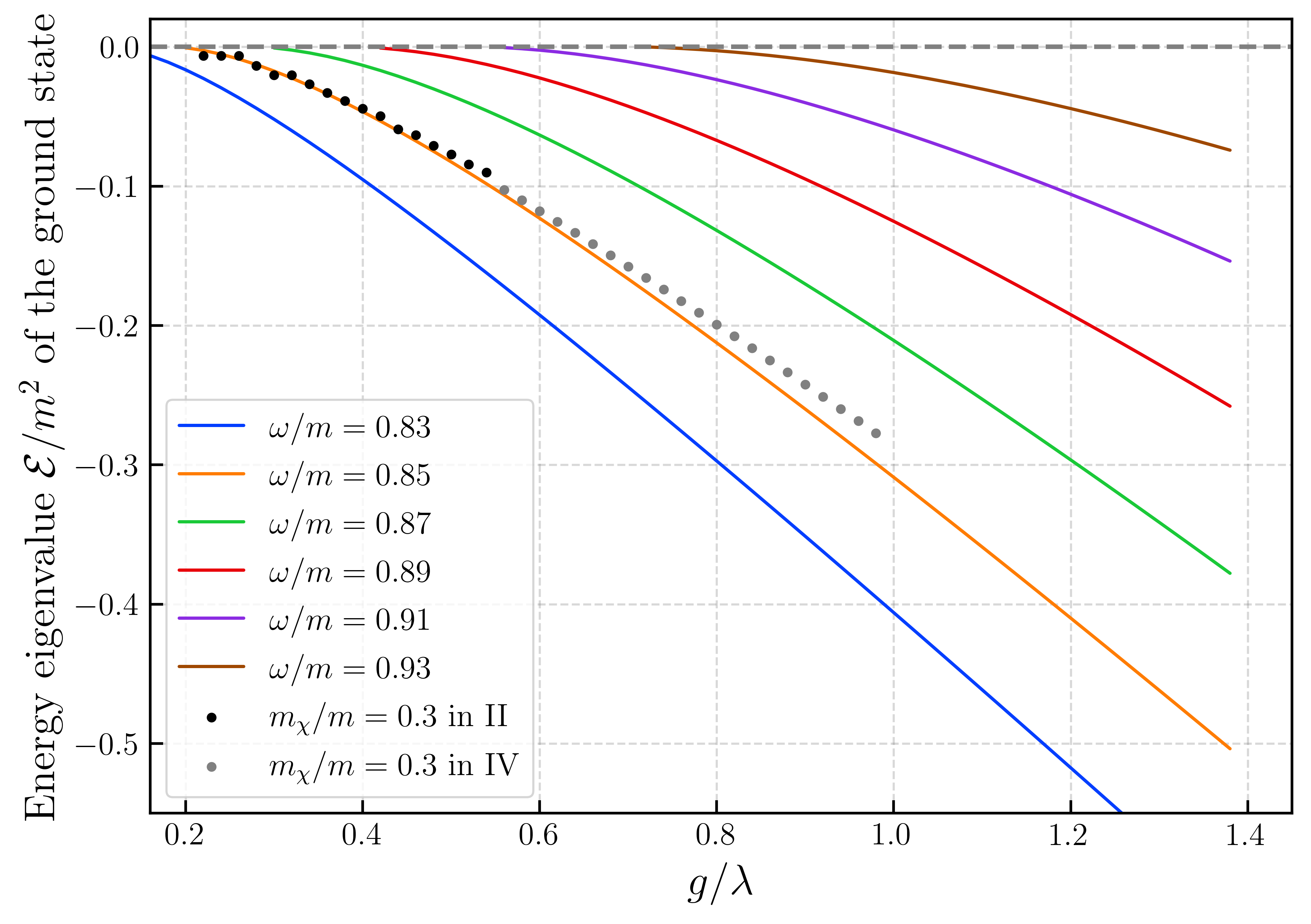}
    \end{subfigure}
    \hfill
    \begin{subfigure}[t]{0.48\textwidth}
        \centering        \includegraphics[width=\textwidth]{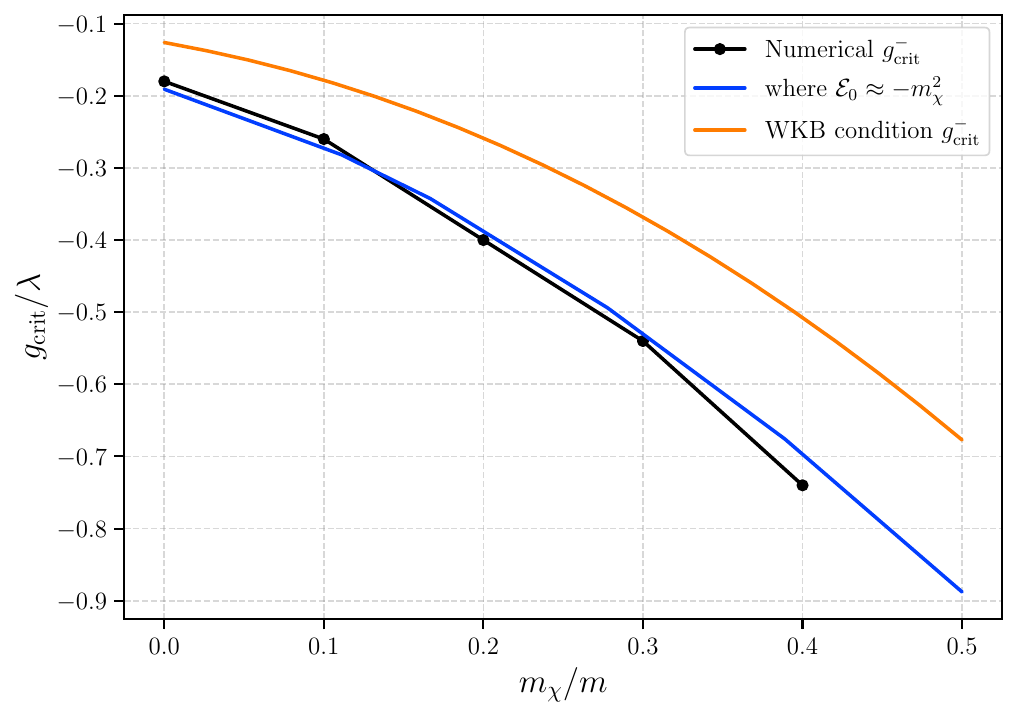}
    \end{subfigure}
    \caption{Left panel: the dependence of ground-state energy eigenvalue  $\mathcal{E}$ for Eq.~\eqref{eq:1d Schrodinger eq} on $g$ obtained by finite difference method numerically, with $\psi(r)$ taken as oscillon profile of various values of $\omega$. The black and gray dots are $\mathcal{E}$ obtained by Eq.~\eqref{eq:energy eigenvalue} in the numerical simulation with background $\phi$ fixed as oscillon configuration with $\omega/m = 0.85$. 
    Right panel: the blue line shows the critical value of $g$ at which $\mathcal{E} \approx -m_\chi^2$, beyond which the tachyonic instability cannot enhance bound states with energy eigenvalues higher than this threshold, for various $m_\chi$ extracted from the curve of $\omega/m = 0.85$ in the left panel. 
    The critical value $g_{\text{cri}}$ predicted by the WKB condition for bound state given in Eq.~\eqref{eq:WKB condition} is shown by the orange line with the oscillon profile with $\omega/m = 0.85$, while the black dots correspond to the $g_{\text{cri}}$ read off from the numerical simulation results shown in Fig.~\ref{fig:massive mu compare}. }\label{fig:WKB g_cri}
\end{figure}

On the other hand, for the tachyonic band in Region $\text{\Rmnum{4}}$ (and Region $\text{\Rmnum{2}}$), the inhomogeneous background has a more significant influence on the band range.
In order to investigate its effect, we analyze the equation of motion of $\chi$ in the coordinate space.
We can decompose $\chi(t,r)$ as the following, assuming it has a dominant mode component,
\begin{align}
    \chi(r,t) = \frac{u_0(r)}{r} \Re(e^{\widetilde{\mu} t}),
\end{align}
where $\widetilde{\mu} = \Re{(\widetilde{\mu})} + i\omega_\chi$ is generally a complex number with $\Re{(\widetilde{\mu})}$ as the exponential growth rate and $\omega_\chi$ as the oscillating frequency.
We note the definition of $\widetilde{\mu}$ is related to the previous $\mu$ defined in Eq.~\eqref{eq:floquet theorem} by $\widetilde{\mu} = \omega \mu$, where $\omega$ is the frequency of $\phi$ oscillon.
Substituting this decomposition into the spherical symmetric equation of motion for $\chi$ in Eq.~\eqref{eq:chi eom},
\begin{align}
    -\frac{d^2 u_0(r)}{dr^2} + [m_\chi^2 + 4g\psi^2(r) (1 +  \cos{(2 \omega t))}] u_0(r) = -(\Re{(\widetilde{\mu})^2} - \omega_\chi^2)  u_0(r).
\end{align}
To investigate the effect of the inhomogeneity, we neglect the oscillating term first.
Then this equation is approximately
\begin{align}
    -\frac{d^2 u_0(r)}{dr^2} +  \mathcal{V}_{\text{eff}}(r) u_0(r) = \mathcal{E} u_0(r), \label{eq:1d Schrodinger eq}
\end{align}
which is a one-dimensional Schrödinger equation with effective potential and energy eigenvalues
\begin{align}
    &\mathcal{V}_{\text{eff}}(r) = 4g\psi^2(r),\label{eq:effective potential}\\
    &\mathcal{E}= -(m_\chi^2 + \Re{(\widetilde{\mu})}^2- \omega_\chi^2).\label{eq:energy eigenvalue}
\end{align}
With the localized inhomogeneous oscillon profile $\psi(r)$ which is close to a Gaussian function, this $\mathcal{V}_{\text{eff}}(r)$ is an attractive finite potential well with negative $g$ and has $\mathcal{V}_{\text{eff}}(\infty) = 0$.
When the equation of motion for $\chi$ is dominated  by the tachyonic term, the oscillation of the solution is negligible compared with tachyonic growth, i.e., $\Re{(\widetilde{\mu})} \gg \omega_\chi$.
Thus, a visible exponential growth of $\chi$ due to the tachyonic instability on a background of an oscillon configuration is equivalent to the existence of a bound state with energy eigenvalue $\mathcal{E} \approx-(m_\chi^2 + \Re{(\widetilde{\mu})}^2) < -m_\chi^2$ for Eq.~\eqref{eq:1d Schrodinger eq}.
We compute the energy eigenvalues of the bound states for Eq.~\eqref{eq:1d Schrodinger eq} by finite differential method with $\psi^2(r)$ substituted as the oscillon profile solved numerically.
The left panel of Figure~\ref{fig:WKB g_cri} shows the bound state energy eigenvalues obtained by finite difference method numerically with an oscillon profile of various $\omega$.
Then the value of $g_\text{cri}^-$ is determined when the ground state energy $\mathcal{E}_0 = -m_\chi^2$, as shown in the right panel of Fig.~~\ref{fig:WKB g_cri}.
Or it can be simply estimated by the WKB approximation,
\begin{align}
    \int_0^{\infty} dr \sqrt{-m_\chi^2 - \mathcal{V}_{\text{eff}}(r)} \approx \frac{\pi}{2}. \label{eq:WKB condition}
\end{align}

\begin{figure}[t]
     \begin{subfigure}[b]{0.48\textwidth}
        \centering        \includegraphics[width=\textwidth]{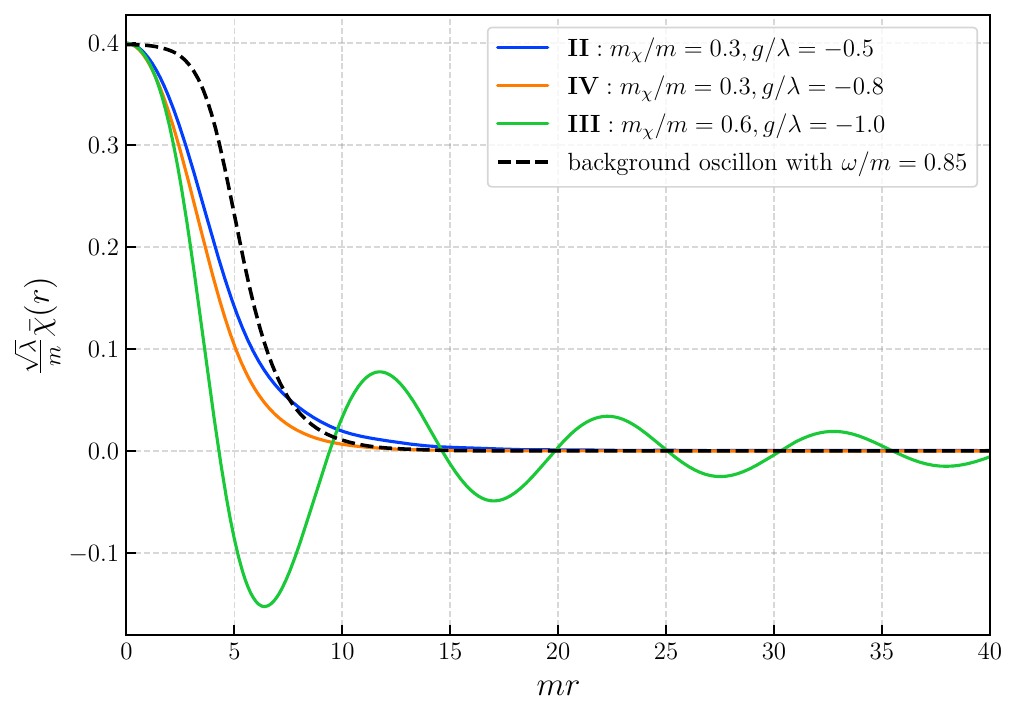}
    \end{subfigure}
    \hfill
    \begin{subfigure}[b]{0.48\textwidth}
        \centering        \includegraphics[width=\textwidth]{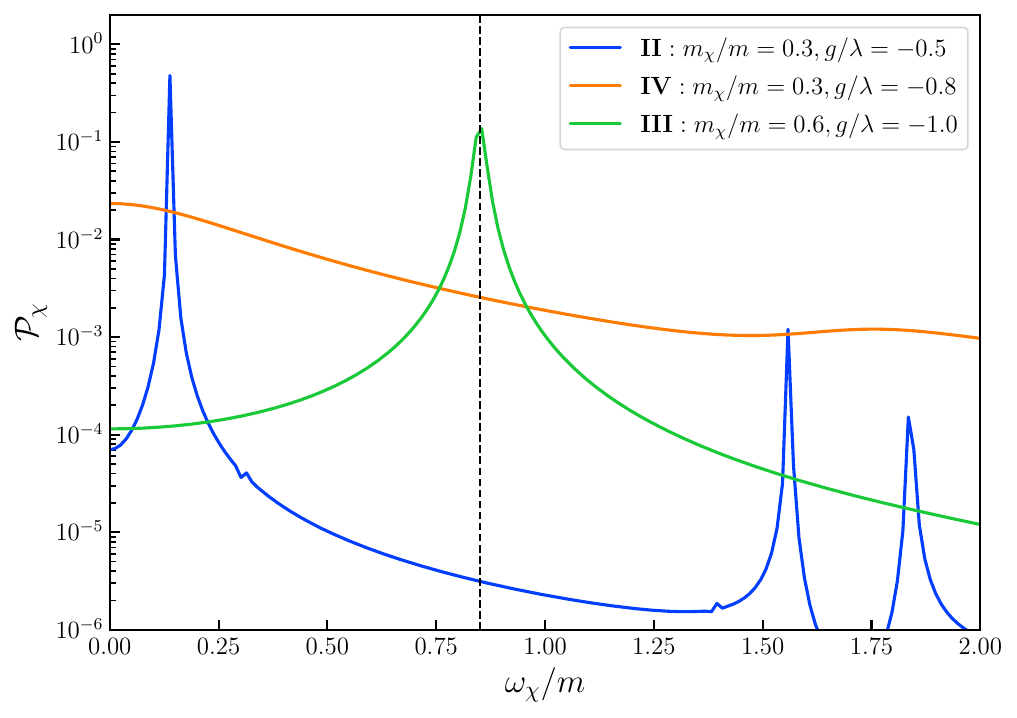}
    \end{subfigure}
    \caption{Time-averaged profile of $\chi$ and the Fourier spectra of its center value, $\chi(t,0)$, obtained from simulations where the background $\phi(t,r)$ fixed as the oscillon configuration with $\omega/m = 0.85$. 
    In order to comparatively demonstrate the shape of the profiles and the spectra,  the profiles are rescaled by the central value $\psi(0)$ of the background field, and the Fourier spectra, defined as $\mathcal{P}\chi \equiv |\chi(\omega_\chi, 0)|^2$, are normalized for illustration purposes.}\label{fig:chi profile & fft}
\end{figure}

We note that the narrow parametric resonance enhances the modes having the same frequency as the background oscillation, i.e., $\omega_\chi = \omega$.
Thus, in our parameter range of interests, $m_\chi < \omega \simeq m$, the modes enhanced in Region $\text{\Rmnum{3}}$ should always be dissipative states of Eq.~\eqref{eq:1d Schrodinger eq} with positive $\mathcal{E}$ (considering $\Re{(\widetilde{\mu})}$ in this region is small), as the green curve shown in Fig.~\ref{fig:chi profile & fft}.

\begin{figure}[t]
     \begin{subfigure}[t]{0.49\textwidth}
        \centering        \includegraphics[width=\textwidth]{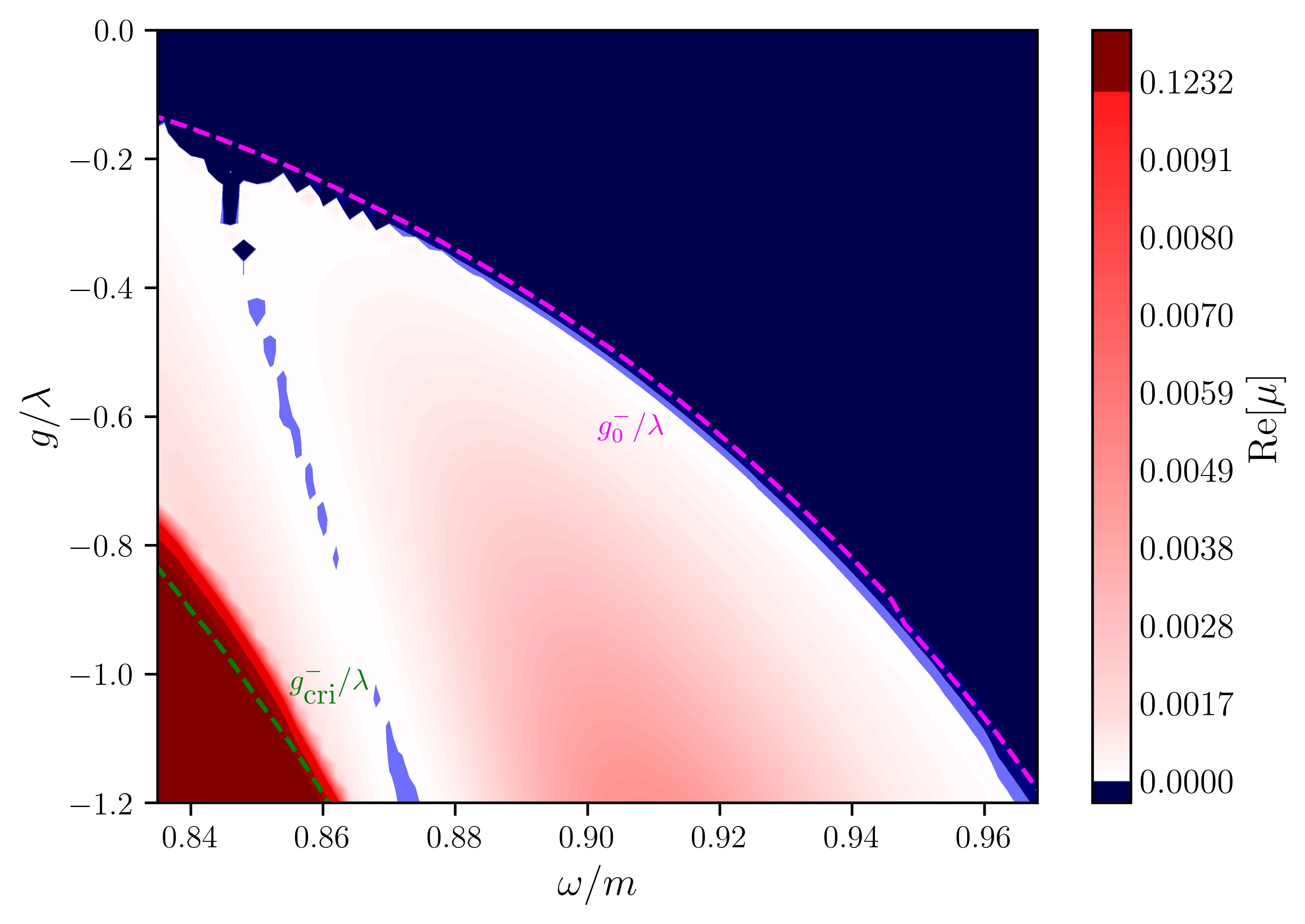}
    \end{subfigure}
    \hfill
    \begin{subfigure}[t]{0.49\textwidth}
        \centering        \includegraphics[width=\textwidth]{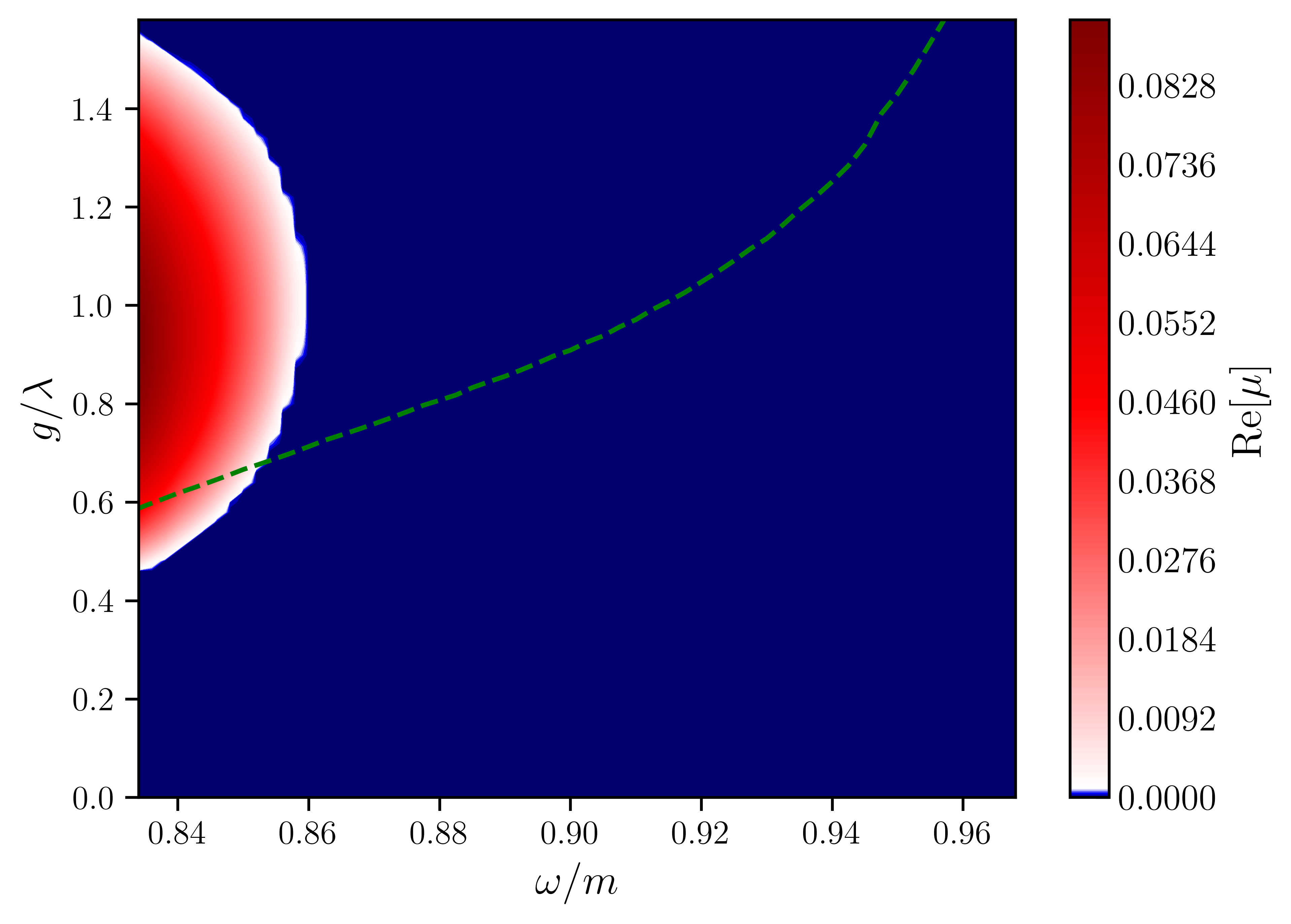}
    \end{subfigure}
    \caption{Contour plots of the exponential growth rate $\Re(\mu)$ of $\chi$ on an oscillon background for various values of $\omega$.
    The left panel corresponds to attractive couplings ($g < 0$) with $m_\chi/m = 0.6$, and the right panel shows results for repulsive couplings ($g > 0$) with $m_\chi/m = 0.3$.
    The blue region is the stable Region $\text{\Rmnum{1}}$ where $\chi$ does not grow exponentially with time. 
    The dark red region corresponds to Region $\text{\Rmnum{3}}$, while the light red region is Region $\text{\Rmnum{2}}$ with $\Re{(\mu)} \sim \mathcal{O}(10^{-3}\sim 10^{-5})$. 
    The green dashed line plots $g_\text{cri}^\pm$ given in Eq.~\eqref{eq:g_cri}, while the magenta dashed line denotes $g_0^-$ obtained numerically by finding the values of $g$ where $\mathcal{V}_{\text{eff}}(r)$ in Eq.~\eqref{eq:effective potential} starts to allow the first bound state. 
    We note that for $g_0^+ = g_\text{cri}^\pm$ for $g>0$.}\label{fig:contour mu}
\end{figure}

Another remarkable phenomenon shown in Fig.~\ref{fig:massive mu compare} and Fig.~\ref{fig:mu_regions} is the appearance of the Region $\text{\Rmnum{2}}$, where a bump of nonzero $\Re{(\mu)} \sim \mathcal{O}(10^{-4})$ occurs for massive $\chi$ at $g_{\text{cri}}^- < g < g_0^-$ but not for repulsive interaction, $g>0$.
Then we define $g_0^\pm$ as another critical value at where $\Re{(\mu)}$ starts to be nonzero, ($g_0^+ = g_{\text{cri}}^+$ since there is no Region $\text{\Rmnum{2}}$ on the positive side).
As we discussed, in this region, neither the attractive coupling should be strong enough to allow any mode of $\chi$ to have tachyonic growth in the finite potential well formed by the background oscillon profile, nor is the narrow parametric resonance fast enough to overcome particle escape rate.
In fact, this region can also be understood by the bound state of Eq.~\eqref{eq:1d Schrodinger eq} with larger energy eigenvalues ($ 0 > \mathcal{E} > -m_\chi^2$) than those that can be enhanced by tachyonic instability when $|g|$ is large.
As shown in the left panel of Fig.~\ref{fig:WKB g_cri}, for $g_{\text{cri}}^- < g < g_0^-$, the bound state still exists with higher energy $-m_\chi^2 < \mathcal{E} < 0$ and their eigenvalues are consistent with the numerical results in Region $\text{\Rmnum{2}}$.
From Eq.~\eqref{eq:energy eigenvalue}, this requires $0 < \omega_\chi^2 - \Re{(\widetilde{\mu})}^2 < m_\chi^2$.
Therefore, the value of $g_0^-$ is determined as the critical value where effective potential $\mathcal{V}_{\text{eff}}(r)$ starts to allow the first bound state, as the magenta dashed line in Fig.~\ref{fig:contour mu}.

To clearly see what happens in this region, we check the specific behavior of $\chi(t,r)$ during the evolution.
We take the time average of $\chi(t,r)$ over a time period starting from $\widetilde{t} = 50/\widetilde{m}_\chi$, which guarantees the relaxation completes, and lasting for $270 \pi/\widetilde{\omega}$.
Figure~\ref{fig:chi profile & fft} shows the normalized time-averaged profile $\bar{\widetilde{\chi}}(\widetilde{r})$ and the normalized Fourier spectra of $\widetilde{\chi}(\widetilde{t},0)$ on the background of an oscillon with $\omega/m = 0.85$ in different regions.
The green lines correspond to a dissipative solution with positive energy eigenvalue in Region $\text{\Rmnum{3}}$ - a spherical Bessel function profile, which corresponds to the mode enhanced by narrow parametric resonance with the same frequency as the background.
The orange line corresponds to the bound state, enhanced by the tachyonic instability in Region $\text{\Rmnum{4}}$, which is clearly localized inside the background oscillon profile in the left panel and has no peak in the frequency domain.
The blue lines demonstrate that in Region $\text{\Rmnum{2}}$, the growing mode is also the bound state, with the profile marginally exceeding the background profile.
The main frequency $\omega_\chi$ in this region is determined by the energy eigenvalues of the bound state, such that there should always be $\omega_\chi \lesssim m_\chi$ to have small negative $\mathcal{E}$ as we confirm in numerical results.
Meanwhile, there are subpeaks at higher frequency on the blue curve in Fig.~\ref{fig:chi profile & fft} corresponding to subdominant dissipative modes.
The source of the enhancement of these modes is considered to be the mixture between the bound state mode and the modes falling in the resonance bands, due to the convolution with inhomogeneous background $\psi(r)$ in Eq.~\eqref{eq:Eom chik}.
This indirect resonance leads only to a very slow production of $\chi$ particles in the bound state modes.
However, since the bound state modes hardly escape, the accumulated particles in the oscillon can still activate Bose enhancement.

Finally, we present the exponential growth rate $\Re{(\mu)}$ of $\chi$ obtained from the results of simulation with various oscillon configurations as a fixed background for negative $g$ with $m_\chi/m = 0.6$ in the left panel of Figure~\ref{fig:contour mu} and for positive $g$ with $m_\chi/m = 0.3$ in the right panel.
The dark red regions indicate the onset of parametric resonance for sufficiently strong coupling, $|g| > |g_{\text{cri}}^\pm|$, 
The green dashed line shows the relation of $g_{\text{cri}}^\pm$ and the oscillon profile shape given in Eq.~\eqref{eq:g_cri}.
We observe that this line coincides with the boundary of the dark red region only at small $\omega$, suggesting that parametric resonance fails to occur for small oscillons, even though their central amplitude is large enough (which we use to get Eq.~\eqref{eq:g_cri}).
This can be interpreted as a consequence of the effective resonance-driving region being too limited in size to sufficiently amplify small-$k$ modes in small oscillons.
Indeed, the instability bands (dark red regions) in Fig.~\ref{fig:contour mu} primarily emerge when the background oscillons have a ``flat-top", which provides a larger effective radius with sufficiently high amplitude to sustain parametric resonance, compared to a Gaussian-like profile whose amplitude rapidly decreases away from the center.
We further confirm that the parametric resonance can happen when we replace the Gaussian-like profile with a ``flat-top" profile that shares the same central amplitude.

In the left panel of Fig.~\ref{fig:contour mu}, the light red region corresponds to the intermediate Region $\text{\Rmnum{2}}$, $g_{\text{cri}} < g < g_0^-$ with small $\Re{(\mu)} \sim \mathcal{O}(10^{-4})$.
The magenta dashed line shows the critical value where the first bound state appears for the one-dimensional Schrödinger equation Eq.~\eqref{eq:1d Schrodinger eq}, which coincides with the boundary of the light red region very well.
We can also observe that the strength of the enhancement in this region is related to the oscillon profile $\psi(r)$ as well.
The oblique cleft in the contour plot around $\omega/m = 0.86 \sim 0.88$ may come from the certain shape of $\psi(r)$ causing weak or no coupling between the mode corresponding to the bound state with the resonance modes.
We leave the precise analysis of $\Re{(\mu)}$ in this region as future work.

From Fig.~\ref{fig:contour mu}, we can define a critical value of oscillon energy $\overline{E}_{0}^{\textrm{osc}}$ (corresponding to a critical frequency $\omega_0$) for a given $g$ and $m_\chi$ as
\begin{align}
    \Re(\tilde{\mu}) \begin{cases}
        >0, \quad \overline{E} > \overline{E}_{0}^{\textrm{osc}} ~(\omega<\omega_0)\\
         = 0, \quad \overline{E} \leq \overline{E}_{0}^{\textrm{osc}} ~(\omega\geq\omega_0)
    \end{cases}
\end{align}
where $\overline{E}$ is the energy of oscillon obtained by Eq.~\eqref{eq:time-average energy from oscillon profile}.
This implies that during the decay of oscillons, the resonance of $\chi$ may stop when the oscillons become small enough, which is confirmed by our simulation in the next section.

\section{Two-field simulation of an oscillon with external coupling}\label{sec:two-field simulation}
So far, we have investigated the exponential growth for $\chi$ field, both in the cases for homogeneous approximation and fixed spatial dependent oscillon profiles in the last section.
However, as $\chi$ grows over time, energy is extracted from the $\phi$ sector, so we expect the background $\phi$ to decay at least at a rate of $\chi$ growth.
This is compounded by the fact that $\phi$ itself decays slowly with time as an oscillon, which in turn affects the instability band of $\chi$.
The inspiration from the instability band of $\chi$ we found in the last section implies that in actual evolution with given coupling strength $g$, the oscillon could decay through both self classical radiation and $\chi$'s resonance until the increasing frequency (corresponding to decreasing energy and charge) reaches the boundary of $\chi$'s instability band plotted by the magenta curves in Fig.~\ref{fig:contour mu}.

In this section, we perform a complete numerical simulation for both time-evolving $\phi$ and $\chi$ fields.
We follow the same numerical setup in Sec.~\ref{subsec:decay and lifetime}.
By fully simulating the time evolution of the equation of motions for both $\phi$ and $\chi$ fields in Eq.~\eqref{eq:full phi eom} \eqref{eq:chi eom}, we investigate the whole decay of the oscillon with external $\chi$ coupled, including the backreaction of the exponentially growing $\chi$ field on the oscillon, and the growth rate of $\chi$ field on the background of evolving $\phi$ oscillons.
The initial conditions are input as the following 
\begin{align}
    \widetilde{\phi}(0, \widetilde{r}) &= 2\widetilde{\psi}(\widetilde{r}),\\
    \widetilde{\chi} (0, \widetilde{r}) &= \widetilde{\chi}_0 \bar{\widetilde{\chi}}(\widetilde{r}),\\
    \dot{\widetilde{\phi}} (0, \widetilde{r})&= \dot{\widetilde{\chi}}(0, \widetilde{r}) = 0,
\end{align}
where $\widetilde{\psi}(\widetilde{r})$ is the single-field oscillon profile solved numerically in Sec.~\ref{subsec:single field oscillon profile} for a given value of $\widetilde{\omega}$, $\bar{\widetilde{\chi}}(\widetilde{r})$ is the normalized time-averaged profile of $\chi$ relaxed on the fixed oscillon configuration background with given values of $\widetilde{\omega}, \widetilde{m_\chi}$ and $\widetilde{g}$, as obtained in Sec.~\ref{subsec: bgfix}.
In all the following simulations, we take $\widetilde{\chi}_0 = 0.01$ unless otherwise specified.
The adiabatic damping boundary condition is imposed on both of the fields on the outer edge of the box.
The energy of both sectors is computed as Eq.~\eqref{eq:time-average energy from oscillon profile},\eqref{eq:chi energy}, and the total energy is computed as 
\begin{align}
    \widetilde{\overline{E}}_{\textrm{tot}} (\widetilde{t}) = \widetilde{\overline{E}}_{\phi} (\widetilde{t})+ \widetilde{\overline{E}}_{\chi}(\widetilde{t}) +\frac{1}{T_{\textrm{ave}}} \int_{\widetilde{t}}^{\widetilde{t}+T_{\textrm{ave}}} d\widetilde{t} \int_0^{\widetilde{r}_{max}} d\widetilde{r}~ 4\pi \widetilde{r}^2 \left(\widetilde{g} \widetilde{\phi}^2\widetilde{\chi}^2 \right),
\end{align}
where we take $T_{\textrm{ave}} = 20 m^{-1}$ and $\widetilde{r}_{max} = 30$.
We note that this implies the total energy should always stay constant at least until $\widetilde{t} = \widetilde{r}_{max}$, after which it may begin to decrease due to dissipative modes propagating beyond the monitored box size $\widetilde{r}_{max}$.

It is intuitive to anticipate that when the growth rate of $\chi$ is relatively large, the backreaction from large field value of $\chi$ will destroy the structure of oscillon in $\phi$ at a timescale about $\tau \sim 1/\Re{(\widetilde{\mu})}$.
We note that the initial amplitude $\chi_0$ changes the factor of the timescale $\tau$ here, but this factor is not essential compared to the exponential time dependence.
In the real universe, we can consider this $\chi_0$ determined as the relative time of the $\chi$ field growth and the oscillon decay.
However, our simulation indicates that the growth of $\chi$ field is not necessarily that fatal for oscillons.

\begin{figure}[b]
     \begin{subfigure}{0.48\textwidth}
        \centering        \includegraphics[width=\textwidth]{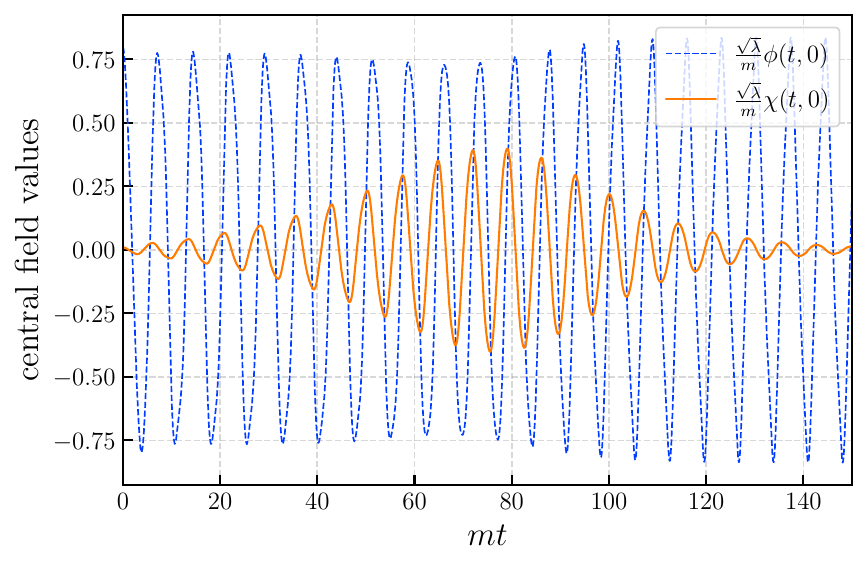}
    \end{subfigure}
    \hfill
    \begin{subfigure}{0.48\textwidth}
        \centering        \includegraphics[width=\textwidth]{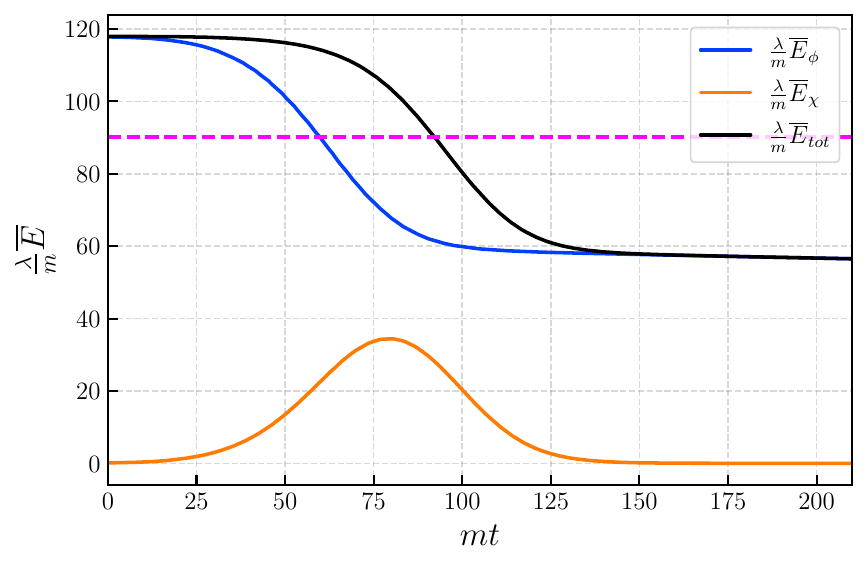}
    \end{subfigure}
    \vfill
    \begin{subfigure}{0.48\textwidth}
        \centering        \includegraphics[width=\textwidth]{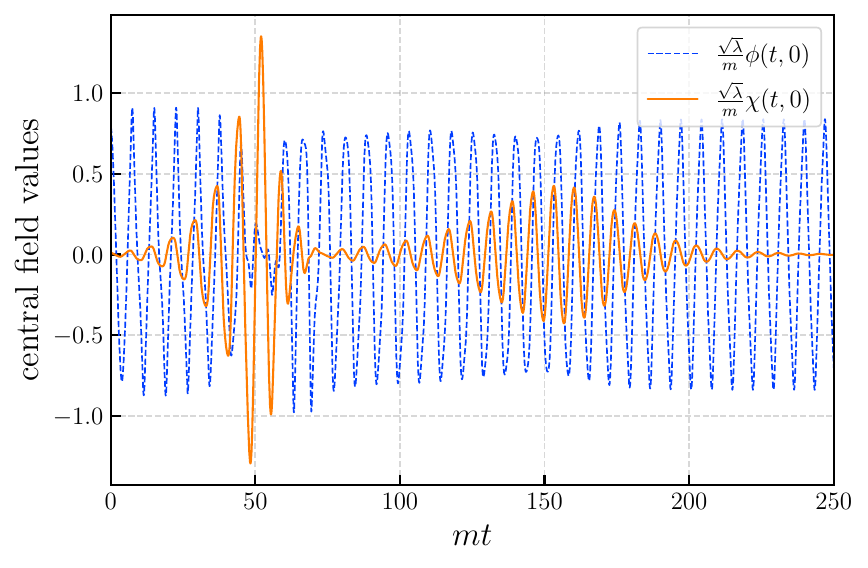}
    \end{subfigure}
    \hfill
    \begin{subfigure}{0.48\textwidth}
        \centering        \includegraphics[width=\textwidth]{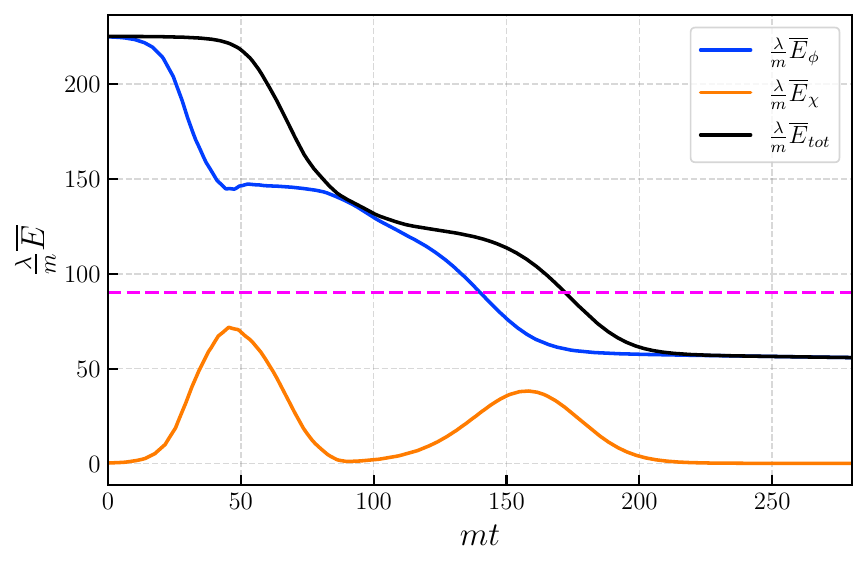}
    \end{subfigure}
    \caption{The time evolution of the center field values of $\phi$ and $\chi$ and the energy in both sectors in a two-field simulation. Top panel: starting from initial oscillon frequency $\omega/m = 0.85$ with $m_\chi/m = 0.3$, $g/\lambda = 1.0$.  Bottom panel: starting from initial oscillon frequency $\omega/m = 0.83$ with $m_\chi/m = 0.3$, $g/\lambda = 1.0$. The magenta dashed line is the corresponding oscillon energy $\overline{E}_{0}^{\textrm{osc}}$ on the boundary of the region plotted in Fig.~\ref{fig:contour mu}, where $\chi$ exits the instability bands and stops growing.}\label{fig:m3g1.0}
\end{figure}

Figure~\ref{fig:m3g1.0} shows the results of simulations with $m_\chi/m = 0.3$, $g/\lambda = 1.0$, starting from the initial profiles of oscillons with $\omega/m = 0.85$ and $\omega/m = 0.83$.
It can be seen from the figure that $\chi$ field experiences an exponential growth due to the parametric resonance for a short period $\lesssim 100 m^{-1}$.
As the energy is taken away from the oscillon by $\chi$, the configuration of $\phi$ changes; namely, the frequency $\omega$ increases and $\psi(r)$ changes shape.
The alteration of oscillon configuration causes the growth rate of $\chi$ to change, as we have shown in the last section, which is why we can observe the growth of $\overline{E}_\chi$ becoming gradually slow from the top right panel of Fig.~\ref{fig:m3g1.0}.
And the resonance eventually stops when the oscillon energy reaches the boundary of the instability band for the certain $g$ value, then the $\chi$ energy propagates away.
After $\chi$ exits the stage, the oscillon regains peace and returns to the original decay trajectory as a single-field oscillon with energy remained.
For more violent growth of $\chi$, for instance, starting from a larger initial oscillon with $\omega/m = 0.83$ as shown in the bottom panel of Fig.~\ref{fig:m3g1.0}, the oscillon amplitude is dramatically suppressed by the backreaction of large $\chi$ field value, which leads to a temporary termination of parametric resonance, at around $t \simeq 50m^{-1}$, but restarts at $t \simeq 90m^{-1}$ when $\phi$ calms down after the most energy of $\chi$ leaves the oscillon region until it stops eventually.

\begin{figure}[b]
    \begin{subfigure}{0.48\textwidth}
        \centering        \includegraphics[width=\textwidth]{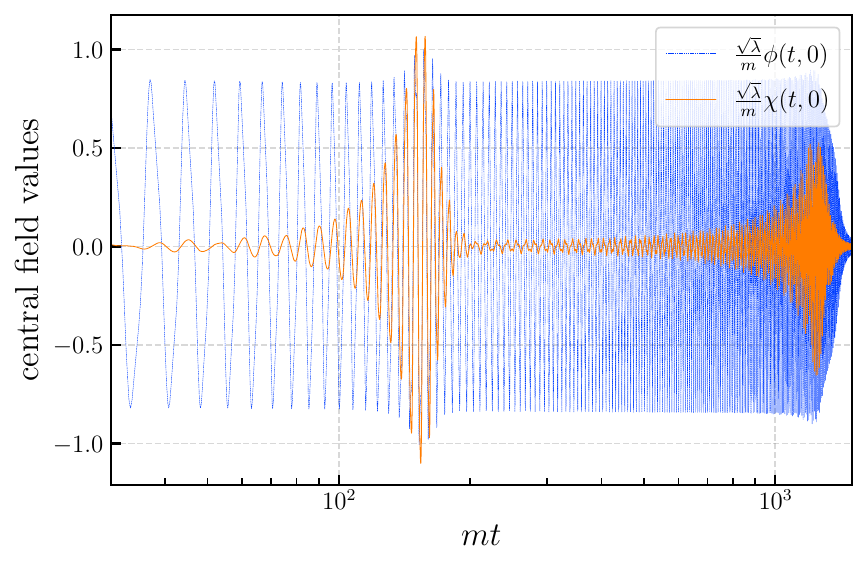}
    \end{subfigure}
    \hfill
    \begin{subfigure}{0.48\textwidth}
        \centering        \includegraphics[width=\textwidth]{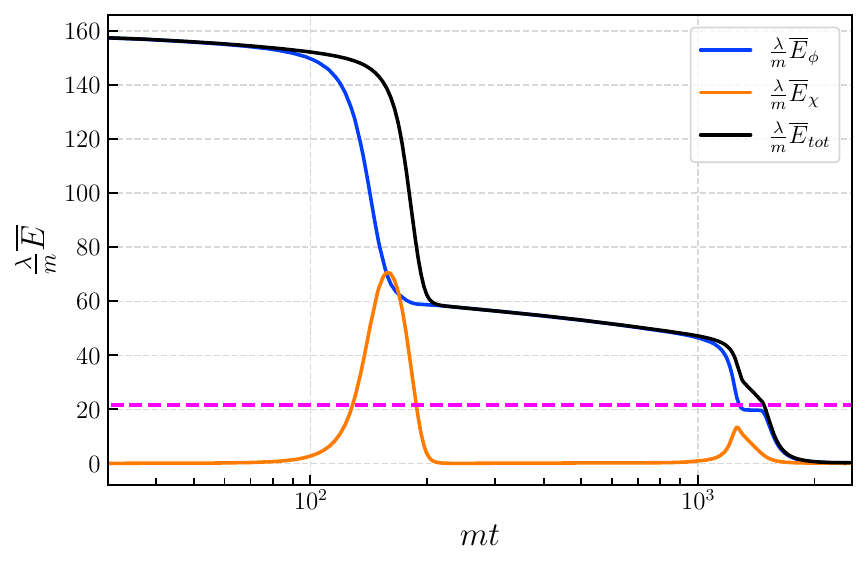}
    \end{subfigure}
    \vfill
    \begin{subfigure}{0.48\textwidth}
        \centering        \includegraphics[width=\textwidth]{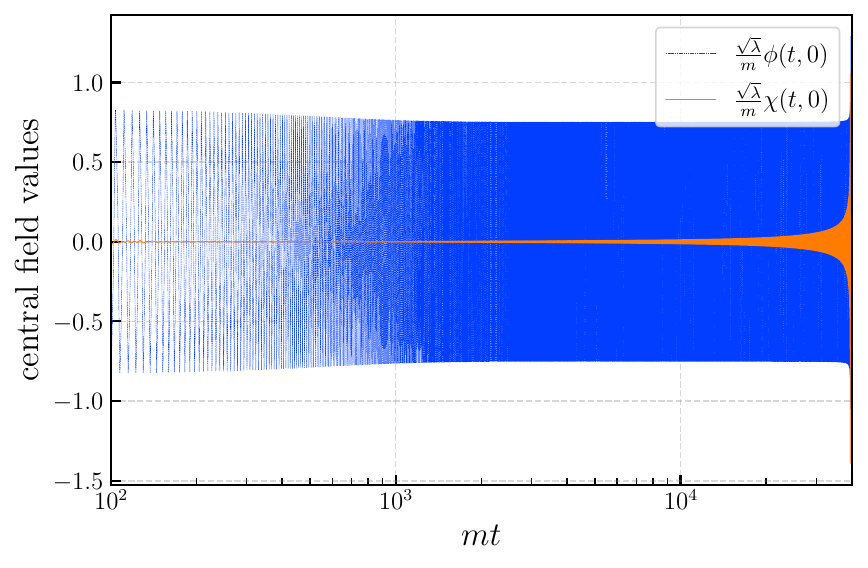}
    \end{subfigure}
    \hfill
    \begin{subfigure}{0.48\textwidth}
        \centering        \includegraphics[width=\textwidth]{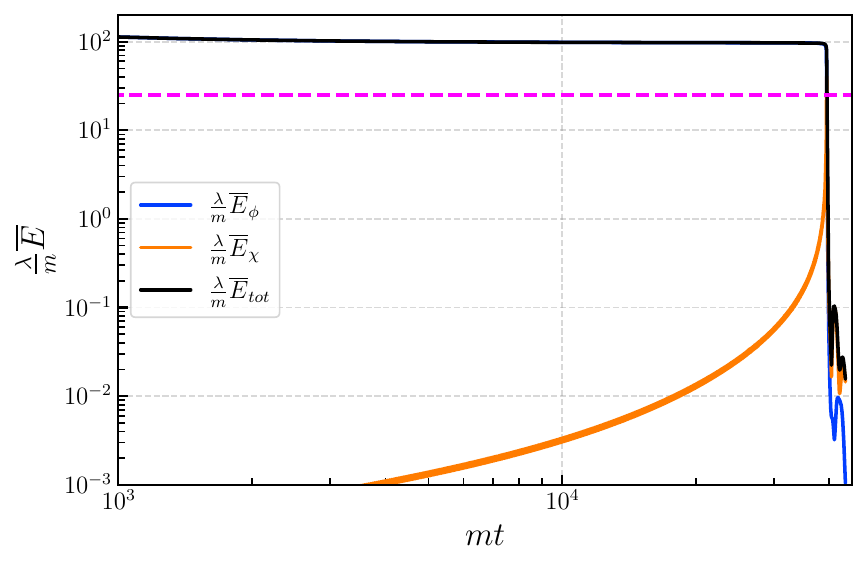}
    \end{subfigure}
    \caption{The time evolution of the center field values of $\phi$ and $\chi$ and the energy in both sectors in a two-field simulation. Top panel: starting from initial oscillon frequency $\omega/m = 0.84$ with $m_\chi/m = 0.6$, $g/\lambda = -1.0$.  Bottom panel: starting from initial oscillon frequency $\omega/m = 0.84$ with $m_\chi/m = 0.6$, $g/\lambda = -0.8$. The magenta dashed line is the corresponding oscillon energy $\overline{E}_{0}^{\textrm{osc}}$ on the boundary of the instable region plotted in Fig.~\ref{fig:contour mu}, where $\chi$ stops growing.}\label{fig:m6negG}
\end{figure}

For negative $g$, because of the complexity of the instability bands, the behavior can be slightly different.
In Figure~\ref{fig:m6negG}, we give results of simulations, starting from initial oscillon frequency $\omega/m = 0.84$, with $m_\chi/m = 0.6$ for attractive coupling $g=-1.0$ and $g=-0.8$.
The top panel also shows two stages of $\chi$ growth, but the first stage is driven by narrow parametric resonance in Region $\text{\Rmnum{3}}$, while the second stage corresponds to Region $\text{\Rmnum{2}}$.
The different mechanisms of the two stages excite different modes as we discussed in the last section, which makes the two stages appear to be not consecutive.
Furthermore, due to Region $\text{\Rmnum{2}}$ in the negative $g$ range, $\overline{E}_{0}^{\textrm{osc}}$ can be very close to (or beyond) the critical energy of the single-field oscillon destruction energy, i.e., the 'energy death'.
The bottom panel of Fig.~\ref{fig:m6negG} demonstrates a case where $\chi$ causes $\overline{E}_\phi$ to drop below $\overline{E}_{\textrm{death}}$, thereby driving the oscillon to immediate destruction.

\begin{figure}[t]
    \centering        \includegraphics[width=\textwidth]{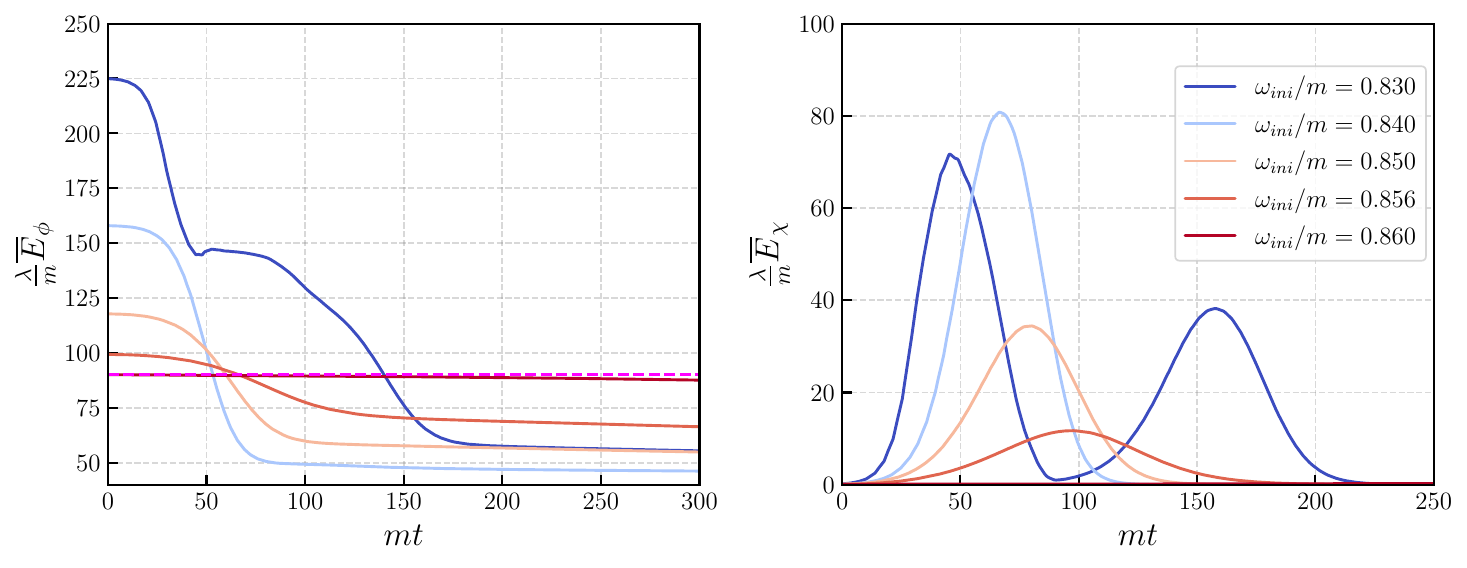}
    \caption{The time evolution of the energy of $\phi$ and $\chi$ in the simulation starting from initial oscillons with various frequencies (and energies) for $g/\lambda = 1.0$ and $m_\chi/m = 0.3$. The magenta dashed line is the corresponding oscillon energy $\overline{E}_{0}^{\textrm{osc}}$ on the boundary of the region plotted in Fig.~\ref{fig:contour mu}, where $\chi$ exits the instability bands and stops growing.} \label{fig:m3g1.0_various_omega}
\end{figure}

More specifically, we see from the two-field simulation, $\chi$ often stops growing slightly later than when the oscillon reaches $\overline{E}_{0}^{\textrm{osc}}$. 
This likely results from the nonlinear backreaction $\phi$, which distorts its spatial profile and slightly changes the instability band compared to a fixed oscillon background.
Besides, even after the termination of parametric resonance, the backreaction from $\chi$ can still excite dissipative modes of $\phi$ before $\chi$ fully disappears.
Figure~\ref{fig:m3g1.0_various_omega} shows the evolution of $\overline{E}_\phi$ and $\overline{E}_\chi$ starting from different initial oscillon frequencies $\omega$ (energy) for $g/\lambda = 1.0$ and $m_\chi/m = 0.3$.
It suggests that more energy stored in the $\chi$ sector at the end of growth leads to greater energy loss through dissipative modes, reducing the final energy remaining as oscillon.
As a result, the residual energy of an oscillon after the entire influence of $\chi$ is always below $\overline{E}_{0}^{\textrm{osc}}$, indicating that $\overline{E}_{0}^{\textrm{osc}}$ from the instability band analysis serves as an upper limit for the remaining energy - beyond where the oscillon decay and lifetime returns identical to the single-field oscillon.
This also explains the lattice results in Ref.~\cite{shafi_formation_2024}, which show the fractional energy contained in oscillons can be partially retained for a $g$ not that large.
Now, we can summarize the energy decay rate and lifetime of an oscillon as the following.
With the presence of the external coupling, the parametric resonance of the external scalar field becomes the primary decay channel, with a decay rate of $\Gamma_{\chi} \sim \Re{(\widetilde{\mu})}$.
However, the resonance rate depends on the oscillon energy $\overline{E}$ given in Eq.~\eqref{eq:time-average energy from oscillon profile}, and vanishes when $\overline{E}$ becomes smaller than a critical value $\overline{E}_{0}^{\textrm{osc}}$ obtained by the instability band analysis for any given coupling $g$ and $m_\chi$.
If $\overline{E}_{0}^{\textrm{osc}} \gtrsim \overline{E}_{\textrm{death}}$, the final stage of oscillons decay proceeds without non-perturbative decay into $\chi$ and has a decay rate dominated  by $\Gamma_\xi$ identical with the single-field case given in Eq.~\eqref{eq:gamma of single field} (assuming perturbative decay through $\phi\phi \to \chi\chi$ is negligible, which is not the main focus of our present work).
Thus, when $\overline{E}_{0}^{\textrm{osc}} \gtrsim \overline{E}_{\textrm{death}}$, the dominant energy decay rate $\Gamma$ for each stage is roughly
\begin{align}
    \Gamma(\overline{E}) \sim 
    \begin{cases}
        \Re{(\tilde{\mu})} , \quad \overline{E} \gtrsim \overline{E}_{0}^{\textrm{osc}}\\
        \Gamma_{\xi}, \quad  \overline{E}_{\textrm{death}}< \overline{E} \lesssim \overline{E}_{0}^{\textrm{osc}}.
    \end{cases}
\end{align}
When $\overline{E}_{0}^{\textrm{osc}} \lesssim \overline{E}_{\textrm{death}}$,
\begin{align}
    \Gamma(\overline{E}) \sim 
        \Re{(\tilde{\mu})} , \quad \overline{E} \gtrsim \overline{E}_{\textrm{death}}.
\end{align}
Now, the total lifetime of an oscillon with initial energy $\overline{E}_{\textrm{ini}}$ can be order estimated by 
\begin{align}
    \tau(\overline{E}_{\textrm{ini}}) \sim \sum_i\frac{1}{\Re(\mu_i)} + \tau_{\textrm{single}}(\min (\overline{E}_{0}^{\textrm{osc}}, \overline{E}_{\textrm{ini}})),
\end{align}
where the first term is the summation over the duration of $\chi$ resonance in different Region $i$ crossed from initial $\overline{E}_{\textrm{ini}}$ to $\overline{E}_{0}^{\textrm{osc}}$, which is zero if $\overline{E}_{\textrm{ini}} \leq \overline{E}_{0}^{\textrm{osc}}$ and the second term is estimated as the lifetime of a single-field oscillon given in Eq.~\eqref{eq:single field tau}, which has $\tau_{\textrm{single}}(\overline{E}_{0}^{\textrm{osc}}) = 0$ if $\overline{E}_{0}^{\textrm{osc}} \lesssim \overline{E}_{\textrm{death}}$.

In the real universe, a population of oscillons can be formed by preheating with a range of initial energies, as demonstrated by the lattice simulations in Ref.~\cite{shafi_formation_2024}.
The energy distribution of oscillon at formation is not well investigated, but our finding suggests that the coupling to another scalar field $\chi$ can later align the oscillon energy to a narrower range with an upper limit determined by the boundary of the instability band of $\chi$.
This is consistent with the results in Ref.~\cite{shafi_formation_2024}, which indicate that the energy fraction retained in oscillons within an expanding universe is not entirely depleted by the external field when the coupling is not extremely large.
Since this work focuses on individual oscillons, we do not account for effects arising from a large population of oscillons, such as the large value $\chi$ field propagating outward in our simulation may encounter and affect nearby oscillons in the universe.
Such collective effects may underlie the findings in Ref.~\cite{shafi_formation_2024}, where almost no energy remains in oscillons for sufficiently strong repulsive external interactions.
Furthermore, an analysis that includes the energy distribution of oscillons at formation may be essential for linking our findings based on individual oscillons, to the conclusions regarding lifetime defined by the spatially averaged behavior in the universe as presented in Ref.~\cite{shafi_formation_2024}.

\section{Discussions and conclusions}\label{sec:conclusion}

In this work, we investigate the decay and lifetime of oscillons consisting of a $\phi$ field in a sextic potential with an external four-point coupling to a scalar field $\chi$ by numerically simulating an individual oscillon under spherical symmetry.
The oscillating oscillon configuration can initiate parametric resonance of the $\chi$ field, which leads to its exponential growth.
We compute the instability bands for $\chi$ in numerical simulations by both approximately neglecting spatial profile of the oscillon and fixing the oscillon profile by hand.
We find, across different ranges of coupling constant $g$, the main mechanism driving the exponential growth varies, leading to different magnitudes and dependencies of the growth exponents $\Re{(\mu)}$ of $\chi$.
Specifically, for repulsive interaction ($g>0$), the primary mechanism is given by the net effect of parametric resonance reduced by the particle escaping rate of the external field from the localized resonance region related to the finite radius of the oscillon. 
For $g<0$, in addition to the parametric resonance and tachyonic instability that occur at large $|g|$, we observe that modes corresponding to the bound state in the finite well constituted by oscillon profile can acquire a tiny exponential growth (of order $\mathcal{O}(10^{-4})$ in our model) through mode-mixing with the resonance modes.
Moreover, we discover that small oscillons are unable to sustain parametric resonance due to their limited size, especially when the instability bands primarily involve small-$k$ modes of the external field.
This implies that the external field may cease to grow once the oscillon becomes sufficiently small.
Also, the explosive decay into the external field particles through parametric resonance may fail to occur effectively if the oscillons formed during preheating in certain models predominantly have low energies.

Further, we conduct full numerical simulations under spherical symmetry for two fields evolving together.
And we find that in spite of the quick exponential growth of $\chi$ field, oscillons are not destroyed immediately by the large $\chi$ field.
Rather, the backreaction on the oscillon may terminate the resonance of $\chi$ by suppressing the oscillon amplitude temporarily.
After the large $\chi$ field propagates away, oscillons configurations can recover with the remaining energy and charge, which may result in resonance of $\chi$ again if they are still large enough.
Thus, the boundary of the instability bands of $\chi$, corresponding to the energy of the smallest oscillon that can activate the enhancement for $\chi$ no matter what the mechanism is, places an upper limit on the residual energy in the oscillon after the entire evolution of $\chi$.
If the residual energy is still larger than the critical energy of oscillon destruction, the oscillon decays as a single-field oscillon thereafter.
Otherwise, if the resonance terminates when the oscillon energy is too close to the critical energy of ``energetic death'' (usually at strong coupling), the oscillon can be destroyed before $\chi$ propagates away completely.
Therefore, the oscillon lifetime can be estimated by a summation of two stages: the $\chi$ growing phase lasting for $1/\Re{(\mu)}$, and the single-field oscillon decay phase after $\chi$ ceases to grow and propagates out, whose duration depends on the remaining energy of the oscillon.

\section*{Acknowledgments}
We would like to thank Gen Sekita for his contributions at the initial stage of this project.
We are grateful for valuable discussions with Edmund J. Copeland, Kaloian D. Lozanov, Swagat S. Mishra, and Paul M. Saffin.
S.L. is supported by JSPS Grant-in-Aid for Research Fellows Grant No.23KJ0936, and by IBS under the project code, IBS-R018-D3.
M.Y. is supported by IBS under the project code, IBS-R018-D3, and by JSPS Grant-in-Aid for Scientific Research Number JP23K20843.
Y.Z. is supported by the Fundamental Research Funds for the Central Universities, and by the Project 12475060 and 12047503 supported by NSFC, Project 24ZR1472400 sponsored by Natural Science Foundation of Shanghai, and Shanghai Pujiang Program 24PJA134.

\appendix

\section{Normalization}\label{app:normalization}
We write down the normalization used in all the numerical calculations of this work in this appendix.
The action of the oscillon consisting of $\phi$ with external coupling to field $\chi$ is 
\begin{align}
    \mathcal{S} = \int d^4 x \left( \frac{1}{2}\partial_\mu \phi \partial^\mu \phi - \frac{1}{2} m^2 \phi^2 + \lambda \phi^4 - g_6\phi^6 + \frac{1}{2}\partial_\mu \chi \partial^\mu \chi - \frac{1}{2} m_
    \chi^2 \chi^2 - g\phi^2 \chi^2\right),
\end{align}
Defining the dimensionless variables:
\begin{align}
    \widetilde{x}^\mu = m x^\mu, ~ \widetilde{\phi} = \frac{\sqrt{\lambda} \phi}{m},~\widetilde{g_6} = \frac{m^2 g_6}{\lambda^2}, ~\widetilde{\chi} = \frac{\sqrt{\lambda} \chi}{m}, ~\widetilde{m}_\chi = \frac{m_\chi}{m}, ~\widetilde{g} = \frac{g}{\lambda}.
\end{align}
Then the action in terms of the above quantities become
\begin{align}
    \mathcal{S} &= \frac{1}{\lambda}\int d^4\widetilde{x}\left( \frac{1}{2} \partial_{\widetilde{\mu}} \widetilde{\phi} \partial^{\widetilde{\mu}} \widetilde{\phi} + \frac{1}{2} \partial_\mu \widetilde{\chi} \partial^\mu \widetilde{\chi} - V(\widetilde{\phi}) - \mathcal{V}(\widetilde{\chi}) - \widetilde{g} \widetilde{\phi}^2 \widetilde{\chi}^2\right),\\
    V(\widetilde{\phi}) &= \frac{1}{2} \widetilde{\phi}^2 - \widetilde{\phi}^4 + \widetilde{g_6} \widetilde{\phi}^6,~~\mathcal{V}(\widetilde{\chi}) =  \frac{1}{2} \widetilde{m}_\chi^2 \widetilde{\chi}^2.
\end{align}
The dimensionless equations of motion, which we actually solve in the numerical simulation is
\begin{align}
    \Ddot{\widetilde{\phi}} - \nabla^2 \widetilde{\phi} + \widetilde{\phi} - 4 \widetilde{\phi}^3 + 6 \widetilde{g_6} \widetilde{\phi}^5 + 2\widetilde{g} \widetilde{\chi}^2\widetilde{\phi}= 0,\\
    \Ddot{\widetilde{\chi}} - \nabla^2 \widetilde{\chi} + \widetilde{m}_\chi^2 \widetilde{\chi} + 2\widetilde{g} \widetilde{\phi}^2\widetilde{\chi}= 0.
\end{align}
Now the free parameters in the model are $\widetilde{g_6}$, $\widetilde{m}_\chi$ and $\widetilde{g}$.
Under the single-frequency approximation , $\psi$ is normalized the same as $\phi$, $\widetilde{\psi} = \frac{\sqrt{\lambda} \psi}{m}$, then the dimensionless equation giving the oscillon profile is
\begin{align}
    \frac{\partial^2 \widetilde{\psi}}{\partial \widetilde{r}^2} + \frac{2}{r}\frac{\partial \widetilde{\psi}}{\partial \widetilde{r}} = (1-\widetilde{\omega}^2)\widetilde{\psi} -12\widetilde{\psi}^3 + 60\widetilde{g_6}\widetilde{\psi}^5,\label{eq:normalized profile eom}
\end{align}
where $\widetilde{\omega} = \omega/m$, $\widetilde{V_{\textrm{eff}}}(\widetilde{\psi}) = -6 \widetilde{\psi}^4 + 20 \widetilde{g_6}\widetilde{\psi}^6$.

\section{Adiabatic damping boundary condition} \label{app:ADB}

\begin{figure}[bth]
\centering
\includegraphics[width=8cm]{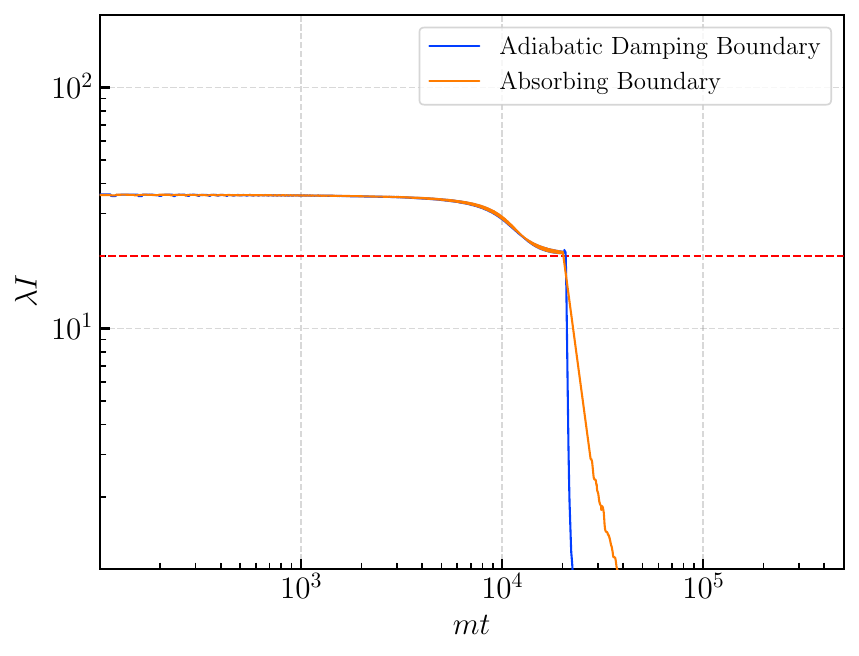}
\caption{Comparison of adiabatic damping boundary result and absorbing boundary condition result of single-field oscillon decay with $\omega_{\textrm{ini}}/m = 0.91$.}\label{fig:ADBvsABC}
\end{figure}

We use the adiabatic damping method \cite{gleiser_long-lived_2000}, to eliminate the effect of reflection wave.
This method is also used in Ref.~\cite{gleiser_generation_2011} for a two-field oscillon with Higgs potential.
We confirm that the lifetime obtained by using adiabatic damping boundary condition is consistent with absorbing boundary condition widely used in Ref.~\cite{salmi_radiation_2012, ibe_fragileness_2019}, which is also consistent with the result of a large enough box size.
We introduce a damping term that is turned on outside the box we used for computing all the quantities, $r > r_{max}$, by
\begin{align}
    \Ddot{X} + \gamma(r)\dot{X} - \nabla^2 X + V'(X) = 0,
\end{align}
where $X$ represents all the fields we evolve in the simulation, namely, $\phi$ and $\chi$.
$\gamma(r)$ is set as $\gamma = 0$ at $r \leq r_{\eta}$ and $\gamma = \eta(r - r_{\eta})^2$ at $r > r_{\eta}$, where $r_{\eta}$ is taken as $r_{\eta} = 64 m^{-1} > r_{max}$, and $\eta$ is a tiny constant taken as $\eta = 0.007$.
Figure~\ref{fig:ADBvsABC} compares the results of single-field oscillon decay under these two types of boundary conditions. 
We see that the difference between them only becomes apparent when the oscillon begins to break down.
Moreover, we have verified that our simulations using the adiabatic damping boundary yield the same results as those with a box size larger than the simulation time.

\section{Two variable analysis of Mathieu's equation} \label{app:mathieu's eq}

The standard Mathieu's equation is:
\begin{equation}
    y''(z) + \left[ A + 2q \cos(2z) \right] y(z) = 0.\label{eq:standard Mathieu's eq App}
\end{equation}

We can use the two variable expansion method \cite{kovacic_mathieus_2018,kevorkian_perturbation_1981} to find a general solution for $q \ll 1 $. 
We introduce a slow time variable $ \eta = qz$ and regard $y$ as a function of $z$ and $\eta$, that is $y=y(z,\eta)$. 
Using the chain rule, $\frac{dy}{dz} = \frac{\partial y}{\partial z} + \frac{d\eta}{dz} \frac{\partial y}{\partial \eta} = \frac{\partial y}{\partial z} + q \frac{\partial y}{\partial \eta}$,
$\frac{d^2 y}{dz^2} = \left( \frac{\partial}{\partial z} + q \frac{\partial}{\partial \eta} \right) \left( \frac{\partial y}{\partial z} + q \frac{\partial y}{\partial \eta} \right)$.

Then the equation can be rewritten as
\begin{equation}
    \frac{\partial^2 y}{\partial z^2} + 2q \frac{\partial^2 y}{\partial z \partial \eta} + q^2 \frac{\partial^2 y}{\partial \eta^2} + \left[ A + 2q \cos(2z) \right] y = 0.
\end{equation}
Expanding \( y(z, \eta) \) in terms of \( q \):

\begin{equation}
    y(z, \eta) = y_0 (z, \eta) + q y_1 (z, \eta) + \dots
\end{equation}
Then the equation in various orders of $q$ give
\begin{align}
    &\mathcal{O}(q^0): \quad \frac{\partial^2 y_0}{\partial z^2} + A y_0 = 0, \\
    &\mathcal{O}(q^1): \quad \frac{\partial^2 y_1}{\partial z^2} + A y_1 = -2 \frac{\partial^2 y_0}{\partial z \partial \eta} - 2 \cos(2z) y_0.
\end{align}

For $A > 0$, the solution to equation of $\mathcal{O}(q^0)$ in terms of $z$ is:
\begin{equation}
    y_0 (z, \eta) = C(\eta) \cos(\sqrt{A}z) + D(\eta) \sin(\sqrt{A}z),
\end{equation}
where $C$ and $D$ are coefficient functions of $\eta$.
Substituting this solution into the equation of $\mathcal{O}(q^1)$,
\begin{align}
    \frac{\partial^2 y_1}{\partial z^2} + A y_1 &= 2\sqrt{A} C'(\eta) \sin(\sqrt{A}z) - 2\sqrt{A} D'(\eta) \cos(\sqrt{A}z) \nonumber \\
    & \quad - C \left[ \cos{\left( (\sqrt{A} + 2)z)\right)} + \cos{\left( (\sqrt{A} - 2)z)\right)} \right] \nonumber \\
    & \quad - D \left[\sin{\left( (\sqrt{A} + 2)z)\right)} + \sin{\left( (\sqrt{A} - 2)z)\right)}\right].
\end{align}
The first two terms on the right-hand side are resonant terms, which yield solutions of the form \( y_1 \sim z [\cos(\sqrt{A}z) + \sin(\sqrt{A}z)] \) growing linearly and dominate the leading order unless they vanish. 
For consistency, this requires
\begin{equation}
    C'(\eta) = 0, \quad D'(\eta) = 0,
\end{equation}
whose solutions are trivially constants.
Thus, for general $A$, there is no effect from the oscillating term, $\cos(2z)$, in the original Mathieu's equation. 
However, for a certain value $ \sqrt{A} - 2 = -\sqrt{A}$, i.e., $ A = 1 $, the equation of $y_1$ becomes
\begin{align}
    \frac{\partial^2 y_1}{\partial z^2} + y_1 &= 2 C'(\eta) \sin z - 2D'(\eta) \cos z \nonumber \\
    & \quad - C ( \cos 3z + \cos z ) - D (\sin 3z - \sin z ).
\end{align}
Then the vanishing resonant terms require matching the coefficients,
\begin{align}
    2 C' + D &= 0, \quad 2D' + C = 0,
\end{align}
which gives the equation of $C$, $4 \frac{d^2 C}{d\eta^2} - C = 0.$
The general solution is
\begin{equation}
    C = c_1 e^{\eta/2} + c_2 e^{-\eta/2}, \quad D = -c_1 e^{\eta/2} + c_2 e^{-\eta/2},
\end{equation}
where $c_1$ and $c_2$ are constants.
Thus, the solution to the standard Mathieu's equation in the leading order when $q \ll 1, A \simeq 1$ is 
\begin{align}
    y_0 &= c_1 e^{\eta/2}(\cos z - \sin z) + c_2 e^{-\eta/2} (\cos z + \sin z) \nonumber \\
    &\simeq \sqrt{2}c_1 e^{\frac{q}{2}z} \cos \left(z + \frac{\pi}{4} \right) + \sqrt{2}c_1 e^{-\frac{q}{2}z} \sin \left(z + \frac{\pi}{4} \right).
\end{align}
This corresponds to the first narrow band on the Floquet chart, where the maximum value of the exponential growth rate $\mu_{\textrm{max}} \equiv \Re{(\mu)}\approx |q|/2$.

\begin{figure}[t]
\centering
\includegraphics[width=8cm]{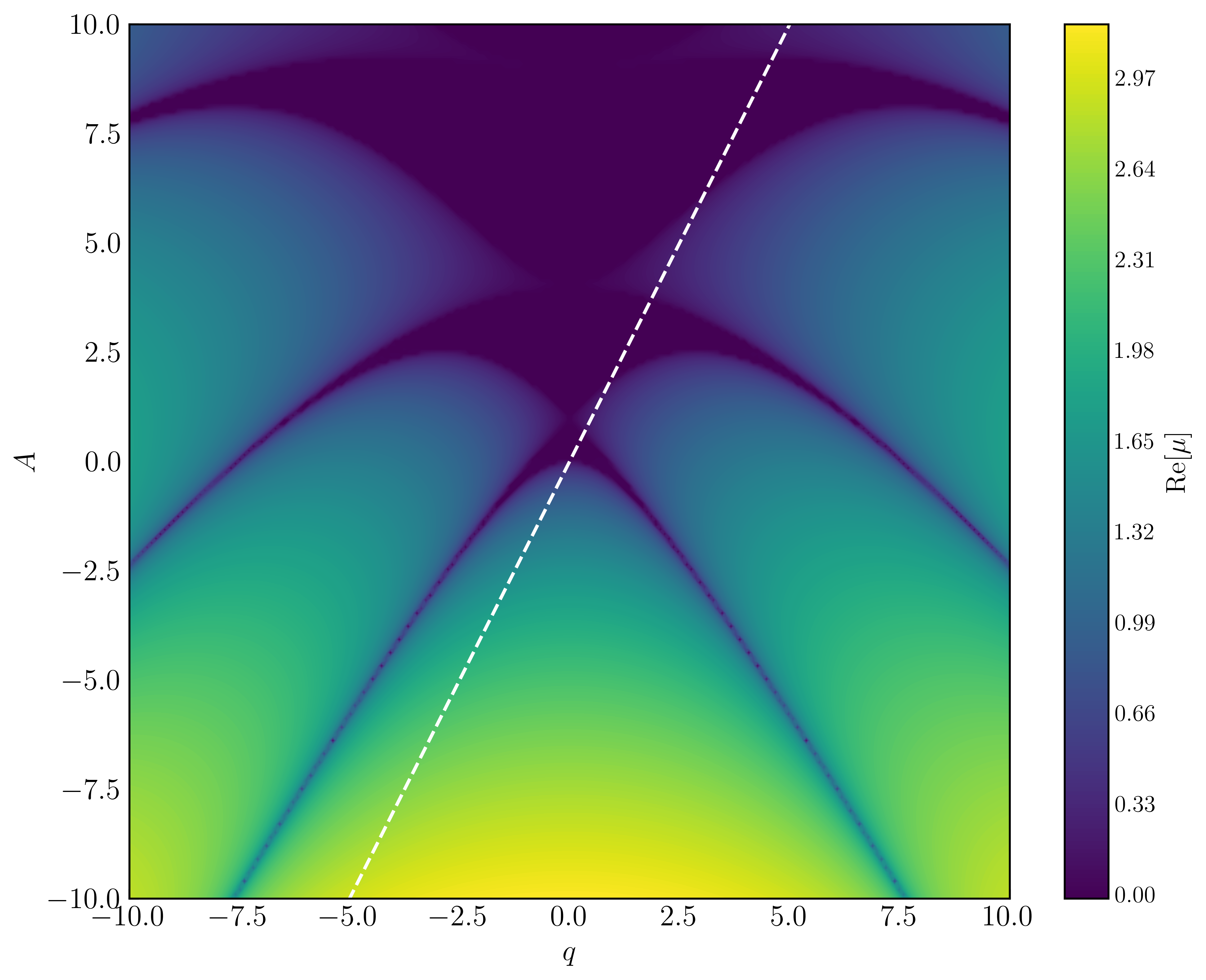}
\caption{Floquet chart obtained numerically for the standard Mathieu's equation, Eq.~\eqref{eq:standard Mathieu's eq App}. The dashed white line is $A=2q$ dividing the narrow bands and broad bands.}\label{fig:standard Floquet chart}
\end{figure}
On the other hand, for $A<0$, the solution to the equation of $\mathcal{O}(q^0)$ in terms of $z$ now becomes
\begin{align}
     y_0 (z, \eta) = C(\eta) e^{\sqrt{|A|}z} + D(\eta) e^{-\sqrt{|A|}z}.
\end{align}
And in all the equations in different orders, the negative $A$ term on the left-hand side dominantly leads the solution to $y_n(z, \eta) \propto e^{\pm \sqrt{|A|}z}$.
Thus, the solution to the standard Mathieu's equation when $q \ll 1, A < 0$ is 
\begin{align}
    y(z) \propto e^{\sqrt{|A|}z},
\end{align}
where Floquet exponent is $\mu \approx \sqrt{|A|}$.

\bibliographystyle{JHEP}
\bibliography{oscillon}

@article{piani_ephemeral_2025,
doi = {10.1088/1475-7516/2025/06/024},
url = {https://dx.doi.org/10.1088/1475-7516/2025/06/024},
year = {2025},
month = {jun},
publisher = {IOP Publishing},
volume = {2025},
number = {06},
pages = {024},
author = {Piani, Matteo and Rubio, Javier and Torrenti, Francisco},
title = {Ephemeral oscillons in scalar-tensor theories: the Higgs-like case},
journal = {Journal of Cosmology and Astroparticle Physics}
}

@article{blaschke_oscillons_2025-1,
  title = {Oscillons from $Q$-balls},
  author = {Blaschke, F. and Roma\ifmmode \acute{n}\else \'{n}\fi{}czukiewicz, T. and S\l{}awi\ifmmode \acute{n}\else \'{n}\fi{}ska, K. and Wereszczy\ifmmode \acute{n}\else \'{n}\fi{}ski, A.},
  journal = {Phys. Rev. D},
  volume = {111},
  issue = {3},
  pages = {036034},
  numpages = {23},
  year = {2025},
  month = {Feb},
  publisher = {American Physical Society},
  doi = {10.1103/PhysRevD.111.036034},
  url = {https://link.aps.org/doi/10.1103/PhysRevD.111.036034}
}

@article{blaschke_oscillons_2025,
  title = {Oscillons from $Q$-Balls through Renormalization},
  author = {Blaschke, F. and Roma\ifmmode \acute{n}\else \'{n}\fi{}czukiewicz, T. and S\l{}awi\ifmmode \acute{n}\else \'{n}\fi{}ska, K. and Wereszczy\ifmmode \acute{n}\else \'{n}\fi{}ski, A.},
  journal = {Phys. Rev. Lett.},
  volume = {134},
  issue = {8},
  pages = {081601},
  numpages = {6},
  year = {2025},
  month = {Feb},
  publisher = {American Physical Society},
  doi = {10.1103/PhysRevLett.134.081601},
  url = {https://link.aps.org/doi/10.1103/PhysRevLett.134.081601}
}

@article{cyncynates_nonperturbative_2022,
  title = {Nonperturbative structure in coupled axion sectors and implications for direct detection},
  author = {Cyncynates, David and Simon, Olivier and Thompson, Jedidiah O. and Weiner, Zachary J.},
  journal = {Phys. Rev. D},
  volume = {106},
  issue = {8},
  pages = {083503},
  numpages = {20},
  year = {2022},
  month = {Oct},
  publisher = {American Physical Society},
  doi = {10.1103/PhysRevD.106.083503},
  url = {https://link.aps.org/doi/10.1103/PhysRevD.106.083503}
}

@article{antusch_parametric_2016,
doi = {10.1088/1475-7516/2016/02/044},
url = {https://dx.doi.org/10.1088/1475-7516/2016/02/044},
year = {2016},
month = {feb},
publisher = {},
volume = {2016},
number = {02},
pages = {044},
author = {Antusch, Stefan and Cefalà, Francesco and Nolde, David and Orani, Stefano},
title = {Parametric resonance after hilltop inflation caused by an inhomogeneous inflaton field},
journal = {Journal of Cosmology and Astroparticle Physics}
}

@article{kawasaki_i-ball_2014,
	title = {I-ball formation with logarithmic potential},
	volume = {2014},
	issn = {1475-7516},
	url = {https://dx.doi.org/10.1088/1475-7516/2014/07/038},
	doi = {10.1088/1475-7516/2014/07/038},
	language = {en},
	number = {07},
	urldate = {2025-07-11},
	journal = {Journal of Cosmology and Astroparticle Physics},
	author = {Kawasaki, Masahiro and Takeda, Naoyuki},
	month = jul,
	year = {2014},
	pages = {038},
}

@article{cotner_primordial_2018,
	title = {Primordial black holes from inflaton fragmentation into oscillons},
	volume = {98},
	url = {https://link.aps.org/doi/10.1103/PhysRevD.98.083513},
	doi = {10.1103/PhysRevD.98.083513},
	number = {8},
	urldate = {2024-05-15},
	journal = {Physical Review D},
	author = {Cotner, Eric and Kusenko, Alexander and Takhistov, Volodymyr},
	month = oct,
	year = {2018},
	note = {Publisher: American Physical Society},
	pages = {083513},
}

@article{cotner_analytic_2019,
	title = {Analytic description of primordial black hole formation from scalar field fragmentation},
	volume = {2019},
	issn = {1475-7516},
	url = {https://dx.doi.org/10.1088/1475-7516/2019/10/077},
	doi = {10.1088/1475-7516/2019/10/077},
	language = {en},
	number = {10},
	urldate = {2024-05-15},
	journal = {Journal of Cosmology and Astroparticle Physics},
	author = {Cotner, Eric and Kusenko, Alexander and Sasaki, Misao and Takhistov, Volodymyr},
	month = oct,
	year = {2019},
	pages = {077},
}

@article{kawasaki_oscillon_2020,
	title = {Oscillon of ultra-light axion-like particle},
	volume = {2020},
	issn = {1475-7516},
	url = {https://dx.doi.org/10.1088/1475-7516/2020/01/047},
	doi = {10.1088/1475-7516/2020/01/047},
	language = {en},
	number = {01},
	urldate = {2025-07-06},
	journal = {Journal of Cosmology and Astroparticle Physics},
	author = {Kawasaki, Masahiro and Nakano, Wakutaka and Sonomoto, Eisuke},
	month = jan,
	year = {2020},
	pages = {047},
}

@article{arvanitaki_large-misalignment_2020,
	title = {Large-misalignment mechanism for the formation of compact axion structures: {Signatures} from the {QCD} axion to fuzzy dark matter},
	volume = {101},
	shorttitle = {Large-misalignment mechanism for the formation of compact axion structures},
	url = {https://link.aps.org/doi/10.1103/PhysRevD.101.083014},
	doi = {10.1103/PhysRevD.101.083014},
	number = {8},
	urldate = {2025-07-06},
	journal = {Physical Review D},
	author = {Arvanitaki, Asimina and Dimopoulos, Savas and Galanis, Marios and Lehner, Luis and Thompson, Jedidiah O. and Van Tilburg, Ken},
	month = apr,
	year = {2020},
	note = {Publisher: American Physical Society},
	pages = {083014},
}

@article{kolb_nonlinear_1994,
	title = {Nonlinear axion dynamics and the formation of cosmological pseudosolitons},
	volume = {49},
	url = {https://link.aps.org/doi/10.1103/PhysRevD.49.5040},
	doi = {10.1103/PhysRevD.49.5040},
	number = {10},
	urldate = {2025-07-06},
	journal = {Physical Review D},
	author = {Kolb, Edward W. and Tkachev, Igor I.},
	month = may,
	year = {1994},
	note = {Publisher: American Physical Society},
	pages = {5040--5051},
}

@article{nazari_oscillon_2021,
	title = {Oscillon collapse to black holes},
	volume = {2021},
	issn = {1475-7516},
	url = {https://dx.doi.org/10.1088/1475-7516/2021/05/027},
	doi = {10.1088/1475-7516/2021/05/027},
	language = {en},
	number = {05},
	urldate = {2025-07-06},
	journal = {Journal of Cosmology and Astroparticle Physics},
	author = {Nazari, Zainab and Cicoli, Michele and Clough, Katy and Muia, Francesco},
	month = may,
	year = {2021},
	note = {Publisher: IOP Publishing},
	pages = {027},
}

@article{widdicombe_black_2020,
	title = {Black hole formation in relativistic {Oscillaton} collisions},
	volume = {2020},
	issn = {1475-7516},
	url = {https://dx.doi.org/10.1088/1475-7516/2020/01/027},
	doi = {10.1088/1475-7516/2020/01/027},
	language = {en},
	number = {01},
	urldate = {2025-07-06},
	journal = {Journal of Cosmology and Astroparticle Physics},
	author = {Widdicombe, James Y. and Helfer, Thomas and Lim, Eugene A.},
	month = jan,
	year = {2020},
	pages = {027},
}

@article{adib_long-lived_2002,
	title = {Long-lived oscillons from asymmetric bubbles: {Existence} and stability},
	volume = {66},
	shorttitle = {Long-lived oscillons from asymmetric bubbles},
	url = {https://link.aps.org/doi/10.1103/PhysRevD.66.085011},
	doi = {10.1103/PhysRevD.66.085011},
	number = {8},
	urldate = {2025-07-06},
	journal = {Physical Review D},
	author = {Adib, Artur B. and Gleiser, Marcelo and Almeida, Carlos A. S.},
	month = oct,
	year = {2002},
	note = {Publisher: American Physical Society},
	pages = {085011},
}

@article{shafi_formation_2024,
	title = {Formation and decay of oscillons after inflation in the presence of an external coupling. {Part} {I}. {Lattice} simulations},
	volume = {2024},
	issn = {1475-7516},
	url = {https://dx.doi.org/10.1088/1475-7516/2024/10/082},
	doi = {10.1088/1475-7516/2024/10/082},
	language = {en},
	number = {10},
	urldate = {2025-07-05},
	journal = {Journal of Cosmology and Astroparticle Physics},
	author = {Shafi, Mohammed and Copeland, Edmund J. and Mahbub, Rafid and Mishra, Swagat S. and Basak, Soumen},
	month = oct,
	year = {2024},
	note = {Publisher: IOP Publishing},
	pages = {082},
}

@article{amin_gravitational_2018,
	title = {Gravitational waves from asymmetric oscillon dynamics?},
	volume = {98},
	url = {https://link.aps.org/doi/10.1103/PhysRevD.98.024040},
	doi = {10.1103/PhysRevD.98.024040},
	number = {2},
	urldate = {2025-07-03},
	journal = {Physical Review D},
	author = {Amin, Mustafa A. and Braden, Jonathan and Copeland, Edmund J. and Giblin, John T. and Solorio, Christian and Weiner, Zachary J. and Zhou, Shuang-Yong},
	month = jul,
	year = {2018},
	note = {Publisher: American Physical Society},
	pages = {024040},
}

@article{antusch_gravitational_2017,
	title = {Gravitational {Waves} from {Oscillons} after {Inflation}},
	volume = {118},
	url = {https://link.aps.org/doi/10.1103/PhysRevLett.118.011303},
	doi = {10.1103/PhysRevLett.118.011303},
	number = {1},
	urldate = {2025-07-03},
	journal = {Physical Review Letters},
	author = {Antusch, Stefan and Cefalà, Francesco and Orani, Stefano},
	month = jan,
	year = {2017},
	note = {Publisher: American Physical Society},
	pages = {011303},
}

@article{piani_preheating_2023,
	title = {Preheating in {Einstein}-{Cartan} {Higgs} {Inflation}: {Oscillon} formation},
	volume = {2023},
	issn = {1475-7516},
	shorttitle = {Preheating in {Einstein}-{Cartan} {Higgs} {Inflation}},
	url = {http://arxiv.org/abs/2304.13056},
	doi = {10.1088/1475-7516/2023/12/002},
	number = {12},
	urldate = {2025-07-03},
	journal = {Journal of Cosmology and Astroparticle Physics},
	author = {Piani, Matteo and Rubio, Javier},
	month = dec,
	year = {2023},
	note = {arXiv:2304.13056 [hep-ph]},
	pages = {002},
}

@article{graham_electroweak_2007,
	title = {An {Electroweak} {Oscillon}},
	volume = {98},
	url = {https://link.aps.org/doi/10.1103/PhysRevLett.98.101801},
	doi = {10.1103/PhysRevLett.98.101801},
	number = {10},
	urldate = {2025-07-03},
	journal = {Physical Review Letters},
	author = {Graham, N.},
	month = mar,
	year = {2007},
	note = {Publisher: American Physical Society},
	pages = {101801},
}

@article{gleiser_analytical_2008,
	title = {Analytical {Characterization} of {Oscillon} {Energy} and {Lifetime}},
	volume = {101},
	url = {https://link.aps.org/doi/10.1103/PhysRevLett.101.011602},
	doi = {10.1103/PhysRevLett.101.011602},
	number = {1},
	urldate = {2025-07-03},
	journal = {Physical Review Letters},
	author = {Gleiser, Marcelo and Sicilia, David},
	month = jul,
	year = {2008},
	note = {Publisher: American Physical Society},
	pages = {011602},
}

@article{fodor_computation_2009,
	title = {Computation of the radiation amplitude of oscillons},
	volume = {79},
	url = {https://link.aps.org/doi/10.1103/PhysRevD.79.065002},
	doi = {10.1103/PhysRevD.79.065002},
	number = {6},
	urldate = {2025-07-03},
	journal = {Physical Review D},
	author = {Fodor, Gyula and Forgács, Péter and Horváth, Zalán and Mezei, Márk},
	month = mar,
	year = {2009},
	note = {Publisher: American Physical Society},
	pages = {065002},
}

@article{dimopoulos_n-flation_2008,
	title = {N-flation},
	volume = {2008},
	issn = {1475-7516},
	url = {https://dx.doi.org/10.1088/1475-7516/2008/08/003},
	doi = {10.1088/1475-7516/2008/08/003},
	language = {en},
	number = {08},
	urldate = {2025-07-03},
	journal = {Journal of Cosmology and Astroparticle Physics},
	author = {Dimopoulos, S and Kachru, S and McGreevy, J and Wacker, J G},
	month = aug,
	year = {2008},
	pages = {003},
}

@book{baumann_inflation_2015,
	address = {Cambridge},
	series = {Cambridge {Monographs} on {Mathematical} {Physics}},
	title = {Inflation and {String} {Theory}},
	isbn = {978-1-107-08969-3},
	url = {https://www.cambridge.org/core/books/inflation-and-string-theory/BBEE95783A1CB58FC0E8D5D8360AEBB2},
	urldate = {2025-07-03},
	publisher = {Cambridge University Press},
	author = {Baumann, Daniel and McAllister, Liam},
	year = {2015},
	doi = {10.1017/CBO9781316105733},
}

@article{linde_chaotic_1983,
	title = {Chaotic inflation},
	volume = {129},
	issn = {0370-2693},
	url = {https://www.sciencedirect.com/science/article/pii/0370269383908377},
	doi = {10.1016/0370-2693(83)90837-7},
	number = {3},
	urldate = {2025-07-03},
	journal = {Physics Letters B},
	author = {Linde, A. D.},
	month = sep,
	year = {1983},
	pages = {177--181},
}

@article{kofman_reheating_1994,
	title = {Reheating after {Inflation}},
	volume = {73},
	copyright = {http://link.aps.org/licenses/aps-default-license},
	issn = {0031-9007},
	url = {https://link.aps.org/doi/10.1103/PhysRevLett.73.3195},
	doi = {10.1103/PhysRevLett.73.3195},
	language = {en},
	number = {24},
	urldate = {2025-07-03},
	journal = {Physical Review Letters},
	author = {Kofman, Lev and Linde, Andrei and Starobinsky, Alexei A.},
	month = dec,
	year = {1994},
	pages = {3195--3198},
}

@article{lozanov_enhanced_2023,
	title = {Enhanced {Gravitational} {Waves} from {Inflaton} {Oscillons}},
	volume = {130},
	url = {https://link.aps.org/doi/10.1103/PhysRevLett.130.181002},
	doi = {10.1103/PhysRevLett.130.181002},
	number = {18},
	urldate = {2024-04-17},
	journal = {Physical Review Letters},
	author = {Lozanov, Kaloian D. and Takhistov, Volodymyr},
	month = may,
	year = {2023},
	note = {Publisher: American Physical Society},
	pages = {181002},
}

@article{amin_nonperturbative_2015,
	title = {Nonperturbative dynamics of reheating after inflation: {A} review},
	volume = {24},
	issn = {0218-2718},
	shorttitle = {Nonperturbative dynamics of reheating after inflation},
	url = {https://www.worldscientific.com/doi/abs/10.1142/S0218271815300037},
	doi = {10.1142/S0218271815300037},
	number = {01},
	urldate = {2024-10-11},
	journal = {International Journal of Modern Physics D},
	author = {Amin, Mustafa A. and Hertzberg, Mark P. and Kaiser, David I. and Karouby, Johanna},
	month = jan,
	year = {2015},
	note = {Publisher: World Scientific Publishing Co.},
	pages = {1530003},
}

@article{allahverdi_reheating_2010,
	title = {Reheating in {Inflationary} {Cosmology}: {Theory} and {Applications}},
	volume = {60},
	issn = {0163-8998, 1545-4134},
	shorttitle = {Reheating in {Inflationary} {Cosmology}},
	url = {https://www.annualreviews.org/content/journals/10.1146/annurev.nucl.012809.104511},
	doi = {10.1146/annurev.nucl.012809.104511},
	language = {en},
	number = {Volume 60, 2010},
	urldate = {2024-04-12},
	journal = {Annual Review of Nuclear and Particle Science},
	author = {Allahverdi, Rouzbeh and Brandenberger, Robert and Cyr-Racine, Francis-Yan and Mazumdar, Anupam},
	month = nov,
	year = {2010},
	note = {Publisher: Annual Reviews},
	pages = {27--51},
}

@article{garcia-bellido_preheating_1998,
	title = {Preheating in hybrid inflation},
	volume = {57},
	url = {https://link.aps.org/doi/10.1103/PhysRevD.57.6075},
	doi = {10.1103/PhysRevD.57.6075},
	number = {10},
	urldate = {2025-07-03},
	journal = {Physical Review D},
	author = {García-Bellido, Juan and Linde, Andrei},
	month = may,
	year = {1998},
	note = {Publisher: American Physical Society},
	pages = {6075--6088},
}

@article{linde_hybrid_1994,
	title = {Hybrid inflation},
	volume = {49},
	url = {https://link.aps.org/doi/10.1103/PhysRevD.49.748},
	doi = {10.1103/PhysRevD.49.748},
	number = {2},
	urldate = {2025-07-02},
	journal = {Physical Review D},
	author = {Linde, Andrei},
	month = jan,
	year = {1994},
	note = {Publisher: American Physical Society},
	pages = {748--754},
}

@article{gleiser_pseudostable_1994,
	title = {Pseudostable bubbles},
	volume = {49},
	copyright = {http://link.aps.org/licenses/aps-default-license},
	issn = {0556-2821},
	url = {https://link.aps.org/doi/10.1103/PhysRevD.49.2978},
	doi = {10.1103/PhysRevD.49.2978},
	language = {en},
	number = {6},
	urldate = {2025-06-20},
	journal = {Physical Review D},
	author = {Gleiser, Marcelo},
	month = mar,
	year = {1994},
	pages = {2978--2981},
}

@article{micha_turbulent_2004,
	title = {Turbulent thermalization},
	volume = {70},
	url = {https://link.aps.org/doi/10.1103/PhysRevD.70.043538},
	doi = {10.1103/PhysRevD.70.043538},
	number = {4},
	urldate = {2025-06-20},
	journal = {Physical Review D},
	author = {Micha, Raphael and Tkachev, Igor I.},
	month = aug,
	year = {2004},
	note = {Publisher: American Physical Society},
	pages = {043538},
}

@article{shtanov_universe_1995,
	title = {Universe reheating after inflation},
	volume = {51},
	url = {https://link.aps.org/doi/10.1103/PhysRevD.51.5438},
	doi = {10.1103/PhysRevD.51.5438},
	number = {10},
	urldate = {2025-06-20},
	journal = {Physical Review D},
	author = {Shtanov, Y. and Traschen, J. and Brandenberger, R.},
	month = may,
	year = {1995},
	note = {Publisher: American Physical Society},
	pages = {5438--5455},
}

@article{albrecht_cosmology_1982,
	title = {Cosmology for {Grand} {Unified} {Theories} with {Radiatively} {Induced} {Symmetry} {Breaking}},
	volume = {48},
	url = {https://link.aps.org/doi/10.1103/PhysRevLett.48.1220},
	doi = {10.1103/PhysRevLett.48.1220},
	number = {17},
	urldate = {2025-06-20},
	journal = {Physical Review Letters},
	author = {Albrecht, Andreas and Steinhardt, Paul J.},
	month = apr,
	year = {1982},
	note = {Publisher: American Physical Society},
	pages = {1220--1223},
}

@article{bassett_inflation_2006,
	title = {Inflation dynamics and reheating},
	volume = {78},
	url = {https://link.aps.org/doi/10.1103/RevModPhys.78.537},
	doi = {10.1103/RevModPhys.78.537},
	number = {2},
	urldate = {2025-06-20},
	journal = {Reviews of Modern Physics},
	author = {Bassett, Bruce A. and Tsujikawa, Shinji and Wands, David},
	month = may,
	year = {2006},
	note = {Publisher: American Physical Society},
	pages = {537--589},
}

@article{linde_new_1982,
	title = {A new inflationary universe scenario: {A} possible solution of the horizon, flatness, homogeneity, isotropy and primordial monopole problems},
	volume = {108},
	issn = {0370-2693},
	shorttitle = {A new inflationary universe scenario},
	url = {https://www.sciencedirect.com/science/article/pii/0370269382912199},
	doi = {10.1016/0370-2693(82)91219-9},
	number = {6},
	urldate = {2025-06-20},
	journal = {Physics Letters B},
	author = {Linde, A. D.},
	month = feb,
	year = {1982},
	pages = {389--393},
}

@article{guth_inflationary_1981,
	title = {Inflationary universe: {A} possible solution to the horizon and flatness problems},
	volume = {23},
	shorttitle = {Inflationary universe},
	url = {https://link.aps.org/doi/10.1103/PhysRevD.23.347},
	doi = {10.1103/PhysRevD.23.347},
	number = {2},
	urldate = {2025-06-20},
	journal = {Physical Review D},
	author = {Guth, Alan H.},
	month = jan,
	year = {1981},
	note = {Publisher: American Physical Society},
	pages = {347--356},
}

@article{kawasaki_oscillons_2021,
	title = {Oscillons of axion-like particle: mass distribution and power spectrum},
	volume = {2021},
	issn = {1475-7516},
	shorttitle = {Oscillons of axion-like particle},
	url = {https://dx.doi.org/10.1088/1475-7516/2021/01/061},
	doi = {10.1088/1475-7516/2021/01/061},
	language = {en},
	number = {01},
	urldate = {2025-06-04},
	journal = {Journal of Cosmology and Astroparticle Physics},
	author = {Kawasaki, Masahiro and Nakano, Wakutaka and Nakatsuka, Hiromasa and Sonomoto, Eisuke},
	month = jan,
	year = {2021},
	pages = {061},
}

@article{lozanov_universal_2025,
	title = {Universal gravitational wave signatures of cosmological solitons},
	volume = {2025},
	issn = {1475-7516},
	url = {https://dx.doi.org/10.1088/1475-7516/2025/01/094},
	doi = {10.1088/1475-7516/2025/01/094},
	language = {en},
	number = {01},
	urldate = {2025-04-04},
	journal = {Journal of Cosmology and Astroparticle Physics},
	author = {Lozanov, Kaloian D. and Sasaki, Misao and Takhistov, Volodymyr},
	month = jan,
	year = {2025},
	note = {Publisher: IOP Publishing},
	pages = {094},
}

@book{kevorkian_perturbation_1981,
	title = {Perturbation {Methods} in {Applied} {Mathematics}},
	isbn = {978-0-387-90507-5},
	language = {en},
	publisher = {Springer},
	author = {Kevorkian, J. and Cole, J. D.},
	year = {1981},
	note = {Google-Books-ID: pT3vAAAAMAAJ},
}

@book{mclachlan_theory_1947,
	title = {Theory and {Application} of {Mathieu} {Functions}},
	language = {en},
	publisher = {Clarendon Press},
	author = {McLachlan, Norman William},
	year = {1947},
	note = {Google-Books-ID: xErX0AEACAAJ},
}

@article{floquet_sur_1883,
	title = {Sur les équations différentielles linéaires à coefficients périodiques},
	volume = {12},
	issn = {1873-2151},
	url = {http://www.numdam.org/item/?id=ASENS_1883_2_12__47_0},
	doi = {10.24033/asens.220},
	language = {en},
	urldate = {2024-12-22},
	journal = {Annales scientifiques de l'École Normale Supérieure},
	author = {Floquet, G.},
	year = {1883},
	pages = {47--88},
}

@article{yoshida_construction_1990,
	title = {Construction of higher order symplectic integrators},
	volume = {150},
	issn = {0375-9601},
	url = {https://www.sciencedirect.com/science/article/pii/0375960190900923},
	doi = {10.1016/0375-9601(90)90092-3},
	number = {5},
	urldate = {2024-12-19},
	journal = {Physics Letters A},
	author = {Yoshida, Haruo},
	month = nov,
	year = {1990},
	pages = {262--268},
}

@article{kou_oscillon_2020,
	title = {Oscillon preheating in full general relativity},
	volume = {38},
	issn = {0264-9381},
	url = {https://dx.doi.org/10.1088/1361-6382/abd09f},
	doi = {10.1088/1361-6382/abd09f},
	language = {en},
	number = {4},
	urldate = {2024-11-22},
	journal = {Classical and Quantum Gravity},
	author = {Kou, Xiao-Xiao and Tian, Chi and Zhou, Shuang-Yong},
	month = dec,
	year = {2020},
	note = {Publisher: IOP Publishing},
	pages = {045005},
}

@article{lozanov_end_2014,
	title = {End of inflation, oscillons, and matter-antimatter asymmetry},
	volume = {90},
	url = {https://link.aps.org/doi/10.1103/PhysRevD.90.083528},
	doi = {10.1103/PhysRevD.90.083528},
	number = {8},
	urldate = {2024-10-21},
	journal = {Physical Review D},
	author = {Lozanov, Kaloian D. and Amin, Mustafa A.},
	month = oct,
	year = {2014},
	note = {Publisher: American Physical Society},
	pages = {083528},
}

@article{lozanov_expansion_2017,
	title = {{EXPANSION} {AFTER} {INFLATION} {AND} {REHEATING} {WITH} {A} {CHARGED} {INFLATON}},
	url = {https://www.repository.cam.ac.uk/handle/1810/267822},
	doi = {10.17863/CAM.13746},
	language = {en},
	urldate = {2024-10-11},
	author = {Lozanov, Kaloian Dimitrov},
	month = oct,
	year = {2017},
}

@article{farhi_emergence_2008,
	title = {Emergence of oscillons in an expanding background},
	volume = {77},
	url = {https://link.aps.org/doi/10.1103/PhysRevD.77.085019},
	doi = {10.1103/PhysRevD.77.085019},
	number = {8},
	urldate = {2024-10-01},
	journal = {Physical Review D},
	author = {Farhi, E. and Graham, N. and Guth, A. H. and Iqbal, N. and Rosales, R. R. and Stamatopoulos, N.},
	month = apr,
	year = {2008},
	note = {Publisher: American Physical Society},
	pages = {085019},
}

@article{multamaki_analytical_2000,
	title = {Analytical and numerical properties of {Q}-balls},
	volume = {574},
	issn = {0550-3213},
	url = {https://www.sciencedirect.com/science/article/pii/S0550321399008275},
	doi = {10.1016/S0550-3213(99)00827-5},
	number = {1},
	urldate = {2024-09-25},
	journal = {Nuclear Physics B},
	author = {Multamäki, Tuomas and Vilja, Iiro},
	month = may,
	year = {2000},
	pages = {130--152},
}

@article{zhang_gravitational_2021,
	title = {Gravitational effects on oscillon lifetimes},
	volume = {2021},
	issn = {1475-7516},
	url = {https://dx.doi.org/10.1088/1475-7516/2021/03/102},
	doi = {10.1088/1475-7516/2021/03/102},
	language = {en},
	number = {03},
	urldate = {2024-07-30},
	journal = {Journal of Cosmology and Astroparticle Physics},
	author = {Zhang, Hong-Yi},
	month = mar,
	year = {2021},
	note = {Publisher: IOP Publishing},
	pages = {102},
}

@article{kawasaki_decay_2014,
	title = {Decay rates of {Gaussian}-type {I}-balls and {Bose}-enhancement effects in 3+1 dimensions},
	volume = {2014},
	issn = {1475-7516},
	url = {https://dx.doi.org/10.1088/1475-7516/2014/02/001},
	doi = {10.1088/1475-7516/2014/02/001},
	language = {en},
	number = {02},
	urldate = {2024-07-25},
	journal = {Journal of Cosmology and Astroparticle Physics},
	author = {Kawasaki, Masahiro and Yamada, Masaki},
	month = feb,
	year = {2014},
	pages = {001},
}

@article{zhou_gravitational_2013,
	title = {Gravitational waves from oscillon preheating},
	volume = {2013},
	issn = {1029-8479},
	url = {https://doi.org/10.1007/JHEP10(2013)026},
	doi = {10.1007/JHEP10(2013)026},
	language = {en},
	number = {10},
	urldate = {2024-07-25},
	journal = {Journal of High Energy Physics},
	author = {Zhou, Shuang-Yong and Copeland, Edmund J. and Easther, Richard and Finkel, Hal and Mou, Zong-Gang and Saffin, Paul M.},
	month = oct,
	year = {2013},
	pages = {26},
}

@article{cyncynates_structure_2021,
	title = {Structure of the oscillon: {The} dynamics of attractive self-interaction},
	volume = {103},
	shorttitle = {Structure of the oscillon},
	url = {https://link.aps.org/doi/10.1103/PhysRevD.103.116011},
	doi = {10.1103/PhysRevD.103.116011},
	number = {11},
	urldate = {2024-07-03},
	journal = {Physical Review D},
	author = {Cyncynates, David and Giurgica-Tiron, Tudor},
	month = jun,
	year = {2021},
	note = {Publisher: American Physical Society},
	pages = {116011},
}

@article{copeland_oscillons_1995,
	title = {Oscillons: {Resonant} configurations during bubble collapse},
	volume = {52},
	shorttitle = {Oscillons},
	url = {https://link.aps.org/doi/10.1103/PhysRevD.52.1920},
	doi = {10.1103/PhysRevD.52.1920},
	number = {4},
	urldate = {2024-07-02},
	journal = {Physical Review D},
	author = {Copeland, E. J. and Gleiser, M. and Müller, H.-R.},
	month = aug,
	year = {1995},
	note = {Publisher: American Physical Society},
	pages = {1920--1933},
}

@article{gleiser_long-lived_2000,
	title = {Long-lived localized field configurations in small lattices: {Application} to oscillons},
	volume = {62},
	shorttitle = {Long-lived localized field configurations in small lattices},
	url = {https://link.aps.org/doi/10.1103/PhysRevE.62.1368},
	doi = {10.1103/PhysRevE.62.1368},
	number = {1},
	urldate = {2024-07-01},
	journal = {Physical Review E},
	author = {Gleiser, M. and Sornborger, A.},
	month = jul,
	year = {2000},
	note = {Publisher: American Physical Society},
	pages = {1368--1374},
}

@article{lee_nontopological_1992,
	title = {Nontopological solitons},
	volume = {221},
	issn = {0370-1573},
	url = {https://www.sciencedirect.com/science/article/pii/0370157392900647},
	doi = {10.1016/0370-1573(92)90064-7},
	number = {5},
	urldate = {2024-06-13},
	journal = {Physics Reports},
	author = {Lee, T. D and Pang, Y},
	month = nov,
	year = {1992},
	pages = {251--350},
}

@article{kovacic_mathieus_2018,
	title = {Mathieu's {Equation} and {Its} {Generalizations}: {Overview} of {Stability} {Charts} and {Their} {Features}},
	volume = {70},
	issn = {0003-6900},
	shorttitle = {Mathieu's {Equation} and {Its} {Generalizations}},
	url = {https://doi.org/10.1115/1.4039144},
	doi = {10.1115/1.4039144},
	number = {020802},
	urldate = {2024-05-24},
	journal = {Applied Mechanics Reviews},
	author = {Kovacic, Ivana and Rand, Richard and Mohamed Sah, Si},
	month = feb,
	year = {2018},
}

@article{amin_flat-top_2010,
	title = {Flat-top oscillons in an expanding universe},
	volume = {81},
	url = {https://link.aps.org/doi/10.1103/PhysRevD.81.085045},
	doi = {10.1103/PhysRevD.81.085045},
	number = {8},
	urldate = {2024-05-09},
	journal = {Physical Review D},
	author = {Amin, Mustafa A. and Shirokoff, David},
	month = apr,
	year = {2010},
	note = {Publisher: American Physical Society},
	pages = {085045},
}

@article{aurrekoetxea_oscillon_2023,
	title = {Oscillon formation during inflationary preheating with general relativity},
	volume = {108},
	url = {https://link.aps.org/doi/10.1103/PhysRevD.108.023501},
	doi = {10.1103/PhysRevD.108.023501},
	number = {2},
	urldate = {2024-04-17},
	journal = {Physical Review D},
	author = {Aurrekoetxea, Josu C. and Clough, Katy and Muia, Francesco},
	month = jul,
	year = {2023},
	note = {Publisher: American Physical Society},
	pages = {023501},
}

@article{lozanov_gravitational_2019,
	title = {Gravitational perturbations from oscillons and transients after inflation},
	volume = {99},
	issn = {2470-0010, 2470-0029},
	url = {https://link.aps.org/doi/10.1103/PhysRevD.99.123504},
	doi = {10.1103/PhysRevD.99.123504},
	language = {en},
	number = {12},
	urldate = {2024-04-13},
	journal = {Physical Review D},
	author = {Lozanov, Kaloian D. and Amin, Mustafa A.},
	month = jun,
	year = {2019},
	pages = {123504},
}

@article{olle_oscillons_2020,
	title = {Oscillons and dark matter},
	volume = {2020},
	issn = {1475-7516},
	url = {https://dx.doi.org/10.1088/1475-7516/2020/02/006},
	doi = {10.1088/1475-7516/2020/02/006},
	language = {en},
	number = {02},
	urldate = {2024-04-13},
	journal = {Journal of Cosmology and Astroparticle Physics},
	author = {Ollé, Jan and Pujolàs, Oriol and Rompineve, Fabrizio},
	month = feb,
	year = {2020},
	pages = {006},
}

@article{antusch_properties_2019,
	title = {Properties of oscillons in hilltop potentials: energies, shapes, and lifetimes},
	volume = {2019},
	issn = {1475-7516},
	shorttitle = {Properties of oscillons in hilltop potentials},
	url = {https://dx.doi.org/10.1088/1475-7516/2019/10/002},
	doi = {10.1088/1475-7516/2019/10/002},
	language = {en},
	number = {10},
	urldate = {2024-04-13},
	journal = {Journal of Cosmology and Astroparticle Physics},
	author = {Antusch, Stefan and Cefalà, Francesco and Torrentí, Francisco},
	month = oct,
	year = {2019},
	pages = {002},
}

@article{kofman_towards_1997,
	title = {Towards the theory of reheating after inflation},
	volume = {56},
	url = {https://link.aps.org/doi/10.1103/PhysRevD.56.3258},
	doi = {10.1103/PhysRevD.56.3258},
	number = {6},
	urldate = {2024-03-29},
	journal = {Physical Review D},
	author = {Kofman, Lev and Linde, Andrei and Starobinsky, Alexei A.},
	month = sep,
	year = {1997},
	note = {Publisher: American Physical Society},
	pages = {3258--3295},
}

@misc{amin_inflaton_2010,
	title = {Inflaton fragmentation: {Emergence} of pseudo-stable inflaton lumps (oscillons) after inflation},
	shorttitle = {Inflaton fragmentation},
	url = {http://arxiv.org/abs/1006.3075},
	urldate = {2024-03-28},
	publisher = {arXiv},
	author = {Amin, Mustafa A.},
	month = sep,
	year = {2010},
	note = {arXiv:1006.3075 [astro-ph, physics:hep-ph, physics:hep-th]},
}

@article{vakhitov_stationary_1973,
	title = {Stationary solutions of the wave equation in a medium with nonlinearity saturation},
	volume = {16},
	issn = {1573-9120},
	url = {https://doi.org/10.1007/BF01031343},
	doi = {10.1007/BF01031343},
	language = {en},
	number = {7},
	urldate = {2024-03-15},
	journal = {Radiophysics and Quantum Electronics},
	author = {Vakhitov, N. G. and Kolokolov, A. A.},
	month = jul,
	year = {1973},
	pages = {783--789},
}

@article{fodor_radiation_2009,
	title = {Radiation of scalar oscillons in 2 and 3 dimensions},
	volume = {674},
	issn = {0370-2693},
	url = {https://www.sciencedirect.com/science/article/pii/S0370269309003426},
	doi = {10.1016/j.physletb.2009.03.054},
	number = {4},
	urldate = {2024-03-12},
	journal = {Physics Letters B},
	author = {Fodor, Gyula and Forgacs, Péter and Horvath, Zalan and Mezei, Mark},
	month = apr,
	year = {2009},
	pages = {319--324},
}

@article{lozanov_self-resonance_2018,
	title = {Self-resonance after inflation: {Oscillons}, transients, and radiation domination},
	volume = {97},
	shorttitle = {Self-resonance after inflation},
	url = {https://link.aps.org/doi/10.1103/PhysRevD.97.023533},
	doi = {10.1103/PhysRevD.97.023533},
	number = {2},
	urldate = {2024-03-12},
	journal = {Physical Review D},
	author = {Lozanov, Kaloian D. and Amin, Mustafa A.},
	month = jan,
	year = {2018},
	note = {Publisher: American Physical Society},
	pages = {023533},
}

@article{kasuya_i-balls_2003,
	title = {I-balls},
	volume = {559},
	issn = {03702693},
	url = {https://linkinghub.elsevier.com/retrieve/pii/S0370269303003447},
	doi = {10.1016/S0370-2693(03)00344-7},
	language = {en},
	number = {3-4},
	urldate = {2023-09-08},
	journal = {Physics Letters B},
	author = {Kasuya, S. and Kawasaki, M. and Takahashi, Fuminobu},
	month = may,
	year = {2003},
	pages = {99--106},
}

@article{gleiser_generation_2011,
	title = {Generation of coherent structures after cosmic inflation},
	volume = {83},
	issn = {1550-7998, 1550-2368},
	url = {https://link.aps.org/doi/10.1103/PhysRevD.83.096010},
	doi = {10.1103/PhysRevD.83.096010},
	language = {en},
	number = {9},
	urldate = {2024-02-02},
	journal = {Physical Review D},
	author = {Gleiser, Marcelo and Graham, Noah and Stamatopoulos, Nikitas},
	month = may,
	year = {2011},
	pages = {096010},
}

@article{evslin_quantum_2023,
	title = {Quantum oscillons may be long-lived},
	volume = {2023},
	issn = {1029-8479},
	url = {https://link.springer.com/10.1007/JHEP08(2023)182},
	doi = {10.1007/JHEP08(2023)182},
	language = {en},
	number = {8},
	urldate = {2024-01-09},
	journal = {Journal of High Energy Physics},
	author = {Evslin, Jarah and Romańczukiewicz, Tomasz and Wereszczyński, Andrzej},
	month = aug,
	year = {2023},
	pages = {182},
}

@article{farhi_oscillon_2005,
	title = {An oscillon in the {S} {U} ( 2 ) gauged {Higgs} model},
	volume = {72},
	issn = {1550-7998, 1550-2368},
	url = {https://link.aps.org/doi/10.1103/PhysRevD.72.101701},
	doi = {10.1103/PhysRevD.72.101701},
	language = {en},
	number = {10},
	urldate = {2023-12-21},
	journal = {Physical Review D},
	author = {Farhi, E. and Graham, N. and Khemani, V. and Markov, R. and Rosales, R.},
	month = nov,
	year = {2005},
	pages = {101701},
}

@article{amin_oscillons_2012,
	title = {Oscillons after {Inflation}},
	volume = {108},
	issn = {0031-9007, 1079-7114},
	url = {https://link.aps.org/doi/10.1103/PhysRevLett.108.241302},
	doi = {10.1103/PhysRevLett.108.241302},
	language = {en},
	number = {24},
	urldate = {2024-01-08},
	journal = {Physical Review Letters},
	author = {Amin, Mustafa A. and Easther, Richard and Finkel, Hal and Flauger, Raphael and Hertzberg, Mark P.},
	month = jun,
	year = {2012},
	pages = {241302},
}

@article{salmi_radiation_2012,
	title = {Radiation and relaxation of oscillons},
	volume = {85},
	issn = {1550-7998, 1550-2368},
	url = {https://link.aps.org/doi/10.1103/PhysRevD.85.085033},
	doi = {10.1103/PhysRevD.85.085033},
	language = {en},
	number = {8},
	urldate = {2023-11-27},
	journal = {Physical Review D},
	author = {Salmi, Petja and Hindmarsh, Mark},
	month = apr,
	year = {2012},
	pages = {085033},
}

@article{mukaida_longevity_2017,
	title = {On longevity of {I}-ball/oscillon},
	volume = {2017},
	issn = {1029-8479},
	url = {https://doi.org/10.1007/JHEP03(2017)122},
	doi = {10.1007/JHEP03(2017)122},
	language = {en},
	number = {3},
	urldate = {2023-11-29},
	journal = {Journal of High Energy Physics},
	author = {Mukaida, Kyohei and Takimoto, Masahiro and Yamada, Masaki},
	month = mar,
	year = {2017},
	pages = {122},
}

@article{mukaida_correspondence_2014,
	title = {Correspondence of {I}- and {Q}-balls as {Non}-relativistic {Condensates}},
	volume = {2014},
	issn = {1475-7516},
	url = {http://arxiv.org/abs/1405.3233},
	doi = {10.1088/1475-7516/2014/08/051},
	number = {08},
	urldate = {2023-09-13},
	journal = {Journal of Cosmology and Astroparticle Physics},
	author = {Mukaida, Kyohei and Takimoto, Masahiro},
	month = aug,
	year = {2014},
	note = {arXiv:1405.3233 [astro-ph, physics:hep-ph, physics:hep-th]},
	pages = {051--051},
}

@article{olle_recipes_2021,
	title = {Recipes for {Oscillon} {Longevity}},
	volume = {2021},
	issn = {1475-7516},
	url = {http://arxiv.org/abs/2012.13409},
	doi = {10.1088/1475-7516/2021/09/015},
	number = {09},
	urldate = {2024-03-07},
	journal = {Journal of Cosmology and Astroparticle Physics},
	author = {Olle, Jan and Pujolas, Oriol and Rompineve, Fabrizio},
	month = sep,
	year = {2021},
	note = {arXiv:2012.13409 [astro-ph, physics:hep-ph, physics:hep-th]},
	pages = {015},
}

@article{van_dissel_symmetric_2022,
	title = {Symmetric multifield oscillons},
	volume = {106},
	issn = {2470-0010, 2470-0029},
	url = {https://link.aps.org/doi/10.1103/PhysRevD.106.096018},
	doi = {10.1103/PhysRevD.106.096018},
	language = {en},
	number = {9},
	urldate = {2024-02-02},
	journal = {Physical Review D},
	author = {Van Dissel, Fabio and Sfakianakis, Evangelos I.},
	month = nov,
	year = {2022},
	pages = {096018},
}

@article{hertzberg_quantum_2010,
	title = {Quantum radiation of oscillons},
	volume = {82},
	issn = {1550-7998, 1550-2368},
	url = {https://link.aps.org/doi/10.1103/PhysRevD.82.045022},
	doi = {10.1103/PhysRevD.82.045022},
	language = {en},
	number = {4},
	urldate = {2023-12-19},
	journal = {Physical Review D},
	author = {Hertzberg, Mark P.},
	month = aug,
	year = {2010},
	pages = {045022},
}

@article{van_dissel_oscillon_2023,
	title = {Oscillon spectroscopy},
	volume = {2023},
	issn = {1029-8479},
	url = {https://doi.org/10.1007/JHEP07(2023)194},
	doi = {10.1007/JHEP07(2023)194},
	language = {en},
	number = {7},
	urldate = {2024-02-02},
	journal = {Journal of High Energy Physics},
	author = {van Dissel, Fabio and Pujolàs, Oriol and Sfakianakis, Evangelos I.},
	month = jul,
	year = {2023},
	pages = {194},
}

@article{ibe_fragileness_2019,
	title = {Fragileness of exact {I} -ball/oscillon},
	volume = {100},
	issn = {2470-0010, 2470-0029},
	url = {https://link.aps.org/doi/10.1103/PhysRevD.100.125021},
	doi = {10.1103/PhysRevD.100.125021},
	language = {en},
	number = {12},
	urldate = {2024-03-11},
	journal = {Physical Review D},
	author = {Ibe, Masahiro and Kawasaki, Masahiro and Nakano, Wakutaka and Sonomoto, Eisuke},
	month = dec,
	year = {2019},
	pages = {125021},
}

@article{ibe_decay_2019,
	title = {Decay of {I}-ball/oscillon in classical field theory},
	volume = {2019},
	issn = {1029-8479},
	url = {https://doi.org/10.1007/JHEP04(2019)030},
	doi = {10.1007/JHEP04(2019)030},
	language = {en},
	number = {4},
	urldate = {2023-09-13},
	journal = {Journal of High Energy Physics},
	author = {Ibe, Masahiro and Kawasaki, Masahiro and Nakano, Wakutaka and Sonomoto, Eisuke},
	month = apr,
	year = {2019},
	pages = {30},
}

@article{zhang_classical_2020,
	title = {Classical decay rates of oscillons},
	volume = {2020},
	issn = {1475-7516},
	url = {https://iopscience.iop.org/article/10.1088/1475-7516/2020/07/055},
	doi = {10.1088/1475-7516/2020/07/055},
	language = {en},
	number = {07},
	urldate = {2024-02-02},
	journal = {Journal of Cosmology and Astroparticle Physics},
	author = {Zhang, Hong-Yi and Amin, Mustafa A. and Copeland, Edmund J. and Saffin, Paul M. and Lozanov, Kaloian D.},
	month = jul,
	year = {2020},
	pages = {055--055},
}

@article{hiramatsu_gravitational_2021,
	title = {Gravitational wave spectra from oscillon formation after inflation},
	volume = {2021},
	issn = {1029-8479},
	url = {http://link.springer.com/10.1007/JHEP03(2021)021},
	doi = {10.1007/JHEP03(2021)021},
	language = {en},
	number = {3},
	urldate = {2024-01-08},
	journal = {Journal of High Energy Physics},
	author = {Hiramatsu, Takashi and Sfakianakis, Evangelos I. and Yamaguchi, Masahide},
	month = mar,
	year = {2021},
	pages = {21},
}

\end{document}